**Emergence time, curvature, space, causality, and complexity in encoding a discrete impulse information process**

Vladimir S. Lerner, USA, lernervs@gmail.com

**Abstract**

Interactions build structure of Universe, and impulse is elementary interactive *discrete* action ↓↑ modeling standard unit of information Bit in any natural process, including variety of micro and macro objects and systems.

We study how this *recursive inter-actions, independently of its physical nature*, particles, and interacting objects, becomes an information Bit during observation of random process of interacting impulses, and how in the observation emerges the impulse time and space intervals, a microprocess, and information macroprocess of the multiple bits.

Interacting random field of Kolmogorov probabilities links Kolmogorov law's 0-1 probabilities and Bayesian probabilities observing Markov diffusion process under Yes-No actions of random impulse. These objective Yes-No probabilities measure virtual probing impulses which, processing the interactions, generate an idealized (virtual) probability measurement in observable process. The impulse observation increases each posteriori correlation reducing conditional entropy measures from a finite uncertainty up to certainty of real impulse. The reduced entropy conveys probabilistic causality with time course along the process and temporal memory in collecting correlations, which interactive impulse innately cuts exposing hidden process' entropy. The natural cut of this entropy reveals information hidden in the correlation connections. Inside the merging probing impulse emerges reversible microprocess with yes-no conjugated entangled entropy, curvature and logical complexity. Within the impulse time interval, *entanglement starts before its space is formed and ends with beginning the space during reversible relative time interval of* $0.015625\pi$ part of $\pi$ with time interval $\tau = 1nat$.

The merging impulse curves and rotates the impulse interactive actions in the microprocess whose space interval measures this transitive movement. The cutting entropy sequentially converting to information memorizes the probes logic in Bit, participating in next probe conversions and encoding which memorizes information causality. The curvature, enclosing the entropy of interacting impulses, enables converting energy of an external process to Bit of the interacting process, which memorizes the bit by delivering the Landauer energy, and/or memorizing correlation of entangled qubits. In quantum mechanics any encoding random qubits requires retrieval, erasure or copying, measurements, leading to energy cost for getting physical information, while logic in reversible computation does not cost energy. Sequential interactive cuts integrate the cutting information in information macroprocess with irreversible time course. The topological transitive gap separates the micro- and macroprocess on an edge of reality. The memorized information binds reversible microprocess within impulse with irreversible information macroprocess cooperating the triple bits units. The emerging curved space encloses cooperative complexity of the macroprocess' triple bits rotating in attractive movement whose complexity measures information mass. The complexity and mass appears after the space emerges from the entanglement. The observing logical operations with triplet information units achieve a goal, integrating the discrete information hidden in the cutting correlations in information structure of Observer, enclosing nested information networks (IN) with the hierarchy of quality information, encoded in triplet code. Emerging self-organization of the observer information self-creates law of evolution dynamics toward cognition and intelligence. The observer's cognition assembles the common units through the multiple attraction and resonances at forming the IN-triplet hierarchy, which accept only units that concentrates and recognizes each IN node.

The ended triplet of a cooperating hierarchical INs measures level of the observer intelligence. The synthesized optimal process minimizes the observations time in Artificial designed information Observer with intellectual searching logic.

The described information equations finalize main results, validate them analytically and numerically.

**Introduction**

Origin of time, space and information is actual scientific problem in physics and philosophy that has not solved yet.

Since information originates in quantum process with conjugated probabilities, its study, focusing not on physics of observing process particles but on information-theoretical essence, allows formal description of information process from probabilistic observation, emerging time, microprocess, entanglement, space, qubits, encoding, and evolving macroprocess.

This information approach develops Wheeler's concept of Bit as Observer-Participator, which extends it from probabilistic logic, emerging space up to Observer information geometrical structure toward it cognition and intellect.

Mathematical analysis of a basic "anatomy" of the interacting impulse reveals scientifically conclusive notion of information as phenomenon, originating in probabilistic entropy of hidden correlation whose interactive cut-erasure produces physical information Bit without necessity of particles in Physics.

The paper organizes in following parts, sections, and subsections:

**Part I. Emerging time, curvature, space, causality, information, and complexity in observing process of interactive impulses**

Section 1. Connects the problems in quantum mechanics and the impulse information process.

Section 2. Explains essentials of emerging time, curvature, space, causality, information, and complexity.

Section 3. Analyses an "anatomy" of the impulse $\downarrow\uparrow$ as model of information bit and introduces basic mathematical foundation with entropy integral functional (EF) on random trajectories, information path functional (IPF), and minimax variation principle, which includes subsection:

3.1. The entropy regular integral functional.

3.2. Application of the minimax variation principle to this functional.

3.3. Dirac's delta-function' impulse action on the entropy integral cutting it on discreet information measures.

3.4. Introduces class of the impulse opposite discrete functions, which preserve the Markov diffusion process' additive and multiplicative measures within the cutting process of each impulse. Actions these functions on the entropy functional determine the impulse finite entropy measures 1 Nat, invariant size of the impulse $\pi$, and evaluates the entropy functional's contributions collected on the intervals actions, satisfying variation principle.

The contributions at each cutting interval determine the time distance interval, when each multiplicative entropy increment supplies each impulse information increment, and the density information measure of each impulse.

3.5. Determines Information Path functional in $n$-dimensional Markov process under $n$-cutoff discrete impulses, which integrates sequence of the discrete impulse with distinct information densities.

The IPF concentrates the measured information in the integrated Bit, and measures time intervals of integrations. It evaluates discrete correlation functions arising between the cutting actions, increments of the impulse correlations, the entropy functional contributions between the interacting impulses on an edge of each impulse, and information compensating the interactive entropy.

The results specify the impulse process of capturing an external entropy influx in interaction of the nearest impulses.

3.6. Introduces microprocess within a merging impulse, which includes

3.6.1. The entropy increments under starting step-functions initiating the microprocess;

3.6.2. The microprocess' conjugated dynamics within the impulse satisfying the additive and/or multiplicative measures. Entangling the conjugated entropy increments and its analytical evaluation in the microprocess;

3.6.3. Probabilities functions of the microprocess;

3.6.4. The relation between the curved time intervals and equivalent space length within an impulse;

3.6.4.1. Curvature measures of the impulse;

3.6.5. The observation edge at the end of uncertainty within the entropy-information gap;

3. 6.5.1. Random, quantum, and a mean time intervals;

**Part II. The emerging self-organization of observer information in evolution dynamics toward intelligence**

Section 1. The stages and levels of the emerging observer self-organization of information and the evolving dynamic regularities, showing that the emergence of time, space, and information follows the emerging evolution information dynamics creating multiple evolving observers with intellect and mechanism of cognition.

Section 2. Analytical and numerical attributes distinguishing the main stages of the evolving regularities, their thresholds and constraints. Evaluates Kolmogorov-Bayes probability links connecting Markov correlations, identifies the probabilistic location of emerging microprocess within observing Markov process.

Math description of the observation process, the emerging processes' regularities, and the evolution stages formulizes and logically verifies the approach.

Summary of the emerging nested evolutionary levels in impulse observation self-creating law of evolution.

**I. Emerging time, curvature, space, causality, information, and complexity in observing process of interactive impulses**

**1. Quantum mechanics and impulse information process**

Since information initially originates in quantum process with conjugated probabilities, its study should focusing not on physics of observing process' particles but on its information-theoretical essence. The impulses *discrete* Yes-No or 1-0 actions enable model standard unit of information a Bit without necessity of any particles in QM.

Multiple impulses interact in an observing process generating a *random process of interactive impulses*.

The series of observing random impulses are uncertain before measuring and acquisition.

We study the observation of random impulses using probabilistic approach beginning with interacting random field of Kolmogorov probabilities, linking Kolmogorov 0-1 law's probabilities and Bayesian probabilities which are observing Markov diffusion process. The Yes-No impulse observation increases each posteriori correlation and reducing the conditional entropy measures. These objective Yes-No probabilities measure virtual probing impulses which, processing the interactions, generate an idealized (virtual) probability measurement from a finite uncertainty measuring the conditional entropy minimizing up to certainty in the observable process. This introduces *notion of certainty* as opposite to uncertainty, which measures the posteriori probability approaching 1 when the entropy erases.

A path of the observation from different sources may contain various *physical phenomena*, whose specifics are still mostly unknown, while the probabilistic observation can bring information about these phenomena by observing just their interactions and creating information process on the path.

The *quantum mechanics* (QM) of moving particles [2-3] govern E.Schrodinger equation, which R. Feynman latter confirmed analytically from variation principle for Feynman's path functional on trajectories of the particles [4].

The observable movement before measurement (interaction) describes complex pair of conjugated wave functions with the probability amplitude of position, momentum, and other physical properties of a particle. The superimposing conjugated probability *amplitudes* describe the wave packet with the particle states' probable superposition.

The probabilistic observable particles have *uncertain* values of a particle's position and momentum or its energy and time up to the measurement which *defines their* values. The Bayes probability approach to QM is introduced in [5].

*In the QM entanglement* phenomenon, the quantum states (objects) connect the mutual conjugated complex probability amplitudes*;* the probabilistic particles cohere in the superposition via the quantum correlation.

The entangle pair may produce random objects with equal probability ½ as qubit, or the correlation is source of multiple qubits, bits at other probabilities.

*Computing* is various encoding the information in the state of a physical system through a physically realizable device.

Encoding a Bit extracts an asymmetrical time-space position of its no-yes actions, which erases their position and produces the memorized Bit of information.

According to Landauer principle [6], any logically irreversible operation such as encoding leads to irreversible process which should spend entropy $S = k \ln 2$ on the erasure.

Maxwell demon [7] associates the entropy of Landauer's principle with the acquisition of one information Bit equal to that entropy-as an inevitable cost for the erasure. The acquisition of one bit requires at least work $W = kT \ln 2$ which theoretically *limits the finite energy recourses or the time of performing such operation.*

Bennett [8] found that any computation can be performed using only reversible steps, and in principle it requires no dissipation and no power spending. However, specific reversible computer needs to reproduce the map of inputs to outputs, erasing everything else that requires the energy cost.

Quantum information cannot be copied with perfect accuracy by no-cloning principle of Wootters, Zurek, Dieks [9].

Making perfect copy of a quantum state measures observable in the copy, which disturbs the original states of the system.

Acquiring information and creating the disturbance relates to quantum randomness and the errors requiring erasing.

Bell [10, 11] shows that predictions of quantum mechanics cannot be reproduced by any local hidden variable theory, but quantum information can be encoded in nonlocal correlations between the different parts of a physical system, which have interacted and then separated.

In Shor algorithm [12], the properties of the correlations between the "input register" and "output register" of quantum computer functions require huge memory to store. This nonlocal information is hard to decode, time of $n$ operations increase grows faster than any power of $\ln(n)$. Practical computations either *run out of time or run out of memory*.

Are there a way to merge memory with the time of memorizing information?

Since any encoding random (uncertain) qubits requires retrieval-erasure and/or copying, measurements, it leads to the energy cost for getting physical information. While a logic in a reversible computation does not cost energy.

Transitive movement of correlated qubits during a finite cutting off tunnel breaks their partial symmetry and reversibility, producing strongly quantum entangled information qubits.

*Information* being physical, needs definition which will connect its origin with encoding, memory and energy cost.

That leads to *notion* of information as phenomenon originating in probabilistic entropy hidden in a time-space correlation whose cut-erasure produces physical information without necessity of any particles in QM.

Since all real particles carry information, recovering the particle information origin presents actual scientific problem.

The observer microprocess emerges as part of information theory within the path from probabilistic observation to certainty of physical substances.

**2. Emerging time, curvature, space, causality, information, and complexity**

The correlation, emerging in the observable process, initially holds the related *time interval* [13] of the events hidden within the correlation.

Within the merging impulse ↓ *yes* ↑ *no* actions, a jump ↓ *yes* squeezes its time interval (Fig.1), curving time coordinate measure $1/2p[\tau]$ in transitional orthogonal driven rotation up to the emergence impulse space coordinate measure $h[l]$.

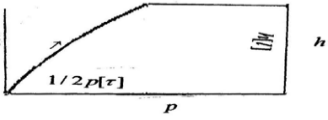

**Fig.1.Illustration origin of the impulse space coordinate measure $h[l]$ at curving time coordinate measure $1/2p[\tau]$ in transitional movement.**

In the end of rotation at $2\pi h[l]/4 = 1/2p[\tau]$, the coordinate/time ratio $h/p$ leads the ratio of measuring units: $[\tau]/[l] = \pi/2$ with elementary space curvature equals to inverse radius $K_s = h[l]^{-1}$. (Details in Sec.3.6.4.1).

Thus, the squeezed time interval at interactions *originates both curvature and space coordinate.*

Kolmogorov's symmetry of the probabilities [14] leads to statistical symmetry of the probabilistic causality.

The sequentially reduced Bayesian relational entropy conveys probabilistic causality along the process with the entropy *objective measure*. Each current correlation temporally memorizes a quantity of uncertainty of the probabilistic causality.

The correlation of the time- space impulse holds the related entropy of a probabilistic observation.

When the probabilistic observation is approaching certainty, the impulse ↑ *no* action cuts (erases) the impulse entropy. That provides both the reduced process' entropy and discrete unit of the cutting entropy, which the cut memorizes in information Bit.

From that follows definition: *Information is memorized entropy (uncertainty) cutting from high probable correlation hidden in random observations which process interactions.*

The Bits are nonrandom units emerging in random process of interacting impulses. Multiple bits cooperate information process which assembles a code. The Bits nonrandom interaction brings errors when they structure a code, requiring the error corrections, which study communication theory.

On the observing path of uncertainty to certainty, the certainty reveals through the observing impulse actions directed on reductions of the entropy-uncertainty. Initial uncertainty gradually transforms to more probable process and finally to certainty-as information about reality, terminating the probabilistic Bayesian entropy causality *measure*.

Cutting equivalent entropy of the uncertain causality originates reality in information Bit which measures transfer to certain actual causality (Details in Sec 3.6.5.4).

In a natural random process, impulse interactions *innately* cut the process correlations' entropy with possible *natural* origin of information.

Any possessor of information in the interactive observations, naturally or artificially created in observations (communications), we call Information Observer, emerging as indicator of the reality appearing from the actual causality.

A view of natural origin of information from general time-space continuum does not lead to discrete nature of information. Impulse interactions lead to discrete time–space, where the impulse *discrete* contributions reduce the entropy, which inverses the time course in an Information Observer. According to Second Law, in real observing process, both time course its entropy increases. The Observer's growing correlations automatically increases its life time. Definition and measuring information are concepts subsequent to origin of information a *phenomenon* of interactions from which all Universe composes.

When ↑ *no* action of the impulse (Fig.1) cuts its entropy, the ending state of transitional orthogonal movement fixes the space coordinate, indicating that actual-certain space appears simultaneously with information.

The curving time compresses and hides the space curvature during the transition, and finalizes transforming uncertain causality to certain causality.

The transitional space geometry encloses a space Bit, or the impulse information Bit enfolds a space geometry and space curvature. The curving impulse ↓↑ gets form (Fig.1a) whose curvature holds hidden information and complexity.

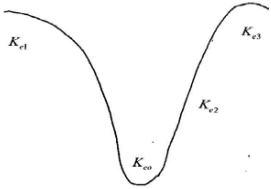

**Fig.1a. Curving impulse with curvature $K_{e1}$ of the impulse step-down part, curvature $K_{eo}$ of the cutting part, curvature $K_{e2}$ of the entropy transitive part, and curvature $K_{e3}$ of the final part cutting all impulse entropy** (Sec.3.6.4)**.**

Ch. Bennett has found time-space trade-offs in the reversible computation and the connection to complexity [15].
A logical geometrical structure of mutually connected information symbols *defines* information *process*, which starts with observing a random process via sequential logic of probing impulses that cuts the process and creates a microprocess of a merging interactive actions ↓↑, the discrete information units and their inter-dependencies, then integrates and enfolds them in an information geometrical structure of an Observer.
*The microprocess entanglement starts before the space (Fig.1) is formed and ends with beginning the space during reversible relative time interval* of the impulse $0.03125\pi$ part of $\tau = \pi$. (Secs. 3.6.2-3.6.4).
The specific impulses logic determines its algorithmic complexity measured by the required quantity of information.
The impulse' (Fig.1) No-cutting action precedes some probing logics with its complexity, which the transitional space geometry encloses. Since time coordinate $1/2p[\tau]$ is given by a latest correlation which encloses the previous logics, that time interval compresses these logics. Whereas the jump-wise action squeezes this time interval, creating the curvature of transitional space geometry, that intermediate curvature enfolds logical complexity of the time interval. Or, impulse Bit's ↓ *yes* ↑ *no* actions includes logical complexity of its prehistory, enclosed in the impulse space position topology, encoding it in each cut. The same 1-0 impulse with different space topology contains different logical complexity. Or vice versa, space structure of elementary Bit conveys is logic which evaluates the curvature complexity.
Cutting the process correlation retrieves hidden information, which the cut-erasure memorizes in the bit encoding.
The currently observed correlation temporary memorizes whole pre-history from the starting observation, including the summarized (integrated) entropy. The random impulses hold virtually observing random time intervals.
The origin of information associates with impulse ability to cut the correlation which depends on its probabilistic certainty. Such cut generates information Bit whose memory holds the impulse' cutting time interval.
The impulse observation, immanently collecting the growing time interval which temporally integrates and memorizes a final posterior correlation, formalizes entropy functional (EF) measuring uncertainty along the random process (formulas (I-III)), which is proportional to the running time intervals (operating according to formulas (1.1), (2.1))**.**
Multiple cuts of sequential posterior correlations during the interactive multi-dimensional observation compose multiple Bits with memories of the collecting impulse' cutting time intervals, which freeze the observing events dynamics in information processes. These cutting bits integrate information path functional (IPF) (Sec.3.5).
The integration of cutting Bits time intervals along the observing time course coverts it to the Information Observer inner time course that is opposite to the observable time course in which the process entropy increases.
The memory encloses the memorized time interval of the bits, which the cutting process' correlation already possesses.
*The time interval running the erasure of entropy is a measure of the memory of the hidden correlation that impulse cuts. During this time, the impulse encodes the information bit of cutting correlation and memorizes it*. (Sec.3.6.5.4).

The memory holds the bit, its curvature, topology, complexity, and also actual causation of the observing prehistory.

The impulse encoding holds the irreversible origin of information through erasure and memorizing the asymmetric position of the impulse opposite actions. Virtual probing impulses model a reversible encoding. Thereafter, the above definition of information leads to Natural encoding [47] which includes origination of reality in the actual causality.

The jumping action on a border of the impulse creates space-time conjugated entropy functions, rotating within space-dimensional impulse $h_s[l]$ (Secs. 3.4, 3.6), which are processing spiral geometry with optimal logic.

The multi-impulse processes hold geometry of double spiral structure that encodes the bits information with its logic, applicable to individual DNA sequence. The information double spiral geometry encodes Biosystems organization [16].

The information Bit's geometry encloses the impulse time, curvature, space coordinates, casualty, and complexity.

The rotating movement may capture the correlated entangled qubits and adjoin them in triple information structure (Fig. 2).

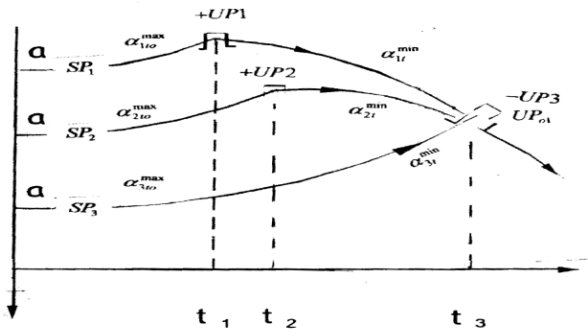

**Fig.2.** Illustration of assembling a triple qubit units $+UP1, +UP2, -UP3$ and adjoining them to the composed triplet unit $UP_{o1}$ at changing information speeds on the space-time trajectory from $\alpha_{1to}^{max}$, $\alpha_{2to}^{max}$, $-\alpha_{3to}^{max}$ to $\alpha_{1t}^{min}$, $\alpha_{2t}^{min}$, $-\alpha_{3t}^{min}$ accordingly; **a** is dynamic information invariant of an impulse.

The opposite directional speeds of the rotating correlated units minimize their ending speeds down: $\alpha_{1t}^{min} \to \alpha_{2t}^{min} \to \alpha_{3t}^{min}$, and the arising entropy forces attract them up to equalization $\alpha_{1t}^{min} = |\alpha_{2t}^{min}| = \alpha_{3t}^{min} = \alpha_{uo1}$, bind and memorize the joint triplet unit $UP_{o1}$ with common information $\alpha_{1to}^{max} t_1 = \alpha_{2to}^{max} t_2 = \alpha_{3to}^{max} t_3 = \mathbf{a}_o \cong 3\mathbf{a}$ formed during time intervals $t_1, t_2, t_3$ of the movement. The impulse information of 1 Nat includes bit information $\ln 2 Nat$ and "free" information equal to information invariant $\mathbf{a} = 1/3 bit \approx 0.23 Nat$ which determines the attracting forces bringing the joint triplet information $\mathbf{a}_o \cong \ln 2$.

### 3. The analytical "anatomy" of a discrete impulse and creation of a Bit

*Basic mathematical foundation*

The *integral measure* of the observing *process* trajectories are formalized by an *Entropy Functional* (EF), which is expressed through the regular and stochastic components of Markov diffusion process $\tilde{x}_t$ [18-20]:

$$\Delta S[\tilde{x}_t]|_s^T = 1/2 E_{s,x} \{\int_s^T a^u(t,\tilde{x}_t)^T (2b(t,\tilde{x}_t))^{-1} a^u(t,\tilde{x}_t) dt\} = \int_{\tilde{x}(t) \in B} -\ln[p(\omega)] P_{s,x}(d\omega) = -E_{s,x}[\ln p(\omega)], \quad (I)$$

where $a^u(t,\tilde{x}_t) = a(t,\tilde{x}_t,u_t)$ is a drift function, depending on control $u_t$, and $b(t,\tilde{x}_t)$ is a covariation function, describing its diffusion component in Ito Eqs.[21,22]; $E_{s,x}$ is a conditional to the initial states $(s,x)$ math expectation, taken along the $\tilde{x}_t = \tilde{x}(t)$ trajectories.

The right side of (I) is the EF equivalent formula, expressed via probability density $p(\omega)$ of random events $\omega$, integrated with probability measure $P_{s,x}(d\omega)$ along the process trajectories $\tilde{x}(t) \in B$, defined at the set $B$ by averaging additive functional $\varphi = \varphi[a^u,b]$ (Ia).

This functional describes transformation of the Markov processes' random time traversing the various sections of a trajectory. Formula (I) measures integral entropy which is directly connected to the diffusion random process' defining functions. At the known functions, measuring the process' entropy through the probability is not necessary. That "master" formula, published in [24], since provides the integral measuring of the multiple physical processes, which Markov diffusion process models, including Schrodinger quantum process [41].

Generally, random process (as a continuous or discrete function $x(\omega,s)$ of random variable $\omega$ and time $s$), describes an elementary *changes* of its probabilities from one distribution (a *priori*) $P^a_{s,x}(d\omega)$ to another distribution (*a posteriory*) $P^p_{s,x}(d\omega)$ in the form of their transformation [23]:

$$p(\omega) = \frac{P^a_{s,x}(d\omega)}{P^p_{s,x}(d\omega)}. \qquad (II)$$

Integral (I) embeds Bayes probabilities [33, 33a] and connects them directly to particular drift and diffusions of Markov process functional that allows measuring the probabilities through these functional' Markov components, and vice versa. Sequence of these probabilities ratios *generalizes* diverse forms of specific functional relations, represented by a series of different transformations.

The probability ratio in the form of natural logarithms:

$$-\ln p(\omega) = -\ln P^a_{s,x}(d\omega) - (-\ln P^p_{s,x}(d\omega)) = s_a - s_p = \Delta s_{ap}, \qquad (IIa)$$

describes the difference of a *priory* $s_a > 0$ and a *posteriori* $s_p > 0$ *random entropies, which* measure *uncertainty,* resulting from the transformation of probabilities for the process events, satisfying the entropies additivity. A *change* brings a certainty or *information* if uncertainty $\Delta s_{ap}$ is removed by some equivalent entity call information $\Delta i_{ap}: \Delta s_{ap} - \Delta i_{ap} = 0$. Thus, information is delivered if $\Delta s_{ap} = \Delta i_{ap} > 0$, which requires $s_p < s_a$ and a positive logarithmic measure (II) with $0 < p(\omega) < 1$. Condition of zero information: $\Delta i_{ap} = 0$ describes a redundant change, transforming a priori probability to equal a posteriory probability, or this transformation is identical–informational undistinguished. The removal of uncertainty $s_a$ by $i_a: s_a - i_a = 0$ brings an equivalent certainty or *information* $i_a$ about entropy $s_a$.

Logarithmic measure (IIa) of Markov diffusion process' probabilities approximates the probability ratios for other random processes. Mathematical expectation of the random probabilities and entropies:

$$E_{s,x}\{-\ln[p(\omega)]\} = E_{s,x}[\Delta s_{ap}] = \Delta S_{ap} \Rightarrow I_{ap} \neq 0, \qquad (III)$$

defines mean entropy $\Delta S_{ap}$ as the equivalent of nonrandom *information* $I_{ap}$ of a random source.

Averaging in (III) and (I) by probabilities (II) includes Shannon's formula for relative entropy-information of the states (events), depending on the probabilities of a source: *processes* or its events.

The EF (I) measures the entropy distance between distributions $\tilde{P}_{s,x}, P_{s,x}$ along the process [23, 24].

Minimum of this functional measures closeness above distributions [26]:

$$\min_{\varphi_s^t} S[\tilde{x}_t / \varsigma_t] = S^o. \tag{IV}$$

Conditional Kolmogorov probability: $P(A_i / B_k) = [P(A_i)P(B_k / A_i)] / P(B_k)$ (IVa)

after substituting an average probability $P(B_k) = \sum_{i=1}^{n} P(B_k / A_i) P(A_i)$ defines Bayes probability by averaging this finite sum or integrating [1]. For each $i, k$ random events $A_i, B_k$ along the observing process, each conditional a priori probability $P(A_i / B_k)$ follows conditional a posteriory probability $P(B_k / A_{i+1})$. Conditional entropy

$$S[A_i / B_k)] = E[-\ln P(A_i / B_k))] = [-\ln \sum_{i,k=1}^{n} P(A_i / B_k)] P(B_k) \tag{V}$$

averages the conditional Kolmogorov- Bayes probability for multiple events along the observing process.

Random current conditional entropy is

$$\tilde{S}_{ik} = -\ln[P(A_i / B_k) P(B_k)]. \tag{VI}$$

The experimental probability measure predicts axiomatic Kolmogorov probability if the experiment satisfies condition of *symmetry* of the equal probable events in its axiomatic probability [1].

Conditional probabilities satisfies Kolmogorov's 1-0 law [1, p.117] for function $f(x) | \xi$, of $\xi, x$ infinite sequence of independent random variables:

$$P_\delta(f(x) | \xi) = \begin{cases} 1, f(x) | \xi) \geq 0 \\ 0, f(x) | \xi) < 0 \end{cases}. \tag{VII}$$

This probability measure has applied for the impulse probing of an observable random process, which holds opposite Yes-No probabilities – as a unit of the probability step-function.

### 3.1. The entropy regular integral functional

A Markov diffusion process, with its statistical interconnections of states, represents the most adequate formal model of the information process, we are considering.

For a single dimensional Ito Eq. with drift function $a = c\tilde{x}(t)$ at given nonrandom function $c = c(t)$ and diffusion $\sigma = \sigma(t)$, entropy functional EF (I) acquires form

$$S[\tilde{x}_t / \varsigma_t] = 1/2 \int_s^T E[c^2(t) \tilde{x}^2(t) \sigma^{-2}(t)] dt, \tag{1.1}$$

from which, at $\sigma(t)$ and nonrandom function $c(t)$, we get

$$S[\tilde{x}_t / \varsigma_t] = 1/2 \int_s^T [c^2(t) \sigma^{-2}(t) E_{s,x}[x^2(t)] dt = 1/2 \int_s^T c^2 [2b(t)]^{-1} r_s dt, \tag{1.2}$$

where for the diffusion process, the following relations hold true:

$$2b(t) = \sigma(t)^2 = dr/dt = \dot{r}_t, E_{s,x}[x^2(t)] = r_s. \tag{1.3}$$

This *identifies* the entropy functional on observed Markov process $\tilde{x}_t = \tilde{x}(t)$ by measuring above correlation functions, applying positive function $u(t) = c^2(t)$ and *representing* functional (I) through regular integral of non-random functions

$A(t,s) = [2b(t)]^{-1} r_s$ (1.4a) and $\quad u(t) = c^2(t)$ (1.4b)

in form $\quad S[\tilde{x}_t / \varsigma_t] = 1/2 \int_s^T u(t) A(t,s) dt$. (1.4)

The *n*-dimensional functional integrant (1.4) follows from related *n*-dimensional covariations (1.3), dispersion matrix, and by applying *n*-dimensional function $u(t)$. At given nonrandom function $u(t)$, (1.4) is regular integral which measures the EF of Markov process at probability transformation (II) with additive functional (Ia) that averages the functional integrant.

### 3.2. Application of variation condition (IV)
Proposition 2.1.

Integral (1.4), satisfying variation condition (IV) at linear function $c^2(t) = u(t) = c^2 t$, has form

$S[\tilde{x}_t / \varsigma_t] = 1/2 \int_s^T u(t) o(t) dt$, (2.1)

where the extreme of function (1.4a) holds minimum

$A(t, s_k^{+o}) = o(s) b_k(s_k^{+o}) / b_k(t) = o(t)$, (2.2)

which decreases with growing time $t = s_k^{+o} + o(t)$ at $t \to T$ and fixed both $b_k(s_k^{+o})$ and

$o(s) = A(s,s)$. (2.3)

Since satisfaction of this variation condition includes transitive transformation of a current distribution to that of Feller kernel, $b_k(t)$ is transition dispersion at this transformation, which is growing with the time of the transformation.

<u>Proof.</u> Applying Euler's equation to variation condition (IV) of this integral, at simple linear function $c^2(t) = u(t) = c^2 t$ and fixed $c^2$ on starting moment $s_k^{+o}$ leads to

$\partial A(t, s_k^{+o}) t / \partial t = \dot{A}(t, s_k^{+o}) t + A(t, s_k^{+o}) = 0$, (2.4)

$t \partial A(t, s_k^{+o}) / \partial t + A(t, s_k^{+o}) = 0, \partial A(t, s_k^{+o}) / A(t, s_k^{+o}) + \partial t / t = 0, t \neq 0, A(t, s_k^{+o}) \neq 0, \partial t / t \neq 0$. (2.4a)

From these it follows

$\ln A(t, s_k^{+o}) + \ln Ct = \ln[A(t, s_k^{+o}) Ct] = 0, A(t, s_k^{+o}) Ct = 1,$ (2.5a)

$A(s_k^{+o}, t = s_k^{+o})^{-1} = C s_k^{+o}, C = A(s_k^{+o}, s_k^{+o})^{-1} / s$. (2.5b)

On a current time interval $\Delta_t = t - s_k^{+o}$, relations (2.5a,b) determine functions

$A(s_k^{+o}, s_k^{+o}) / A(t, s_k^{+o}) = t / s_k^{+o}$. (2.6)

Relations (1.3) allow representations

$A(s_k^{+o}, s_k^{+o}) = r_k(s_k^{+o}) / 2 b_k(s_k^{+o}), \quad A(s_k^{+o}, t) = r_k(s_k^{+o}) / 2 b_k(t),$ (2.7)

where correlation function

$$r(s) = \int_{s}^{s+o(s)} 2b(t)dt = 2b(s)o(s) \tag{2.8}$$

leads to

$$A(s,s) = [2b(s)]^{-1} 2b(s)o(s) = o(s). \tag{2.9}$$

Substitution (2.9) and (2.7) in (2.6) brings

$$t = s_k^{+o} b_k(t) / b_k(s_k^{+o}). \tag{2.10}$$

This shows that with growing time $t = s_k^{+o} + o(t)$ dispersion $b_k(t)$ has tendency to grow.

Substituting (2.10) and (2.9) in (2.5a) determines function

$$A(t, s_k^{+o}) = o(s)b_k(s_k^{+o}) / b_k(t) = o(t), \quad \Delta_t = o(t) \tag{2.11}$$

which decreases with growing time $t \to T$.

The extreme holds minimum, since second derivative of function (2.5a) is positive. •

Comments 2.1. On interval $\Delta_t$, at moment $t = t_k$ in equation (2.10) in form

$$t = s_k^{+o} \dot{r}(t) / \dot{r}(s_k), \tag{2.12}$$

at $b(t) = 1/2\dot{r}(t), b(s_k) = 1/2\dot{r}(s_k)$, \hfill (2.12a)

can be identified by controlling correlation:

$$t_k \cong s_k r_k(t_k) / r_k(s_k). \quad • \tag{2.12b}$$

Thus, EF (2.10) integrates time of correlation which function $u(t) = u_t$ cuts.

### 3.3. Impulse action on the entropy integral

Let's define $u(t) = u_t$ on space $KC(\Delta, U)$ of a piece-wise continuous step-functions $u_t(u_-^t, u_+^t)$ at $t \in \Delta$, applying to integrant of (2.1) the difference of step-down $u_-^t = u_-(t)/\delta_o$ and step-up $u_+^t = u_+(t+\delta_o)/\delta_o$ functions at fixed interval

$$\delta_o : u_t^{\delta_o} = [u_-(t) - u_+(t+\delta_o)]/\delta_o = u_t^{\delta_o} \tag{3.1}$$

which forms Dirac's delta-function at

$$\lim_{\delta_o \to 0} = \delta u_t. \tag{3.2}$$

Proposition 3.1.

Entropy integral (2.1) under impulse control (3.2) in form of delta-function $c^2(t, \tau_k) = \delta u_t(t - \tau_k)$:

$$\Delta S[\tilde{x}_t / \varsigma_t]\Big|_{\tau_k^{-o}}^{\tau_k^{+o}} = 1/2 \int_{\tau_k^-}^{\tau_k^+} \delta u_t(t - \tau_k) o(t) dt, \quad \tau_k^{-o} < \tau_k < \tau_k^{+o} \tag{3.3}$$

takes the information values:

$$\Delta S[\tilde{x}_t / \varsigma_t]\Big|_{t=\tau_k^{-o}}^{t=\tau_k^{+o}} = \begin{cases} 0, t < \tau_k^{-o} \\ 1/4 o(\tau_k^{-o}), t = \tau_k^{-o} \\ 1/2 o(\tau_k), t = \tau_k, \tau_k^{-o} < \tau_k < \tau_k^{+o} \\ 1/4 o(\tau_k^{+o}), t = \tau_k^{+o} \end{cases}, \quad \tau_k^{-o} < \tau_k < \tau_k^{+o} \tag{3.4}$$

where function (3.4) follows from [25,p.678-681].

The delta-cutoff brings amount of entropy integral $S[\tilde{x}_t/\varsigma_t]_{t=\tau_k} = 1/2o(\tau_k) = 1/2$ Nats, while on borders of interval $o(\tau_k)$, the integral amounts are $S[\tilde{x}_t/\varsigma_t]_{t=\tau_k^{-o}} = 1/4o(\tau_k^{-o})$ Nats and $S[\tilde{x}_t/\varsigma_t]_{t=\tau_k^{+o}} = 1/4o(\tau_k^{+o})$ Nats accordingly. •

### 3.4. Discrete control action on the entropy functional

Let us find a class of step-down $u_-^t = u_-(\tau_k^{-o})$ and step-up $u_+^t = u_+(\tau_k^{+o})$ functions acting on discrete interval $o(\tau_k) = \tau_k^{+o} - \tau_k^{-o}$, which will preserve the Markov diffusion process' additive and multiplicative functions within the cutting process of each impulse.

<u>Lemma 4.1.</u>

1. Opposite discrete functions $u_-^t$ and $u_+^t$ in form

$$u_-(\tau_k^{-o}) = \downarrow_{\tau_k^{-o}} \bar{u}_-, u_+(\tau_k^{+o}) = \uparrow_{\tau_k^{+o}} \bar{u}_+ \tag{4.1}$$

satisfy conditions of additivity

$$[u_+^t - u_-^t] = U_a \text{ (a) or } [u_+^t + u_-^t] = U_a \text{ (b)} \tag{4.1A}$$

and multiplicativity

$$1/2o(\tau_k) \, 1/2o(\tau_k) \quad \text{at} \tag{4.1B}$$

$$U_a = U_m = U_{am} = c^2 > 0, \tag{4.1C}$$

where instance-jump $\downarrow_{\tau_k^{-o}}$ has time interval $\bar{u}_-$ and instance jump $\uparrow_{\tau_k^{+o}}$ has high $\bar{u}_+$ for relation (4.1A)(a) at real values

$$\bar{u}_- = 0.5, \bar{u}_+ = 1, \bar{u}_+ = 2\bar{u}_-, \tag{4.2a}$$

and for relation (4.1A) (b) at real values

$$\bar{u}_-^o = \bar{u}_+^o = 2. \tag{4.2b}$$

2. Complex functions

$$u_t(u_\pm^{t1}, u_\pm^{t2}), u_\pm^{t1} = [u_+ = (j-1), u_- = (j+1)], j = \sqrt{-1} \tag{4.2c}$$

satisfy conditions (4.1aA), (4.B) in forms

$$u_+ - u_- = (j-1) - (j+1) = -2, u_+ \times u_- = (j-1) \times (j+1) = (j^2 - 1) = -2,$$

which however do not preserve positive (4.1C).

Opposite complex functions

$$u_t(-u_\pm^{t1}) = u_t(u_\pm^{t2}), u_\pm^{t2} = [u_+ = (j+1), u_- = (j-1)], \tag{4.2d}$$

satisfy (4.1bA)-(4.C). And imaginary functions

$$u_+^t = j\sqrt{2}, u_-^t = -j\sqrt{2}, \tag{4.2d1}$$

when the impulse additive meausure $U_a = 0$ leaves only multiplicative $U_m = u_+^t \times u_-^t = -2$ part, which at equal absolute values of actions $|u_+^t| = |u_-^t|$ determines only (4.2d1).

<u>Proofs</u> are straight forward.

Assuming both opposite functions apply on borders of interval $o(\tau_k) = (\tau_k^{+o}, \tau_k^{-o})$ in forms

$$u_-^{t1} = u_-(\tau_k^{-o}), u_+^{t1} = u_+(\tau_k^{-o}) \text{ and } u_-^{t2} = u_-(\tau_k^{+o}), u_+^{t2} = u_+(\tau_k^{+o}),  \qquad (4.2e)$$

at $u_-^{t1}u_+^{t1} = c^2(\tau_k^{-o}), \ u_-^{t2}u_+^{t2} = c^2(\tau_k^{+o}), \ t = \tau_k^{+o},$ \qquad (4.2f)

it follows that only by end of interval at $t = \tau_k^{+o}$ both Markov properties (4.1A,B) satisfy, while at beginning $t = \tau_k^{-o}$ the starting process does not possess yet these properties. •

Corollary 4.1.

1. Conditions 4.1A-4.1C imply that $c^2(\tau_k^{-o}), c^2(\tau_k^{+o})$ are discrete functions (4.1a),(4.2f) switching on interval $\Delta_\tau = \tau_k^{+o} - \tau_k^{-o}$.

Requiring $\Delta_\tau = \delta_o$ leads to discrete delta-function $\delta^o u_t$ (3.1) which for $\delta_o = (\tau_k^{+o} - \tau_k^{-o})$ holds

$\delta^o u_{t=\tau_k} = [u_-(\tau_k^{-o}) - u_+(\tau_k^{+o})]/(\tau_k^{+o} - \tau_k^{-o})$, that at $\Delta = (\tau_k^{-o} - s_k^{+o})$ brings

$$u_-(\tau_k^{-o}) = -1_{\tau_k^{-o}} \bar{u}_-, u_+(\tau_k^{+o}) = +1_{\tau_k^{+o}} \bar{u}_+, \ \bar{u}_- = 0.5, \bar{u}_+ = 2, \qquad (4.3)$$

when positivity of $c^2 > 0$ implies

$$\delta^o u_{t=\tau_k} = [u_+(\tau_k^{+o}) - u_-(\tau_k^{-o})]/(\tau_k^{+o} - \tau_k^{-o}) > 0. \qquad (4.3a)$$

2. Discrete functions $u_+(s_k^{+o}) = +1_{s_k^{+o}} \bar{u}_+, u_-(\tau_k^o) = -1_{\tau k^{+o}} \bar{u}_-$ \qquad (4.3b)

on $\Delta$ are multiplicative: $(u_-(\tau_k^{-o}) - u_+(s_k^{+o})) \times (u_-(\tau_k^{-o}) - u_+(s_k^{+o})) = [u_-(\tau_k^{-o}) - u_+(s_k^{+o})]^2$.

2a. Discrete functions (4.4e) in form

$$\bar{u}_+ = j\bar{u}, \bar{u}_- = -j\bar{u}, \bar{u} \neq 0 \qquad (4.3c)$$

satisfy only condition (4.1A) which for functions (4.3b) holds

$$[u_-(\tau_k^{-o}) - u_+(s_k^{+o})]^2 = -(j\bar{u})^2[-1_{\tau_k^{-o}} - 1_{s_k^{+o}}]^2 > 0. \qquad • \qquad (4.3d)$$

Let us find discrete analog of the integral increments under *discrete* delta-function (4.3),(4.3a):

$$\delta^o u_{t=\tau} = (u_-(\tau_k^{-o}) - u_+(\tau_k^{+o}))(\tau_k^{-o} - \tau_k^{+o})^{-1}.$$

Proposition 4.2.

1. Applying discrete delta-function (4.3) to integral (2.1) leads to

$$\Delta S[\tilde{x}_t / \varsigma_t]\Big|_{t=\tau_k^{-o}}^{t=\tau_k^{+o}} = \begin{cases} 0, t < \tau_k^{-o} \\ 1/4 u_-(\tau_k^{-o}) o(\tau_k^{-o})/\tau_k^{-o}, t = \tau_k^{-o}, 1/4 \downarrow 1_{\tau_k^{-o}} \bar{u}_{ko} \\ 1/2(u_-(\tau_k^{-o}) - u_+(\tau_k^{+o})) o(\tau_k)/(\tau_k^{+o} - \tau_k^{-o}), t = \tau_k, \tau_k^{-o} < \tau_k < \tau_k^{+o}, 1/2(\downarrow 1_{\tau_k^{-o}} - \uparrow 1_{\tau_k^{+o}}) \bar{u}_{km} \\ 1/4 u_+(\tau_k^{+o}) o(\tau_k^{+o})/\tau_k^{+o}, t = \tau_k^{+o}, 1/4 \uparrow 1_{\tau_k^{+o}} \bar{u}_{k1} \end{cases} \qquad (4.4)$$

which is a discrete analog of (3.4), where

$$\bar{u}_{ko} = \bar{u}_- \times o(\tau_k^{-o})/\tau_k^{-o}, \bar{u}_{km} = (\bar{u}_+ - \bar{u}_-) \times o(\tau_k)/(\tau_k^{+o} - \tau_k^{-o}), \bar{u}_{k1} = \bar{u}_+ \times o(\tau_k^{+o})/\tau_k^{+o},$$
$$\bar{u}_{km} = 1/2(\bar{u}_+ - \bar{u}_-) = 0.75, o(\tau_k) = \tau_k^{+o} - \tau_k^{-o}, o(\tau_k^{-o})/\tau_k^{-o} = 0.5, o(\tau_k^{+o})/\tau_k^{+o} = 0.1875 \qquad (4.5)$$

and $|\bar{u}_- \times \bar{u}_+| = |1/2 \times 2| = |\bar{u}_k| = |1|_k$ is multiplicative measure of impulse $(\downarrow 1_{\tau_k^{-o}} - \uparrow 1_{\tau_k^{+o}}) \bar{u}_k$.

Let's measures middle interval in (4.4) by single impulse information unit $\bar{u}_k = |1|_k$, then functions (4.4) determine finite size of the impulse parameters $\bar{u}_{ko}, \bar{u}_k, \bar{u}_{k1}$ which estimate $\bar{u}_{km}$:

$$\bar{u}_{ko} = 0.25 = 1/3\bar{u}_{km}, \bar{u}_{k1} = 2 \times 0.1875 = 0.375 = 0.5\bar{u}_{km} \ . \qquad \bullet \qquad (4.6)$$

Proofs follows from Proposition 4.3 below.

Introducing entropy unit impulse $\bar{u}_s = |1|_s$ with moments $(s_k^{-o}, s_k^o, s_k^{+o})$ prior to impulse $\bar{u}_k = |1|_k$, which measures middle interval of impulse entropy $\bar{u}_{sm}$, will allow us to find increment of $\Delta S[\tilde{x}_t / \varsigma_t]|_{s_k^+}^{\tau_k^{-o}}$ on border of impulse $\bar{u}_k$ at prior $\Delta_{\tau s+} = \delta_{sk\pm} = (s_k^{+o} - \delta_k^{\tau-})$ and posterior $\Delta_{\tau s-} = \delta_{sk\mp} = (\delta_k^{\tau-} - \delta_k^{\tau+})$ moments under impulse functions of $\bar{u}_s = |1|_s$:

$$\delta^o u_{\tau = (s_k^{+o} - \delta_k^{\tau-})} = (u_+(s_k^{+o}) - u_-(\delta_k^{\tau-}))(s_k^{+o} - \delta_k^{\tau-})^{-1} = \uparrow 1_{s_k^{+o}} \bar{u} - \downarrow 1_{\delta_k^{\tau-}} \bar{u} = [\uparrow 1_{s_k^{+o}} - \downarrow 1_{\delta_k^{\tau-}}]\bar{u}, \qquad (4.7)$$

$$\delta^o u_{\tau = (\delta_k^{\tau-} - \delta_k^{\tau+})} = (u_-(\delta_k^{\tau-}) - u_+(\delta_k^{\tau+}))(\delta_k^{\tau-} - \delta_k^{\tau+})^{-1} = \downarrow 1_{\delta_k^{\tau-}} \bar{u} - \uparrow 1_{\delta_k^{\tau+}} \bar{u} = [\downarrow 1_{\delta_k^{\tau-}} - \uparrow 1_{\delta_k^{\tau+}}]\bar{u}, \qquad (4.8)$$

$$\delta^o u_{\tau = (\delta_k^{\tau+} - \tau_k^{-o})} = (u_+(\delta_k^{\tau+}) - u_-(\tau_k^{-o}))(\delta_k^{\tau+} - \tau_k^{-o})^{-1} = \uparrow 1_{\delta_k^{\tau+}} \bar{u} - \downarrow 1_{\tau_k^{-o}} \bar{u} = [\uparrow 1_{\delta_k^{\tau+}} - \downarrow 1_{\tau_k^{-o}}]\bar{u}. \qquad (4.9)$$

Here $\bar{u}$ evaluates each impulse interval, which according to the optimal principle is an invariant.
Applying functions (4.7)-(4.9) leads to additive sum of each increment of the entropy functional:

$$\Delta S[\tilde{x}_t / \varsigma_t]|_{s_k^+}^{\tau_k^{-o}} = \Delta S[\tilde{x}_t / \varsigma_t]|_{s_k^+}^{\delta_k^{\tau-}} + \Delta S[\tilde{x}_t / \varsigma_t]|_{\delta_k^{\tau-}}^{\delta_k^{\tau+}} + \Delta S[\tilde{x}_t / \varsigma_t]|_{\delta_k^{\tau+}}^{\tau_k^{-o}} \qquad (4.10)$$

along time interval

$$\Delta_{\tau sk\pm} = s_k^{+o} - \delta_k^{\tau-} + \delta_k^{\tau-} - \delta_k^{\tau+} + \delta_k^{\tau+} - \tau_k^{-o} = s_k^{+o} - \tau_k^{-o} = \Delta_{\tau s}. \qquad (4.10a)$$

Proposition 4.3.

A. The increments of entropy functional (4.10) collected on intervals (4.10a), satisfying variation condition (IV), bring the following entropy contributions:

$$\Delta S[\tilde{x}_t / \varsigma_t]|_{s_k^+}^{\delta_k^{\tau-}} = 1/2(u_+(s_k^{+o}) - u_-(\delta_k^{\tau-}))o(s_k^{+o} - \delta_k^{\tau-}))(s_k^{+o} - \delta_k^{\tau-})^{-1} = 1/2[\uparrow 1_{s_k^{+o}} - \downarrow 1_{\delta_k^{\tau-}}]\bar{u}_{ks} \qquad (4.11)$$

at $\bar{u}_{ks} = \bar{u}(o(s_k^{+o} - \delta_k^{\tau-})(s_k^{+o} - \delta_k^{\tau-})^{-1}$; $\qquad (4.11a)$

$$\Delta S[\tilde{x}_t / \varsigma_t]|_{\delta_k^{\tau-}}^{\delta_k^{\tau+}} = 1/2(u_-(\delta_k^{\tau-}) - u_+(\delta_k^{\tau+}))o(\delta_k^{\tau-} - \delta_k^{\tau+}))(\delta_k^{\tau-} - \delta_k^{\tau+})^{-1} = 1/2[\downarrow 1_{\delta_k^{\tau-}} - \uparrow 1_{\delta_k^{\tau+}}]\bar{u}_{k\delta s}, \qquad (4.12)$$

at $\bar{u}_{k\delta s} = \bar{u} \times (o(\delta_k^{\tau-} - \delta_k^{\tau+}))(\delta_k^{\tau-} - \delta_k^{\tau+})^{-1}$ $\qquad$ (4.12a) and

$$\Delta S[\tilde{x}_t / \varsigma_t]|_{\delta_k^{\tau+}}^{\tau_k^{-o}} = 1/2(u_+(\delta_k^{\tau+}) - u_-(\tau_k^{-o}))o(\delta_k^{\tau+} - \tau_k^{-o})(\delta_k^{\tau+} - \tau_k^{-o})^{-1} = 1'/2[\uparrow 1_{\delta_k^{\tau+}} - \downarrow 1_{\tau_k^{-o}}]\bar{u}_{k\delta}, \qquad (4.13)$$

at $[\uparrow 1_{\delta_k^{\tau+}} - \downarrow 1_{\tau_k^{-o}}]\bar{u}_{k\delta} = [\uparrow 1_{\delta_k^{\tau+}} + \uparrow 1_{\tau_k^{-o}}]\bar{u}_{k\delta}$. $\qquad (4.13a)$

Here each impulse interval acquires specific entropy measure:

$$\bar{u}_{k\delta} = \bar{u} \times (o(\delta_k^{\tau+} - \tau_k^{-o}))(\delta_k^{\tau+} - \tau_k^{-o})^{-1} = \bar{u} \times (o(\delta_k^{\tau+})(\delta_k^{\tau+} - \tau_k^{-o})^{-1}) + \bar{u} \times (o(\tau_k^{-o})(\tau_k^{-o})^{-1}(\delta_k^{\tau+} - \tau_k^{-o})^{-1}\tau_k^{-o} \quad (4.14)$$

on the impulse invariant interval $\bar{u}$. Relation (4.14) leads to impulse interval

$$\bar{u}_{k\delta} = \bar{u}_{k\delta o} + \bar{u}_{k\delta 1} \qquad (4.14a)$$

with its parts

$$\bar{u}_{k\delta o} = \bar{u} \times (o(\delta_k^{\tau+}))(\delta_k^{\tau+} - \tau_k^{-o})^{-1}) \ , \ \bar{u}_{k\delta 1} = \bar{u}_{ko1} \times \bar{u}_{ko2}, \qquad (4.14b)$$

$$\bar{u}_{ko1} = \bar{u} \times (o(\tau_k^{-o}))(\tau_k^{-o})^{-1}, \quad \bar{u}_{ko2} = \bar{u}^{-1} \times \tau_k^{-o}(\delta_k^{\tau+} - \tau_k^{-o})^{-1}. \tag{4.14c}$$

B. Intervals $\bar{u}_{ko1}$ and $\bar{u}_{ko2}$ are multiplicative parts of impulse step-up interval $\bar{u}_{k\delta1}$, which satisfies relations

$$\bar{u}_{k\delta o} = \bar{u}_{k\delta 1} = 1/2 \bar{u}_{k\delta}, \quad \bar{u}_{k\delta 1} = \bar{u}_{ko1}, \tag{4.15}$$

where invariant impulse $|\bar{u}_{k\delta}| = |1|_s$, acting on time interval $\delta_k^{\tau+} = 2\tau_k^{-o}$, measures

$$\bar{u}_{k\delta} = \bar{u}_{ks} \text{ at } |\bar{u}_{k\delta1}| = 1/2 \bar{u}_{sm}, \tag{4.15a}$$

and the relative time intervals of $\bar{u}_{ko}$ and $\bar{u}_{k1}$ accordingly are

$$o(\tau_k^{-o})(\tau_k^{-o})^{-1} = 0.5, \quad o(\tau_k^{+o})/\tau_k^{+o}) = 0.1875. \tag{4.15b}$$

Step-controls of impulse $\bar{u}_{k\delta}$ apply on two equal time intervals:

$$(\delta_k^{\tau+} - \tau_k^{-o}) = \delta_k^{\tau+}/2 \text{ (4.16a)} \quad \text{and} \quad \tau_k^{-o} = \delta_k^{\tau+}/2. \tag{4.16b}$$

On first (4.16a), its step-up part $[\uparrow 1_{\delta_k^{\tau+}}]$ captures entropy increment

$$\Delta S[\tilde{x}_t / \varsigma_t]|_{\delta_k^{\tau+}}^{\tau_k^{-o}} = 1`/2[\uparrow 1_{\delta_k^{\tau+}}]\bar{u}_- = 1`/8[\uparrow 1_{\delta_k^{\tau+}}], \tag{4.16}$$

on second (4.16b), its step-down multiplicative part in (4.14b) at $\bar{u}_{ko2} = \bar{u}^{-1}$ transfers entropy (4.16) to starting impulse action $[\downarrow 1_{\tau_k^{-o}}]$ which cuts is in impulse (4.4) at $\bar{u}_{ko1} = 1/2 \bar{u}_{ko}$; $\bar{u}_{k\delta1}$ multiplies $\bar{u}[\uparrow 1_{\delta_k^{\tau+}}] \delta_k^{\tau+}/2 \times \bar{u}^{-1}[\downarrow 1_{\tau_k^{-o}}] \tau_k^{-o}$ (4.16b)

Both equal time intervals in (4.16b) and orthogonal opposite inverse entropy increments are on the impulse border.

C. The applied *extremal* solution (Prop.2.1.), decreasing time intervals (2.11), brings (a)-persistence continuation a sequence of the process impulses; (b)-the balance condition for the entropy contributions; (c)-each impulse invariant unit $\bar{u}_k = |1|_k$, supplied by entropy unit $\bar{u}_s = |1|_s$, *triples* information that increases information density in each following information unit. •

Proofs.

The additive sum of entropy increments under invariant impulses (4.7-4.9) satisfies balance condition:

$$\Delta S[\tilde{x}_t / \varsigma_t]|_{s_k^+}^{\tau_k^{-o}} = \Delta S[\tilde{x}_t / \varsigma_t]|_{s_k^+}^{\delta_k^{\tau-}} + \Delta S[\tilde{x}_t / \varsigma_t]|_{\delta_k^{\tau-}}^{\delta_k^{\tau+}} + \Delta S[\tilde{x}_t / \varsigma_t]|_{\delta_k^{\tau+}}^{\tau_k^{-o}} =$$
$$1/2[\uparrow 1_{s_k^{+o}} - \downarrow 1_{\delta_k^{\tau-}}]\bar{u}_{ks} + 1/2[\uparrow 1_{\delta_k^{\tau-}} - \uparrow 1_{\delta_k^{\tau+}}]\bar{u}_{k\delta s} + 1/2 \uparrow 1_{\delta_k^{\tau+}} \bar{u}_{k\delta o} - 1/2 \downarrow 1_{|\tau|_k^{-o}} \bar{u}_{k\delta 1} = 0 \tag{4.17}$$

where action $1/2 \downarrow 1_{|\tau|_k^{-o}} \bar{u}_{k\delta 1} \Rightarrow 1/4 \downarrow 1_{|\tau|_k^{-o}} \bar{u}_{ko}$ transfers the left to the right entropy increment

$\Delta S[\tilde{x}_t / \varsigma_t](\tau_k^{-o}) = 1/4 \downarrow 1_{|\tau|_k^{-o}} \bar{u}_{ko}$ on discrete locality $|\tau|_k^{-o}$ of step-down action $\downarrow 1_{|\tau|_k^{-o}} \bar{u}_{k\delta 1}$.

Fulfilment of relations

$$[\uparrow 1_{s_k^{+o}} \bar{u}_{ks} - \downarrow 1_{\delta_k^{\tau-}} \bar{u}_{ks} + \uparrow 1_{\delta_k^{\tau-}} \bar{u}_{k\delta s} - \uparrow 1_{\delta_k^{\tau+}} \bar{u}_{k\delta s} + \uparrow 1_{\delta_k^{\tau+}} \bar{u}_{k\delta o} - \downarrow 1_{|\tau|_k^{-o}} \bar{u}_{k\delta 1}] = 0$$
$$[\uparrow 1_{s_k^{+o}} \bar{u}_{ks} + \uparrow 1_{\delta_k^{\tau-}} [\bar{u}_{k\delta s} - \bar{u}_{ks}] + \uparrow 1_{\delta_k^{\tau+}} [\bar{u}_{k\delta o} - \bar{u}_{k\delta s}]] = \downarrow 1_{|\tau|_k^{-o}} \bar{u}_{k\delta 1}, \downarrow 1_{|\tau|_k^{-o}} \bar{u}_{k\delta 1} = -1/2 \downarrow 1_{|\tau|_k^{-o}} \bar{u}_{ko},$$

leads to sum of the impulse intervals:

$$\bar{u}_{ks} - \bar{u}_{ks} + \bar{u}_{k\delta s} + \bar{u}_{k\delta s} - \bar{u}_{k\delta s} + \bar{u}_{k\delta o} - \bar{u}_{k\delta 1} = 0, \text{ and } \bar{u}_{k\delta 1} = -1/2 \bar{u}_{ko},$$

or to $\bar{u}_{k\delta o} = \bar{u}_{k\delta 1}$. \hfill (4.17a)

Impulse $[\uparrow 1_{\delta_k^{\tau+}} \bar{u}_{k\delta o} - \downarrow 1_{|\tau|_k^{-o}} \bar{u}_{k\delta 1}] = [\uparrow 1_{\delta_k^{\tau+}} + \uparrow 1_{|\tau|_k^{-o}}] \bar{u}_{k\delta}$ contains intervals $\bar{u}_{k\delta} = \bar{u}_{k\delta o} + \bar{u}_{k\delta 1}$,

where from (4.9), (4.13a) follows $\bar{u}_{k\delta} = \bar{u}$, and (4.17a) leads to

$$\bar{u}_{k\delta o} = \bar{u}_{k\delta 1} = 1/2\bar{u} . \tag{4.17b}$$

Interval $\bar{u}_{k\delta} = \bar{u} \times [(o(\delta_k^{\tau+}))(\delta_k^{\tau+} - \tau_k^{-o})^{-1}) + (o(\tau_k^{-o}))(\tau_k^{-o})^{-1}(\delta_k^{\tau+} - \tau_k^{-o})^{-1}\tau_k^{-o}]$ (4.17c)

consists of $\bar{u}_{k\delta}$ components:

$$\bar{u}_{k\delta o} = \bar{u} \times (o(\delta_k^{\tau+}))(\delta_k^{\tau+} - \tau_k^{-o})^{-1}) \text{ and } \bar{u}_{k\delta 1} = \bar{u}_{ko1} \times \bar{u}_{ko2} / \bar{u} , \tag{4.17d}$$

where $\bar{u}_{ko1} = \bar{u} \times (o(\tau_k^{-o}))(\tau_k^{-o})^{-1}$, $\bar{u}_{ko2} = \bar{u}^{-1} \times \tau_k^{-o}(\delta_k^{\tau+} - \tau_k^{-o})^{-1}$.

Intervals $\bar{u}_{ko1}$ and $[\bar{u}_{ko2}/\bar{u}]$ are multiplicative parts of impulse interval $\bar{u}_{k\delta 1}$ covered by starting interval $|\tau|_k^{-o}$.
From (4.17b) and relations (4.17d) it follows

$$\bar{u}_{k\delta o} = \bar{u} \times (o(\delta_k^{\tau+}))(\delta_k^{\tau+} - \tau_k^{-o})^{-1}) = 1/2\bar{u} ,$$

$$(o(\delta_k^{\tau+}))(\delta_k^{\tau+} - \tau_k^{-o})^{-1}) = 1/2 \tag{4.18}$$

and $\bar{u}_{ko1} = \bar{u} \times (o(\tau_k^{-o}))(\tau_k^{-o})^{-1} = 1/2\bar{u}$. (4.18a)

That leads to

$$(o(\tau_k^{-o}))(\tau_k^{-o})^{-1} = 1/2 , \tag{4.18b}$$

and from (4.18) to

$$(\delta_k^{\tau+} - \tau_k^{-o})^{-1}) = (\tau_k^{-o})^{-1}, \delta_k^{\tau+} - \tau_k^{-o} = \tau_k^{-o}, \tau_k^{-o}(\delta_k^{\tau+} - \tau_k^{-o})^{-1} = 1, \text{ then to} \tag{4.18c}$$

$$\tau_k^{-o} = 1/2\delta_k^{\tau+} . \tag{4.18d}$$

From (4.18c) it follows

$$\bar{u}_{ko2} = \bar{u}^{-1} . \tag{4.18f}$$

Applying sequence of Eqs (4.7-4.9), at fixed invariant $\bar{u}$, leads to

$$\bar{u} = u_+(s_k^{+o}) - u_-(\tau_k^{-o}) , \tag{4.19}$$

$$\bar{u} = u_+(\delta_k^{\tau+}) - u_-(\tau_k^{-o}) \text{ at } u_+(\delta_k^{\tau+}) - u_-(\tau_k^{-o}) = \bar{u}_{kb} , \tag{4.19a}$$

which brings invariant $|\bar{u}_s| = |1|_s$ to both impulses (4.19) and (4.19a).
Relation

$$u_+(s_k^{+o}) + u_-(\tau_k^{-o}) = 2[u_-(\delta_k^{\tau-}) + (u_+(\delta_k^{\tau+})] = 0$$

following from the sequence of Eqs (4.7-4.9) leads to

$$u_-(\delta_k^{\tau-}) = -u_+(\delta_k^{\tau+}) , \tag{4.19b}$$

or to reversing (mutual neutralizing) these actions on related moments $\delta_k^{\tau-} \cong \delta_k^{\tau+}$.

Impulse interval $\bar{u}_{k\delta}$, with $\bar{u}_{k\delta o}$ and $\bar{u}_{k\delta 1}$, starts interval of applying step-down control $o(\tau_k^{-o})(\tau_k^{-o})^{-1} = 0.5$ in (4.4) at

$$\bar{u}_{k\delta 1} = \bar{u}_{k\delta o} = 1/2\bar{u}_{k\delta} .$$

Invariant impulse $|\bar{u}_s|=|1|_s$ consisting of two step-actions $[\uparrow 1_{\delta_k^{\tau+}} \downarrow 1_{|\tau|_k^{-o}}]\bar{u}_{k\delta}$, measures intervals

$$\bar{u}_{kb} = \bar{u}_{sm} = \bar{u}_s \text{ at } \bar{u}_{k\delta 1} = 1/2\bar{u}_{sm}. \tag{4.19c}$$

At conditions (4.18c, d), limiting time-jump in (4.13a), step-actions of impulse $\bar{u}_{k\delta}$ applies on two equal time intervals following from (4.19c). On the first

$$(\delta_k^{\tau+} - \tau_k^{-o}) = \delta_k^{\tau+}/2$$

step-up part of $\bar{u}_{k\delta}$-action $[\uparrow 1_{\delta_k^{\tau+}}]$ captures entropy increment

$$\Delta S[\tilde{x}_t / \varsigma_t]|_{\delta_k^{\tau+}}^{\tau_k^{-o}} = 1/2[\uparrow 1_{\delta_k^{\tau+}}]\bar{u}_- = 1/8[\uparrow 1_{\delta_k^{\tau+}}]. \tag{4.20}$$

On the second interval $\tau_k^{-o} = \delta_k^{\tau+}/2$, the captured entropy (4.20) through the step-down multiplicative part (4.17c,d) delivers to cutting action $\bar{u}_{ko} = \bar{u}_- \times o(\tau_k^{-o})(\tau_k^{-o})^{-1}$ the equal contribution

$$\Delta S[\tilde{x}_t / \varsigma_t]|_{\delta_k^{\tau+}}^{\tau_k^{-o}} = 1/4[\downarrow 1_{\tau_k^{-o}}]\bar{u}_- = 1/8[\downarrow 1_{\tau_k^{-o}}]. \tag{4.20a}$$

The control action $[\downarrow 1_{\tau_k^{-o}}]$ at $\bar{u}_- = 0.5$ cuts external entropy of correlation in impulse (4.4) at $\bar{u}_{ko1} = 1/2\bar{u}_{ko}$.

<u>Comment</u>. Action $[\uparrow 1_{\delta_k^{\tau+}}]$ cuts the captured entropy from impulse $\bar{u}_s = |1|_s$, while multiplicative step-down part (4.17b) transforms the captured entropy to the cutting action in (4.4) at $\bar{u}_{ko2} = \bar{u}^{-1}$. •

At the end of $k$ impulse, control action $\bar{u}_+$ transforms entropy (4.20) on interval $\bar{u}_{kio} = \bar{u}_- \times (o(\tau_k^{+o})/\tau_k^{+o})$ to information

$$\Delta I[\tilde{x}_t / \varsigma_t]|_{\delta_{k+}^{\tau+}}^{\tau_k^{+o}} = 1/4[\uparrow 1_{\tau_k^{-o}}]\bar{u}_{kio}\bar{u}_+ = 1/4 \times (-2\bar{u}_{kio})[\uparrow 1_{\tau_k^{-o}}] \tag{4.21}$$

and supplies it to $k+1$ impulse.

(If between these impulses, the entropy increments on the process trajectory are absent (cut)).

That leads to balance equation for information contributions of $k$-impulse:

$$\Delta I[\tilde{x}_t / \varsigma_t]|_{\delta_k^{\tau+}}^{\tau_k^{-o}} + \Delta I[\tilde{x}_t / \varsigma_t]|_{\tau_k^{-o}}^{\tau_k} + \Delta I[\tilde{x}_t / \varsigma_t]|_{\tau_k}^{\tau_k^{+o}} = \Delta I[\tilde{x}_t / \varsigma_t]|_{\delta_{k+}^{\tau+}}^{\tau_k^{+o}}, \tag{4.21a}$$

where interval $\bar{u}_{kio}$ holds information contribution $\Delta I[\tilde{x}_t / \varsigma_t]|_{\tau_k}^{\tau_k^{+o}} = 1/4\bar{u}_{km}$ satisfied (4.4) at $\bar{u}_+ = -2$, which is measured by $\bar{u}_{km} = 0.75$ (4.5). That brings relations

$$0.125 + 0.75 + \bar{u}_{kio} = -2\bar{u}_{kio}, 0.125 + 0.75 + 3\bar{u}_{kio} = 0, \bar{u}_{k1} = 3\bar{u}_{kio} = 0.375 = \bar{u}_- \times o(\tau_k^{+o})/\tau_k^{+o}) \tag{4.22}$$

and $o(\tau_k^{+o})/\tau_k^{+o} = 0.1875,$ (4.22a)

$$\bar{u}_{ko} + \bar{u}_{km} + \bar{u}_{k1} = 1.25 = 5/3\bar{u}_{km} \tag{4.22b}$$

from which and (4.21a) it follows

$$\Delta I[\tilde{x}_t / \varsigma_t]|_{\tau_k}^{\tau_k^{+o}} = 3\Delta I[\tilde{x}_t / \varsigma_t]|_{\delta_k^{\tau+}}^{\tau_k^{-o}}. \tag{4.23}$$

Ratio $\bar{u}_{k1}/2\bar{u}_{kio} = 3/2$ at $2\bar{u}_{kio} = 0.25$ evaluates part of $k$ impulse information transferred to $k+1$ impulse.

Relations (4.17b,d), (4.18b,d,f), (4.19c), and (4.22a) <u>Prove</u> the Proposition parts A-B(including (4.16b)). •

Since $\bar{u}_- = 0.5$ is cutting interval of impulse $\bar{u}_k$, it allows evaluate the additive sum of the discrete cutoff entropy contributions (4.4) during entire impulse $(\downarrow 1_{\tau_k^{-o}} - \uparrow 1_{\tau_k^{+o}}) = \delta_k$ using $\bar{u}_- = \bar{u}_k$:

$$\Delta S[\tilde{x}_t / \varsigma_t]|_{\tau_k^{-o}}^{\tau_k^{+o}} = 1/4\bar{u}_k/2 + 1/2\bar{u}_k + 1/4 \times 3/2\bar{u}_k = \bar{u}_k, \qquad (4.24)$$

That determines the impulse cutoff information measure

$$\Delta S[\tilde{x}_t / \varsigma_t]_{\delta_k} = \Delta I[\tilde{x}_t / \varsigma_t]_{\delta_k} = (\downarrow 1_{\tau_k^{-o}} - \uparrow 1_{\tau_k^{+o}})\bar{u}_k = |1|\bar{u}_k, \bar{u}_k = |1|_k \; Nat \qquad (4.24a)$$

equals to $\cong 1.44$ Bit, which the cutting entropy functional of that random process generates.

That single unit impulse $\bar{u}_k = |1|_k$ measures the relative information intervals

$$\bar{u}_{ko} = 1/3\bar{u}_{km}, \; \bar{u}_{km} = 1, \; \bar{u}_{kio} = 1/3\bar{u}_{km} = \bar{u}_{ko}, \text{ and } \tau_k^{+o}/\tau_k^{-o} = 3. \qquad (4.24b)$$

From relations $\bar{u}_{ko1} = 1/2\bar{u}_{sm}$ and $\bar{u}_{ko1} = 1/2\bar{u}_{ko} = 1/6\bar{u}_{km}$ it follows

$$\bar{u}_{km} = 3\bar{u}_{sm}, \qquad (4.25)$$

which shows that impulse unit $\bar{u}_k = |1|_k$ triples information supplied by entropy unit $\bar{u}_s = |1|_s$ or interval $\bar{u}_k$ compresses three intervals $\bar{u}_s$.

At satisfaction of the extremal principle, each impulse holds invariant interval size $|\bar{u}_k| = |1|_k$ proportional to middle impulse interval $o(\tau)$ with information $\bar{u}_{km}$ which measures $o(\tau)$, and vice versa time $o(\tau)$ measure the information.

Condition of decreasing $t - s_k^{+o} = o(t) \to 0$ with growing $t \to T$ and squeezing sequence $s_k^{+o} \to \tau_{m-1}^{+o}, k = 1, 2....m$ leads to persistence continuation of the impulse sequence with transforming previous impulse entropy to information of the following impulse: $\bar{u}_s = |1|_s \to \bar{u}_k = |1|_k$.

The sequence of growing and compressed information increases at

$$\bar{u}_{k+1} = |3\bar{u}_k| = |1|_{k+1}. \qquad (4.25a)$$

The persistence continuation of the impulse sequence links intervals between sequential impulses $(\bar{u}_{ks}, \bar{u}_{k\delta s}, \bar{u}_{k\delta o})$ whose imaginary (virtual) function $[\uparrow 1_{s_k^{+o}} - \downarrow 1_{\delta_k^{\tau-}} + \uparrow 1_{\delta_k^{\tau+}}]u$ prognosis entropies increments (4.11), (4.12), (4.10).

Information contributions at each cutting interval $\delta_{k-1}, \delta_k, \; k, k+1,...,m$: $\Delta I[\tilde{x}_t / \varsigma_t]_{\delta_{k-1}}, \Delta I[\tilde{x}_t / \varsigma_t]_{\delta_k},.....$ determine time distance interval $\tau_k^{-o} - \tau_{k-1}^{+o} = o_s(\tau_k)$, when each entropy increment $\Delta S[\tilde{x}_t / \varsigma_t]|_{\tau_{k-1}^{+o}}^{t \to \tau_k^{-o}} = 1/2uo(\tau_k) = \bar{u}_s \times o_s(\tau_k)$ supplies each $\Delta I[\tilde{x}_t / \varsigma_t]_{\delta_k}$ satisfying $\bar{u} \times o(\tau_k) = \Delta I[\tilde{x}_t / \varsigma_t]_{\delta_k}$ at $\bar{u} \times o(\tau_k) = \bar{u}_k(\tau_k^{+o} - \tau_k^{-o})$.

Hence, impulse interval

$$\bar{u}_k = \Delta I[\tilde{x}_t / \varsigma_t]_{\delta_k} / (\tau_k^{+o} - \tau_k^{-o}) \qquad (4.26)$$

measures density of information at each $\delta_k = \tau_k^{+o} - \tau_k^{-o}$, which is sequentially increases in each following Bit.

Relations (4.25a,b), (4.26) confirm part C of Proposition 4.3. •

Such Bit includes three parts: 1- the delivered by multiplicative action (4.16b) by capturing entropy of random process;

2-that is delivered by the impulse step-down cut of the process entropy; 3-that is information delivered by the impulse step-up control, and then is transferred to the nearest impulse, keeping information connection between the impulses and providing persistence continuation of the impulse sequence during the process time $T$.

Corollaries 4.2.

A. The additive sum of discrete functions (4.4) during the impulse intervals determines the impulse information measure equals to Bit, generated from the cutting entropy functional of random process.

The step-down function generates $1/8 + 0.75 = 0.875 Nat$ from which it spends $1/8$ $Nat$ for cutting correlation while getting $0.75$ $Nat$ from the cut. Step-up function holds $1/8$ $Nat$ while $0.675$ $Nat$ it gets from cutting $0.75$ $Nat$, from which $0.5 Nat$ it transfers to next impulse leaving $0.125$ $Nat$ within $k$ impulse.

The impulse has $1/8 + 0.75 + 1/8 = 1 Nat$ of total $1.25 Nat$ from which $1/8$ $Nat$ is the captured entropy increment from a previous impulse. The impulse actually generates $0.75 Nat \cong 1 Bit$, while the step-up control, using $1/8 Nat$, transfers $2/8 Nat$ information to next $k$ impulse, capturing $1/8 Nat$ from the entropy impulse between $k$ and $k+1$ information impulses (on interval $o_s(\tau_k)$).

B. From total maximum $0.875$ Nat, the impulse cuts minimum of that maximum $0.75$ $Nat$ implementing minimax principle, which validates variation condition (IV).

By transferring overall $0.375 Nat$ to next $k+1$ impulse that $k$ impulse supplies it with its maximum of $1/3 \times 0.75 Nat$ from the cutting information, thereafter implementing principle maximum of minimal cut.

C. Thus, each cutting Bit is *active information unit* delivering information from previous impulse and supplying information to following impulse, which transfers information between impulses. Such Bit includes: the cutting step-down control's information delivered through capturing external entropy of random process; the cutoff information, which the above control cuts from random process; the information delivered by the impulse step-up control, which being transferred to the nearest impulse, keeps information connection between the impulses, providing persistence continuation of the impulse sequence.

D. The amount of information that each second Bit of the cutoff sequence condenses grows in three times, which sequentially increases the Bit information density. At invariant increments of impulse (4.4), every $\bar{u}_k$ compresses three previous intervals $\bar{u}_{k-1}$ thereafter sequentially increases both density of interval $\bar{u}_k$ and density of these increments for each $k+1$ impulse. ∎

### 3.5. Information path functional in $n$-dimensional Markov process under $n$-cutoff discrete impulses

The IPF unites information contributions extracting along $n$ dimensional Markov process:

$$I[\tilde{x}_t / \varsigma_t]\Big|_{s_k^-}^{\tau_n^{+o} \to T} = \lim_{k=n \to \infty} \sum_{k=1}^{k=n} \Delta I[\tilde{x}_t / \varsigma_t]_{\delta_k}, \tag{5.1}$$

where each dimensional information contribution $\Delta I[\tilde{x}_t / \varsigma_t]_{\delta_k} = |1| \bar{u}_k Nat$, satisfying ratios of intervals

$$\bar{u}_{k+1} = |3\bar{u}_k| = |1|_{k+1}, \bar{u}_k = |3\bar{u}_{k-1}|, \tag{5.2}$$

increases in each third interval in $\bar{u}_{(k+1)} / \bar{u}_{(k-1)} = 9$ times, concentrating in impulse $|1|_{k+1}$, at

$$\bar{u}_k = \Delta I[\tilde{x}_t / \varsigma_t]_{\delta_k} / (\tau_k^{+o} - \tau_k^{-o}) \tag{5.3}$$

which measures density of information at each $\delta_k = \tau_k^{+o} - \tau_k^{-o}$.

Since sequence of $\bar{u}_k, k=1,...,\infty$ is limited by

$$\lim_{k \to \infty} |1| \bar{u}_k \leq |1|_{k \to \infty} , \qquad (5.3a)$$

information contributions of the sequence in (5.1) converges to this finite integral.

Each Bit $|1|_k$ distinguishes from other $|1|_{k+1}$ by condensing the impulse space-time geometry depending on $\bar{u}_k$ density. Each one-dimensional cutoff enables converting entropy increment $\Delta S[\tilde{x}_t / \varsigma_t]_{\tau_k}$ to kernel information contribution $\Delta I[\tilde{x}_t / \varsigma_t]_{\tau_k}$ with optimal density (5.3) satisfying (5.2). The following Propositions detail the IPF specifics.

Proposition 5.1.

A. Optimal distance between nearest $k-1, k$ information impulses, measured by the difference between each previous ending finite intervals $\bar{u}_{k-1}$ and following starting finite interval $\bar{u}_k$ relatively to following interval $\bar{u}_k: \Delta_{uk}^* = (\bar{u}_{k-1} - \bar{u}_k)/\bar{u}_k$, decreases twice for each fixed $k-1,k$:

$$\Delta_{uk}^* = 1/2^k . \qquad (5.4)$$

Indeed,

$$\Delta_{uk}^* = (\bar{u}_{k-1} - \bar{u}_k)/\bar{u}_k = (3/2\bar{u}_{k-1m} - 1/3\bar{u}_{km})/1/3\bar{u}_{km} = 1/2 .$$

Here $3/2\bar{u}_{k-1m}$ measures information of $k-1$ impulse's ending interval, $1/3\bar{u}_{km}$ measures information of $k$ impulse' starting interval. Time intervals $o_s(\tau_k)$ are along the EF measure, $\delta_k$ is impulse intervals on lengthways the IPF measure.

B. With growing $k \to n$, presence of each previous impulse decreases the between impulses distance by $\Delta_{un}^* = 1/2^n$, which at very high process dimension $n$, approaches limit:

$$\lim_{n \to \infty}[\Delta_{un}^* = 1/2^n] \to 0 . \qquad (5.4a)$$

The finite impulses entropy increment, located between the information impulses:

$$\Delta S[\tilde{x}_t / \varsigma_t]|_{\tau_{k-1}^{+o}}^{t \to \tau_k^{-o}} = 1/2uo(\tau_k) = \bar{u}_s \times o_s(\tau_k) , \qquad (5.4b)$$

at finite impulses time $o_s(\tau_k)$, determines density measure for the impulse of invariant size of $\bar{u}_s = |1|_s$:

$$\Delta S[\tilde{x}_t / \varsigma_t]|_{\tau_{k-1}^{+o}}^{t \to \tau_k^{-o}} / o_s(\tau_k) = \bar{u}_s . \qquad (5.5)$$

With growing $k \to n$, the decrease of interval $o_s(\tau_k) \to 0$ is limited by minimal physical time interval.

Eq. (2.2.1) shows that a source of entropy increment (5.4b) between impulses is *time course*

$$\Delta_k = (\tau_k^{-o} - \tau_{k-1}^{+o}) \to o_s(\tau_k) , \qquad (5.5a)$$

which moves the nearest impulses closer. Moreover, each moment $\delta_k^{\tau+}$ of this time course $\Delta_k$ pushes for automatic conversion its entropy density to information density $\bar{u}_k = \Delta I[\tilde{x}_t / \varsigma_t]_{\delta_k} / (\tau_k^{+o} - \tau_k^{-o})$ in information impulse $\bar{u}_k = |1|_k$, where the relative time intervals between impulses (5.4) measures also information density (4.26).

Distance between nearest information impulses (5.5a) evaluates interval of forming entropy increments

$$\Delta_{ks} = 2\tau_{k-1}^{-o} \to o_s . \qquad (5.5b)$$

Finite instances (5.5), (5.5a,b) limit both information density and equivalence of the entropy and information functionals.

Time course intervals (5.5a,b) also runs to convert entropy increment (5.5) in kernel information contribution $\Delta I[\tilde{x}_t / \varsigma_t]_{\delta_k}$ for each cutoff dimension and drives the sequential *integration* for all contributions.

C. With decreasing $\Delta_t = t - s_k^{+o} = o(t)$ at $t \to T$, both $\Delta_k$ and $\delta_k$ are reduced in limit to zero:

$$\lim_{k \to \infty} \Delta_k = 1/2 \lim_{k \to \infty} \delta_k \to 0, \tag{5.6}$$

which follows from

$$\Delta_t = t - s_k^{+o} = o(t) \to \Delta_k = o(\tau_k) \tag{5.6a}$$

at $t \to \tau_k^{-o}, \tau_{k-1}^{+o} \to s_k^{+o}$, and reduces $o(t)$ at $t \to T$.

C. Total sum of the descending time distances at satisfaction (5.4-5.6):

$$\lim_{n \to \infty} \sum_{k=1}^{k=n} \Delta_k = 1/2 \lim_{n \to \infty} \sum_{k=1}^{k=n} \delta_k = T - s \tag{5.7}$$

is finite, converging to total interval of integrating entropy functional (2.1).

D. Sum of information contributions $\Delta I[\tilde{x}_t / \varsigma_t]_{\delta_k}$ on whole $(T - s)$ is converging to both path functional integral and the entropy increments of the initial entropy functional:

$$\lim_{k=n \to \infty} \sum_{k=1}^{k \to n} \Delta I[\tilde{x}_t / \varsigma_t]_{\delta_k} \to I[\tilde{x}_t / \varsigma_t]_s^T = S[\tilde{x}_t / \varsigma_t]_s^T, \tag{5.8}$$

limiting the converging integrals at the finite time interval (5.7).

The integrals time course contributions run integration of the impulse contributions in (2.1). Information density $\bar{u}_k$ of each dimensional information contribution $\Delta I[\tilde{x}_t / \varsigma_t]_{\delta_k}$ grows according to (5.3), approaching infinity at limit (5.6). •

Comments

1. Sequence of integrant of EF (2.1):

$$\delta s_k[\tilde{x}_t / \varsigma_t] = 1/2 u_k o(t_k), \mathrm{k} = 1,...., \infty \tag{5.8a}$$

at limited $u_{k \to \infty} = c^2 > 0$ approaches to

$$\lim_{k \to \infty} \delta s_k[\tilde{x}_t / \varsigma_t] = 1/2 u_k o(t_k) = 0, \tag{5.8b}$$

where each integrant (5.8a) is an entropy density, which impulse control $u_k$ converts to information density.

Hence, information density at infinite dimensions is finite.

2. Sum of invariant information contributions on discrete intervals increases, where (5.8) integrates all previous contributions. This allows integrates any number of the process' connected information Bits, providing total process information including both random inter-states' and inter-Bits connections.

The IPF information concentrates the integrated Bit •

Correlation functions increments within optimal interval $\Delta_k = \tau_k^{-o} - s_k^{+o}$ and on cutoff time borders $\tau_k^{-o}, \tau_k^{+o}$ determine

Proposition 5.2.

A. Correlation function on discreet interval $\Delta_t = t - s_k^{+o}$ for the extremal process holds

$$r_k^-(t) = 1/2 r_k (s_k^{+o})[t^2 / (s_k^{+o})^2 + 1]|_{s_k^{+o}}^{t \to \tau_k^{-o}}, \tag{5.9}$$

ending with correlation on the cutoff left border $\tau_k^{-o}$:

$$r_k^-(\tau_k^{-o}) = 1/2r_k(s_k^{+o})[(\tau_k^{-o}/s_k^{+o})^2 + 1].  \tag{5.9a}$$

After the cutoff, correlation function on following time interval $(\tau_{k+1}^{-o} - \tau_k^{+o})$ holds

$$r_k^+(t) = 1/2r_k(\tau_k^{+o})[t^2/(\tau_k^{+o})^2 + 1]|_{\tau_k^{+o}}^{t \to \tau_{k+1}^{-o}}.  \tag{5.10}$$

B. Correlation on right border $\tau_k^{+o}$ of the finite cutoff at $\tau_k^{+o}/\tau_k^{-o} = 3$ holds:

$$r_k^+(\tau_k^{+o}) = 1/2r_k(\tau_k^{-o})[(\tau_k^{+o}/\tau_k^{-o})^2 + 1]|_{\tau_k^{-o}}^{\tau_k^{+o}} = 5r_k(\tau_k^{-o}).  \tag{5.11}$$

C. Difference of these correlations, according to (5.4), at $\delta_k^r = \tau_k^{+o} - \tau_k^{-o} = 1/2o(\tau_k)$ is

$$r_{ko}^+(\tau_k^{+o}) - r_{ko}^-(\tau_k^{-o}) = \Delta r_{ko}(\delta_k), \ \Delta r_{ko}(\delta_k) = 5r_k(\tau_k^{-o}) - r_k(\tau_k^{-o}) = 4r_k(\tau_k^{-o})  \tag{5.12}$$

and its relative value during that finite cutoff holds

$$\Delta r_{ko}(\delta_k)/r_k(\tau_k^{-o}) = 4.  \tag{5.13}$$

Correlation within cutoff moment $\tau_k = 1/2\delta_k^r = 1/4o(\tau_k)$ evaluates

$$r_k^+(\tau_k) \to 0 \text{ at } o(\tau_k) \to 0.  \tag{5.13a}$$

<u>Proof A, B,C.</u> Relation $t = s_k^{+o}b_k(t)/b_k(s_k^{+o})$, at $b_k(t) = 1/2\dot{r}_k(t)$, determines functions

$\dot{r}_k(t) = 2b_k(s_k^{+o})t/s_k^{+o}$ at $b_k(s_k^{+o})s_k^{+o} = 1/2r_k(s_k^{+o})$ and solution

$$r_k(t) = \int_{s \to s_k^{+o}}^{t \to \tau_k^{-o}} 2b_k(s_k^{+o})t/s_k^{+o} = b_k(s_k^{+o})t^2/s_k^{+o} + C_1, C_1 = 1/2r_k(s_k^{+o}).  \tag{5.14}$$

From (5.14) follows correlation function on this interval (5.9) and its end (5.9a) for the extremal process.

After the cutoff, correlation function on the next time interval $(\tau_{k+1}^{-o} - \tau_k^{+o})$ holds (5.10).

The correlation, preceding the current cut on its left border $\tau_k^{-o}$:

$$r_{ko}^-(\tau_k^{-o}) = 1/2r_k(s_k^{+o})[3^2 + 1] = 5r_k(s_k^{+o}),  \tag{5.15}$$

grows in five time of the optimal correlation for previous cutoff at $s_k^{+o}$.

Correlation on right border $\tau_k^{+o}$ of finite cutoff (5.11) allows finding both difference of these correlations on $\delta_k = \tau_k^{+o} - \tau_k^{-o} = 1/2o(\tau_k)$ in (5.12) and its relative value during the finite cutoff in (5.13). •

Let us find the entropy increments under control $u_-(\tau_k^{-o}), u_+(\tau_k^{-o} - \delta_k^{\tau+})$ near a left border of the cut $t = \tau_k^{-o} - \delta_k^{\tau+}$.

Applying (4.10), (4.11) at $t = \tau_k^{-o} - \delta_k^{\tau+}$ leads to

$$\Delta S[\tilde{x}_t/\varsigma_t]|_{\tau_k^{-o} - \delta_k^{\tau+}}^{t \to \tau_k^{-o}} = -1/2(u_-(\tau_k^{-o}) - u_+(\tau_k^{-o} - \delta_k^{\tau+}))(\tau_k^{-o} - \delta_k^{\tau+})(\delta_k^{\tau+})^{-1}(\tau_k^{-o} - \delta_k^{\tau+}) =$$

$$-1/2u_-(\tau_k^{-o})(\tau_k^{-o} - \delta_k^{\tau+}) - u_+(\tau_k^{-o} - \delta_k^{\tau+})(\tau_k^{-o} - \delta_k^{\tau+})(\tau_k^{-o} - \delta_k^{\tau+}))(\delta_k^{\tau+})^{-1}.  \tag{5.16}$$

If both entropy measure of these controls:

$$\Delta S_u = 1/2[u_-(\tau_k^{-o}) - u_+(\tau_k^{-o} - \delta_k^{\tau+})](\tau_k^{-o} - \delta_k^{\tau+})  \tag{5.17}$$

and interval $(\tau_k^{-o} - \delta_k^{\tau+})$ are finite, then entropy increment near the border is infinite:

$$\Delta S[\tilde{x}_t / \varsigma_t]|_{\tau_k^{-o} - \delta_k^{\tau+}}^{t \to \tau_k^{-o}} = \Delta S_u (\tau_k^{-o} - \delta_k^{\tau+})(\delta_k^{\tau+})^{-1} \to \infty, \text{ at } \delta_k^{\tau+} \to 0. \tag{5.18}$$

Entropy of control (5.17):

$$\Delta S_u = 1/2[u_-(\tau_k^{-o}) - u_+(\tau_k^{-o} - \delta_k^{\tau+})](\tau_k^{-o} - \delta_k^{\tau+}) = 1/2(2j[\uparrow 1_{\delta_k^{\tau+}} + \downarrow 1_{\tau_k^{-o}}])(\tau_k^{-o} - \delta_k^{\tau+}) = j[\uparrow 1_{\delta_k^{\tau+}} + \downarrow 1_{\tau_k^{-o}}](\tau_k^{-o} - \delta_k^{\tau+})$$

at $(\tau_k^{-o} - \delta_k^{\tau+}) = \delta_k^{\tau+}$, (5.19)

compensates for the infinity in (5.18), when imaginary Bit of potential control $j[\uparrow 1_{\delta_k^{\tau+}} + \downarrow 1_{\tau_k^{-o}}]$ applied on interval $(\tau_k^{-o} - \delta_k^{\tau+}) = \delta_k^{\tau+}$ compensates for relative interval $(\delta_k^{\tau+})^{-1}(\tau_k^{-o} - \delta_k^{\tau+})$.

Both real controls $u_-(\tau_k^{-o}), u_+(\tau_k^{-o} - \delta_k^{\tau+})$ generates real Bit $(-1_{\tau_k^{-o}} + 1|_{\delta_k^{\tau+}/2})$ which compensates for this infinite increment.

The opposite actions of functions $u_+(\delta_k^{\tau+}/4)$ and $u_-(t = \delta_k^{\tau+}/2) \to u_-(\tau_k^{-o})$ model an interaction on $\delta_k^{\tau+}/2$ of a random process with applied control $u_-(\tau_k^{-o})$, which provides external influx entropy that this control captures.

If interactive action $u_+(\delta_k^{\tau+}/4)$ precedes $u_-(\tau_k^{-o})$, then this control is a reaction on $u_+(\delta_k^{\tau+}/4)$, while the control information covers the influx of entropy within interval

$$\delta_k^{\tau+}/2 = \tau_k^{-o}. \tag{5.20}$$

Opposite symmetric actions $u_+(\delta_k^{\tau+}/4) = j(+1_{\delta_k^{\tau+}/4})$ and $u_-(t = \delta_k^{\tau+}/2) = j(-1_{\delta_k^{\tau+}/2})$, at

$$(t - \delta_k^{\tau+}/4)^{-1}(\delta_k^{\tau+}/4)^2, t = \delta_k^{\tau+}/2, (t - \delta_k^{\tau+}/4)^{-1}(\delta_k^{\tau+}/4)^2 = \delta_k^{\tau+}/4, \tag{5.21}$$

bring total imaginary entropy (potential) influx:

$$\Delta S[\tilde{x}_t / \varsigma_t]|_{\delta_k^{\tau+}/4}^{\delta_k^{\tau+}/2} = -1/2[j(-1_{\delta_k^{\tau+}/2}) - j(+1_{\delta_k^{\tau+}/4})](\delta_k^{\tau+}/4) = 1/4 j[+1_{\delta_k^{\tau+}/4} - 1_{\delta_k^{\tau+}/2}]\bar{u}_k \delta_k^{\tau+} = 1/4 j[1_{\delta_k^{\tau+}/4}^{\delta_k^{\tau+}/2}]\bar{u}_k \delta_k^{\tau+} \tag{5.22}$$

with two opposite imaginary entropies:

$$S_+[\tilde{x}_t / \varsigma_t]|_{\delta_k^{\tau+}/2}^{t \to \tau_k^{-o}} = 1/8 j[1_{\delta_k^{\tau+}/4}^{\delta_k^{\tau+}/2}]\bar{u}_k \delta_k^{\tau+}, \tag{5.23a}$$

$$S_-[\tilde{x}_t / \varsigma_t]|_{s_k^+}^{t \to \tau_k^{-o}} = -1/8 j[1_{\delta_k^{\tau+}/2}^{\tau_k^{-o}}]\bar{u}_k \delta_k^{\tau+}. \tag{5.23b}$$

Action $u_-(t = \delta_k^{\tau+}/2) = j(+1_{\delta_k^{\tau+}/2})$ coincides with start of real control $u_-(\tau_k^{-o})$, while entropy (5.16) with $S_+[\tilde{x}_t / \varsigma_t]|_{\delta_k^{\tau+}/2}^{t \to \tau_k^{-o}} = 1/8$ Nat evaluates difference between interactive action $u_+(\delta_k^{\tau+}/4)$ and potential reaction $u_-(t = \delta_k^{\tau+}/2)$. The equivalent parts of entropy of interaction between these actions evaluates multiplicative relation $S_+ \times S_- = 1/2 S_-^+$ (following (4.14b)), which for (5.23a, b) leads to

$$S_-^+ = -1/32. \tag{5.24}$$

Information of control, starting at $\delta_k^{\tau+}/2 = \tau_k^{-o}$ with its impulse wide $\bar{u}_k \times \delta_k^{\tau+}$, implements this interactive action spending part of its information

$$\Delta I_-^+ = 0.25 \times 1/32 = 0.0078 \cong 0.008 Nat \tag{5.25}$$

on compensating the interaction, while capturing (5.22) in a move to the cut.

Thus, information covering (5.23b), includes $\Delta I_-^+$ with total contribution

$$\Delta I_-^o = \Delta I_- + \Delta I_-^+ = 0.125 + 0.008 = 0.133 Nat \ . \tag{5.26}$$

The conjugated components $S_+, S_-$ start not simultaneously but with equal values (5.23a,b), acquiring by moment $\delta_k^{\tau+}/2 = \tau_k^{-o}$ dissimilarity between the entropy and information parts:

$$S_-^o = 0.117 Nat \text{ and } \Delta I_-^o = 0.133 Nat \tag{5.27}$$

which satisfies minimal difference between direct action and its reaction at time shift

$$\delta_k^{\tau+}/4 \ . \tag{5.28}$$

Real control $u_-(\tau_k^{-o})$, applied instead of imaginary action $u_-(\delta_k^{\tau+}/2)$, converts total entropy (5.16) on interval (5.28) to the equal control information and compensates for (5.22). That includes (5.23b) and (5.24).

Entropy gap between the anti-symmetric actions (5.23a,b) is imaginable, as well as time interval $\Delta_t = (t - s_k^{+o})$, compared with $t = \tau_k^{-o}$, when control $u_-(\tau_k^{-o})$ of real impulse applies and covers the gap.

At satisfaction (5.24), the delivered information compensates for entropy

$$\Delta S[\tilde{x}_t/\varsigma_t]|_{t \to \delta_k^{\tau+}/4}^{\tau_k^{-o}} \to -1/4 Nats \ . \tag{5.29}$$

Results (5.16)-(5.29) *extend Prop.4.3 specifying information process of capturing external entropy influx in interaction.*

### 3.6. The emerging microprocess

At growing Bayes a posteriori probability along observations, neighbor impulses may merge, generating interactive jump on each impulse border. The merge converges action with reaction, superimposing cause and effect.
Mathematically the jump increase Markov drift (speed) up to infinity (5.18) and [42].

A starting jumping action ↑ interacting with opposite ↓ action of the bordered impulses initiates the impulse inner process $\tilde{x}_{otk} = \tilde{x}(t \in o(\tau_k)))$ called a microprocess.

Because the merge squeezes to a micro-minimum the inter-action interval.

#### 3.6.1. *The entropy increments in the microprocess under starting step-functions.*

Within discrete impulse (3.1), cutting Markov diffusion process, arises impulse inner process $\tilde{x}_{otk} = \tilde{x}(t \in o(\tau_k)))$ called

The microprocess arises under action of function $u_t(u_-^t, u_+^t) = c^2(t \in o(\tau_k))$ at each fixed impulse interval $o(\tau_k)$.

Step-down $u_-^t = u_-(\tau_k^{-o})$ and step-up $u_+^t = u_+(\tau_k^{+o})$ functions acting on discrete interval $o(\tau_k) = \tau_k^{+o} - \tau_k^{-o}$, which satisfy (4.1A-4.1C) and (4.2a-4.2d), generates the EF (2.1) increments:

$$\Delta S_- = \Delta S_-[u_-^t], \Delta S_+ = \Delta S_+[u_+^t], \tag{6.1}$$

preserving the additive and multiplicative properties within the microprocess.

*Starting step functions* $u_\pm^{t1}$ (4.2c) (an analog of $\overline{u}_{k\delta 1}$ in (4.16b)) at locality $\delta_k^{\tau+}/2$ of beginning impulse moment $\tau_k^{-o}$ initiates increments of the entropies on interval $o(\tau_k - 0)$ by moment $t = \tau_k - 0$:

$$\delta S_+[u_+^{t1}] = \delta S_+^1(t = \tau_k - 0)) = \delta S_+^1(t = \tau_k^{-o})\uparrow_{\tau_k^{+o}}(j-1), \delta S_-[u_-^{t1}] = \delta S_-^1(t = \tau_k - 0) = \delta S_-^1(t = \tau_k^{-o})\downarrow_{\tau_k^{-o}}(j+1) \quad (6.2)$$

Step functions $u_\pm^{t2}$ (4.2d) starting at $t = \tau_k - 0$ contribute the entropy increments on interval $o(\tau_k)$ by moment $t = \tau_k$:

$$\delta S_+[u_+^{t2}] = \partial S_+^2(t = \tau_k) = \partial S_+^2(t = \tau_k - 0))\uparrow_{\tau_k}(j+1), \delta S_-[u_-^{t2}] = \partial S_-^2(t = \tau_k) = \partial S_-^2(t = \tau_k - 0))\downarrow_{\tau_k}(-j+1). \quad (6.3)$$

Complex function $u_+^{t1}$ turns on the multiplication of functions $\delta S_+^1(t = \tau_k^{-o})$ on angle $\varphi_+^1 = -\pi/4$, and function $u_-^{t1}$ turns on the multiplication function $\partial S_-^1(t = \tau_k^{-o})$ on angle $\varphi_-^1 = \pi/4$ by moment $t = \tau_k - 0$:

$$\delta S_+^1(t = \tau_k - 0)) = \delta S_+^1(t = \tau_k^{-o}) \times \uparrow_{\tau_k^{+o}} -\pi/4, \delta S_-^1(t = \tau_k - 0)) = \delta S_-^1(t = \tau_k^{-o}) \times \downarrow_{\tau_k^{+o}} \pi/4. \quad (6.4)$$

Analogous, step functions $u_\pm^{t2}$, starting at $t = \tau_k - 0$, turn entropy increments (6.4) on angles $\varphi_-^2 = \pi/4$ by moment $t = \tau_k$ and on angle $\varphi_+^2 = -\pi/4$ the following entropy increments by moment $t = \tau_k$:

$$\delta S_-^2(t = \tau_k) = \delta S_-^2(t = \tau_k - 0) \times \downarrow_{\tau_k} \pi/4, \delta S_+^2(t = \tau_k) = \delta S_+^2(t = \tau_k - 0) \times \uparrow_{\tau_k} -\pi/4. \quad (6.5)$$

The difference of angles between the functions in (6.4): $\varphi_+^1 - \varphi_-^1 = -\pi/2$ is overcoming on time interval $o(\tau_k - 0) = \tau_k^{-o} + 1/2o(\tau_k)$. After that control $u_\pm^{t2}$, starting with opposite increments (6.5), turns them on angle $\varphi_-^2 - \varphi_+^2 = \pi/2$, equalizing (6.5).

That launches *entanglement* of entropies increments and their angles *within* interval $o(\tau_k)$ (on a middle of the impulse):

$$\delta S_+^2(t = \tau_k) = \delta S_-^2(t = \tau_k) = \delta S_\mp^2. \quad (6.5a)$$

Turning the initial time-located vector-function $u_-^t = u_-(\tau_k^{-o}): \xleftarrow{\tau_k^{-o}, \bar{u}_- = 0.5,} \delta\varphi_1 = 0$ on angle $\delta\varphi_1 = \varphi_+^1 - \varphi_-^1 = \pi/2$ transforms it to space vector $u_+(\tau_k - 0) = \uparrow_{\tau_k - o} \bar{u}_+ = 1$ during a jump from moment $t = \tau_k^{-o}$ to moment $t = \tau_k - 0$ on interval $o(\tau_k - 0)$ in (6.4).

Then vector-function $\downarrow_{\tau_k} \bar{u}_-^o = 2$, starting on time $t = \tau_k - 0$ in (6.5) by space interval $\bar{u}_-^o = 2$, jumps to vector-function $\uparrow_{\tau_{k+0}} \bar{u}_+^o = 2$ forming on time interval $o(\tau_k + 0) = 1/2o(\tau_k) + \tau_k^+$ of the additive space-time impulse

$$u_\mp = [\downarrow_{\tau_k+0} \bar{u}_-^o] + [\uparrow_{\tau_k^{+o}} \bar{u}_+^o]. \quad (6.6)$$

The first part of (6.6) equalizes (6.5a) within *space-time* interval $\bar{u}_- \times 1/2o(\tau_k)$, then joins, summing them on $\bar{u}_- \times o(\tau_k + 0)$, which finalizes the entanglement. The last part of impulse (6.6) cuts-kills the entangled increments on interval $\bar{u}_+ \times \tau_k^+$ at ending moment $\tau_k^+$. Section 3.6.4 details the time –space relation.

Relations (6.1-6.6) lead to following specifics of the microprocess.

6.1a. Step functions $u_\pm^{t1}$ initiate microprocess $\tilde{x}_{otk1} = \tilde{x}(t \in o(\tau_k - 0))$ on beginning of the impulse discrete interval $o(\tau_k - 0)$ with only additive increments (6.2). Opposite step functions $u_\pm^{t2}$ continue the microprocess within interval $o(\tau_k + 0)$ at $\tilde{x}_{otk2} = \tilde{x}(t \in o(\tau_k + 0))$ with both additive and multiplicative increments (6.3) preserving the process Markov properties.

6.1b. Space-time impulse (6.6) within interval $o(\tau_k + 0)$ processes entanglement of increments (6.5a) of microprocess $\tilde{x}_{otk2} = \tilde{x}(t \in o(\tau_k + 0))$ summing these increments on $o(\tau_k)$ locality $t = \tau_k$:

$$S_{\mp}^o = 2\delta S_{\mp}^2[(o(\tau_k)]. \tag{6.7}$$

Then it kills entropies (6.7) at ending moment $\tau_k^{o+} \to \tau_k^+$:

$$S_{\mp}^o[\tau_k^+] = 0. \tag{6.7a}$$

The microprocess, producing entropy increment (6.7) within the impulse interval, is reversible before killing which converts the increments in equal information contribution $S_{\mp}^o[\tau_k^+] \Rightarrow \Delta I[\tau_k^+]$. (6.7b)

The information emerging at ending impulse time interval accomplices injection of an energy with step-up control $[\uparrow_{\tau_k^{+o}} \bar{u}_+^o]$, which follows the impulse mutual interaction and/or with environment.

From that moment starts an irreversible information process.

6.1c. Transferring the initial time-located vector to equivalent space-vector $\uparrow_{\tau_k - o} \bar{u}_+$ transforms a transition impulse, concentrating within jumping time $\tau_k^{-o}$ of interval of duration $\bar{u}_- = 0.5$, in space interval $\bar{u}_+ = 1$.

The opposite space vector $\downarrow_{\tau_k} \bar{u}_-^o = 2$ acting on relative time interval $1/2o(\tau_k)/(\tau_k^{+o} - \tau_k^{-o}) = 0.5$ forms space-time function $\downarrow_{\tau_k} \bar{u}_-^1, \bar{u}_-^1 = 2 \times 0.5 = 1$, which, as inverse equivalent of opposite function $\uparrow_{\tau_k - o} \bar{u}_+$, neutralizes it to zero cutting both $\bar{u}_+ = 1$ and time duration $\bar{u}_- = 0.5$ while concentrating them during the transition in interval $\tau_k - (\tau_k - 0) = 0_k$. Within the impulse, only step-down functions $[\downarrow_{\tau_k^{-o}} \bar{u}_-]$ on time interval $\bar{u}_- = 0.5$ and step-up function $[\uparrow_{\tau_k^{+o}} \bar{u}_+^1]$ on space-time interval $\bar{u}_+^1 = \bar{u}_+ \times \tau_k^+ = 2_{\tau_k^+}$ are left. That determines size of the discrete $1-0$ impulse by multiplicative measure $U_m = |0.5 \times 2| = |1|_k = \bar{u}_k$ generating an information bit. Therefore, functions $u_+(\tau_k - 0) = \uparrow_{\tau_k - o} \bar{u}_+$ and $\downarrow_{\tau_k} \bar{u}_-^o$ are transitional during formation of that impulse and creation time-space microprocess $\tilde{x}_{otk} = \tilde{x}(t \in 1/2o(\tau_k), h_k \in 2_{\tau_k^+})$ with final entropy increment (6.7), and a virtual logic; functions that starting the microprocess transits from $\tau_k$ connecting it to actual information (6.7b).

### 3.6.2. *Conjugated dynamics of the microprocess within the impulse*

Opposite functions jumps $u_{\pm}^{t1}(t^*)$ starting at beginning of the process with relative time

$$t^* = [\mp \pi/2 \times (\delta t^{\pm *}/o(\tau_k))], \delta t^{\pm *} \in (\delta t_{ok}^{\pm} \to 1/2o(\tau_k)), \tag{6.8}$$

hold directions of opposite impulses

$$u_{\pm}^{t1} = [u_+ = \uparrow_{t_o^{*-}} (j-1), u_- = \downarrow_{t_o^{*+}} (j+1)] \tag{6.8a}$$

on interval $\delta_o[t_o^{*-}, t_o^{*+}] = \delta t^* < o(\tau)$ at a locality of the impulse initial time $\tau_k^{-o}$.

The opposite jumps (6.8a) initiate relative increments of entropy:

$$\frac{\delta S}{S}/\delta t^* = u_{\pm}^{t1}, [u_+ = \uparrow_{t_o^{*-}} (j-1), u_- = \downarrow_{t_o^{*+}} (j+1)], \tag{6.9}$$

which in a limit leads to differential Egs

$\dot{S}_+(t^*) = (j-1)S_+(t^*), \dot{S}_-(t^*) = (j+1)S_-(t^*)$. (6.9a)

Solutions of (6.9a) describe opposite process' entropies-function of relative time $t^*$ and imaginary symbol $j$ of orthogonality:

$S_+(t^*) = [exp(-t^*)(Cos(t^*) - jSin(t^*))]|_{t_o^{*-}}^{1/2o(\tau_k)}, S_-(t^*) = [exp(t^*)(Cos(t^*) + jSin(t^*))]|_{t_o^{*+}}^{1/2o(\tau_k)}$ (6.10)

with initial conditions $S_+(t_o^{*-}), S_-(t_o^{*+})$ at moment $t_o^{*+} = t_o^{*-} = [\mp\pi/2\delta t_{ok}^\pm]$.

The relative wide of step-function $u_\pm^{t1}: \delta t_o^\pm / o(\tau_k) = 0.2 + 0.005 = 0.205$ and the impulse initial relative interval of this function $\tau_k^{-o} / o(\tau_k) = 0.25$ determine starting moment $\delta t_{ok}^\pm = \delta t_o^\pm / \tau_k^{-o} = \pm 0.82$.

From that it follows the solutions (6.10) by moment $\delta t_{ok}^\pm = \pm 0.82$:

$S_+(t_o^+) = exp(-\pi/2 \times 0.82)[(Cos(\pi/2 \times 0.82)) - jSin(\pi/2 \times -0.82))] \approx 0.2758 \times 1,$
$S_-(t_o^-) = exp(\pi/2 \times -0.82)[(Cos(-\pi/2 \times -0.82) + jSin(-\pi/2 \times -0.82))] \approx 0.2758 \times 1$ (6.10a)

The solutions by moment $\delta t^{*\mp} = 1/2o(\tau_k)$ of time

$t^{*-} = -\pi/2 \times 1/2o(\tau_k)/o(\tau_k) = -\pi/4, t^{*+} = \pi/2 \times 1/2o(\tau_k)/o(\tau_k) = \pi/4$ (6.10b)

are $S_+(t^{*-}) = S_+(t_o^{*-}) \times exp(-\pi/4)[Cos(\pi/4) - jSin(\pi/4)],$
$S_-(t^{*+}) = S_-(t_o^{*+}) \times exp(-\pi/4)[Cos(-\pi/4) + jSin(-\pi/4)] = S_-(t_o^{*+}) \times exp(-\pi/4)[Cos(-\pi/4) - jSin(\pi/4)]$ .(6.11)

These vector-functions at opposite moments (6.10b) hold opposite signs of their angles $\mp\pi/4$ with values:

$S_+(t^{*-}) \cong 0.2758 \times 0.455 \cong +0.125, S_-(t^{*+}) \cong 0.2758 \times 0.455 \cong -0.125$. (6.11a)

Function $u_\pm^{t2}$, starting with these opposite increments, turn them on angle $\varphi_-^2 - \varphi_+^2 = \pi/2$ that equalizes the increments and starts entangling both equal increments with their angles within interval $t = \tau_k \mp 0$:

$S_-^2(t = \tau_k + 0) = \delta S_-^1(t = \tau_k - 0) \times \downarrow_{\tau_k+0} \pi/2 = S_-^1(t = \tau_k - 0) \times exp(\pi/2 \times t_{\tau_k+0}^{*+})[Cos(\pi/2 \times t_{\tau_k+0}^{*+}) + jSin(\pi/2 \times t_{\tau_k+0}^{*+})]$, (6.12)
$S_+^2(t = \tau_k + 0) = \delta S_+^1(t = \tau_k - 0) \times \uparrow_{\tau_k+0} \pi/2 = S_-^1(t = \tau_k - 0) \times exp(-\pi/2 \times t_{\tau_k+0}^{*-})[Cos(-\pi/2 \times t_{\tau_k+0}^{*-}) + jSin(-\pi/2 \times t_{\tau_k+0}^{*-})]$

at moments

$t_{\tau_k+0}^{*\pm} = [\mp\pi \times 2\delta t_{1k}^\pm], \delta t_{1k}^\pm = \delta t_1^\pm / 1/2\tau_k \cong 0.4375, \delta t_1^\pm = \pm(0.5 - \delta t_\pm^{k1}), \delta t_\pm^{k1} = \tau_k^{-o}/\tau_k + \delta t_\pm^{ko}/\tau_k = 0.25 + 0.03125 = 0.2895$ (6.12a)

where $\delta t_\pm^{ko}/\tau_k \cong 32^{-1}$ evaluates dissimilarities (following (5.24-5.27)) between functions

$u_\pm^{t2} = [u_+ = (j+1), u_- = (j-1)]$ at switching from $t = \tau_k - 0$ to $t = \tau_k$.

Resulting values at $t = \tau_k + 0$ are

$S_-^2(t = \tau_k + 0) = 0.125 exp(\pi/2 \times 0.4375) \times 1 \cong 0.25, S_+^2(t = \tau_k + 0) = 0.125 exp(\pi/2 \times 0.4375) \times 1 \cong 0.25$ (6.13a)

Which, being in the same direction, are summing in this locality:

$S_\mp^o = 2S_\mp^2[(\delta t_\pm^{ko}/\tau_k)] \cong \mp 0.5$. (6.13b)

The entanglement, starting with (6.13a), continues at (6.13b) and up to cutting all entangled entropy increments. The $t = \tau_k \mp 0$ locality evaluates the $0_k$-vicinity of action of inverse opposite functions (6.9), whose signs imply the signs of increments in (6.13b) and in the following formulas.

The subsequent step-up function changes increment (6.13b) according to Eqs

$$S_{\mp}(\tau_k^{+o}) = S_{\mp}^o(\delta t_{\pm}^{ko}/\tau_k) \times \exp(t_{\tau_k^{+o}}^{*+}), t_{\tau_k^{+o}}^{*+} = [\pi/2\delta t_k^{*o}], \delta t_k^{*o} \in (\delta t_{1k}^{*o} \to \tau_k^{+o}/\tau_k),$$ (6.14) at

$$\delta t_{1k}^{*o} = \delta t_{1k}^{\pm}/1/2\tau, \delta t_{1k}^{\pm} = \pm(0.5 - \delta t_{\pm}^{k1}), \delta t_{\pm}^{k1} = \delta t_{\pm}^{ko}/\tau_k + \tau_k^{+o}/\tau_k = 0.25 + 0.03125 = 0.2895,$$ (6.14a1)

$$\delta t_{1k}^{\pm} = \delta t_1^{\pm}/1/2\tau_k \cong 0.4375 \text{ with resulting value}$$

$$S_{\mp}(\tau_k^{+o}) = \mp 0.5 \exp(\pi/2 \times 0.4375) = \mp 0.5 \times (\cong 2) \cong \mp 1,$$ (6.14a)

which measures total entropy of the impulse $\bar{u}_k = |1|_k = 1 Nat$. (6.14b)

Trajectories (6.10-6.14) describe anti-symmetric conjugated dynamics of the microprocess within the impulse, which up to the cutting moment is reversible, generating the entangled entropy increments (6.14a).

The entanglement starts at angle $(\pi/2) \times 0.4375 < \pi/4$ takes relative time interval of the impulse $\delta t_{\pm}^{ko}/\tau_k \cong 0.03125$ and ends by angle $\pi/2$. Since only by angle $\pi/2$ the space interval within impulse begins, it means that *the entanglement starts before the space is formed and ends with beginning the space.*

Here $\tau_k = 1/2o(\tau), o(\tau) = 1Nat$ and $\delta t_{\pm}^{ko} = 0.03125 \times 1/2o(\tau) = 0.015625 o(\tau) = \varepsilon_{ok}$.

*Moreover, the entanglement creates the space during that time interval which is reversible.*

Comments. A potential path during creation both entanglement and space could be a *wormhole*-a short cut in space-time predicted by general relativity. But real *space curvature do not exists during this time*. It may emerge only after entanglement by a moment of forming bit at the end of the impulse. Hence, space curvature may form at the *end of microprocess-analog of quantum process-* when the bit, as an elementary unit of macroprocess, emerges.

•

Cutting this joint entropy at moment $\tau_k^+ \cong 0_k + \tau_k^{o+}$ coverts it to equal information contribution

$S_{\mp}^o[\tau_k^+] = \Delta I[\tau_k^+] \cong 1.44 \text{ bit}$ (6.14c) that each $\bar{u}_k$ impulse produces.

The cut involves an interaction which imposes irreversibility on information process with multiple cutting bits.

Interacting impulse outside of the impulse microprocess delivers entropy on $0_k$-vicinity of the cutting moment:

$$S_c^*(\tau_k^+) = \exp 0_k = 1.$$ (6.14d)

The $t = \tau_k \mp 0$ locality evaluates the $0_k$-vicinity of action of inverse opposite functions (6.9), whose signs imply the signs of increments in (6.14) and in the following formulas.

Each current impulse requests an interaction for generating information bit from the microprocess reversible entropy, since the impulse contains the requested action $[\uparrow_{\tau_k^{+o}} \bar{u}_+^o]$ (in (6.6)).

*3.6.3. Probabilities functions of the microprocess*

Amplitudes of the process probability functions at $S_{\mp}^*(\tau_k^{+o}) = |S_+^*| = |S_-^*| = 1$ are equal:

$$p_{+a} = 0.3679, p_{-a} = 0.3679.$$ (6.15)

That leads to

$$p_{+a} p_{-a} = p_{\pm a}^2 = 0.1353, S_{\mp a}^* = -\ln p_{a\pm}^2 = 2,$$ (6.15a)

or at $S_{\mp a}^* = 2$, to $p_{a\pm} = \exp(-2) = 0.1353$, (6.15b)

where $S_{\mp a}^* = S_{\mp}^*(\tau_k^{+o}+) + S_c^*(\tau_k^+ +)$ (6.16)

includes the interactive components at $\tau_k^{+o}+$ following $k$ impulse.

Space interval (6.14a1) starts with probability $P_\Delta^*(\delta t_\pm^{k1}) = 0.821214$, (6.16a)

where $P_\Delta^*(\delta t_\pm^{k1}) = \exp(-|S_\mp^*(\delta t_\pm^{k1})|), \delta t_\pm^{k1} = 0.2895, S_\mp^*(\delta t_\pm^{k1}) = \mp 0.125\exp(\pi/2 \times \delta t_\pm^{k1}) = \mp 0.1969415$ (6.16b)

Functions $u_+ = (j-1), u_- = (j+1)$, satisfying (4.IA), fulfill additivity at the impulse starting interval $o[t_o^\mp]$, running the anti-symmetric entropy fractions. Opposite functions $u_+ = (1+j), u_- = (1-j)$, satisfying (4.IB) by the end of impulse at $\uparrow_{\tau_{k+}^{+o}} \bar{u}_\pm$, mount entanglement of these entropy fractions within impulse' $|1/2 \times 2| = |\bar{u}_k| = |1|_k$ space interval $\bar{u}_\pm = \pm 2$.

The entangling fractions hold the equal impulse probabilities (6.15), which indicates appearance of both entangled anti-symmetric fractions simultaneously with starting space interval.

Probabilities $p_{\pm a}$ of interacting probability amplitudes $p_{+a}, p_{-a}$ satisfies multiplicativity $p_{\pm a} = \sqrt{p_{+a} p_{-a}}$, but sum of the non-interacting probabilities does not: $p_+ + p_- = \exp(-S_+^*) + \exp(-S_-^*) = p_\pm \neq p_{a\pm}$, being unequal to both interacting probability $p_{\pm a}$ and the summary probability $p_{\pm am} = 0.7358$ of the non-interacting entropy (6.15).

The interacting probabilities in transitional impulse $[\uparrow 1_{\tau|_{\bar{k}}} \downarrow 1_{\tau|_{\bar{k}}^+}]\bar{u}_k$ on $\tau_k$-locality violates their additivity, but preserves additivity of the entropy increments.

The impulse microprocess on the ending interval preserves both additivity and multiplicativity only for the entropies.
These basic results are the impulse' entropy and probability' equivalents for the quantum mechanics (QM) probability amplitudes relations. However, the impulse cutting probabilities $p_+, p_-$, are probability of random events in the hidden correlations, while probability amplitudes $p_{+a}, p_{-a}$ are attributes of the microprocess starting within the cutting impulse.
That distinguishes the considered microprocess from the related QM equations, considered physical particles.
The entropy of multiple impulses integrate microprocess along the observing random distributions.
With minimal impulse entropy ½ Nat starting a virtual observer, each following impulse' initial entropy $S_\pm(t_o) = 0.25 Nat$ self-generates entropy $S_{\mp a}^* = 0.5 Nat$. Thus, the virtual observer' microprocess starts with probability $p_{a\pm} = \exp(-0.5) = 0.6015$. Probability $p_{a\pm} = 0.1353$, relational to the impulse initial conditions, evaluates appearance of time–space actual impulse (satisfying (6.16)) that *decreases* its initial entropy on $S_{\mp a}^* = 2$ Nat.

The impulse's invariant measure, satisfying the minimax, preserves $p_{a\pm}$ along the time-space microprocess for multiple time-space impulses. Reaching probability of appearance the time-space impulse needs $m_p = 0.6015/0.1353 \cong 4.4457 \approx 5$ multiplications of invariant $p_{a\pm} = 0.1353$.

Hence, each reversible microprocess within the impulse generates invariant increment of entropy, which enables sequentially minimize the starting uncertainty of the observation.

Assigning entropy (6.15b) minimal uncertainty measure $h_\alpha^o = 1/137$ - physical structural parameter of energy, which includes the Plank constant's equivalent of energy, leads to relation:

$S_{\mp a}^* = 2h_\alpha^o, p_{\pm a} = \exp(-2h_\alpha^o) = 0.98555075021 \to 1$, (6.17)

which evaluates probability of real impulse' physical strength of coupling independently chosen entropy fractions.

The initially orthogonal non-interacting entropy fractions $S_{+a}^* = h_\alpha^o, S_{-a}^* = h_\alpha^o$, at potential mutual interactive actions, satisfy multiplicative relation

$S_{\mp a}^* = (h_\alpha^o)^2 [Cos^2(\bar{u}t) + Sin^2(\bar{u}t)]|_{t_o^\mp}^{t=1/2\tau} = (h_\alpha^o)^2 = inv$ (6.18)

which at $S^*_{\mp a} = (h^o_\alpha)^2 \to 0$ approaches $p^*_{\pm a} = \exp[-(h^o_\alpha)^2] \to 1$.

The impulse interaction adjoins the initial orthogonal geometrical sum of entropy fractions in liner sum $2h^o_\alpha$.

Starting physical coupling with double structural $h^o_\alpha$ creates initial information triple with probability (6.17).

<u>Examples</u>. Let us find which of the entropy functional expression meets requirements (4.1A,B) within discrete intervals $\Delta_t = (t-s) \to o(t)$, particularly on $\Delta_k = (\tau_k^{-o} - s_k^{+o}) \to o(\tau_k^{-o})$ under opposite functions $u_+, u_-$:

$$u_+(s_k^{+o}) = +1_{s_k^{+o}} \bar{u} = \bar{u}(s_k^{+o}), u_- = -u_+(s_k^{+o}) = -1_{s_k^{+o}} \bar{u}.$$  (6.19)

Following relations (4.11), we get entropy increments

$$S_+[\tilde{x}_t/\varsigma_t]|_{s_k^+}^{t\to\tau_k^{-o}} = -1/2[u_+(s_k^{+o})](\tau_k^{-o} - s_k^{+o})^{-1}(s_k^{+o})^2 = -1/2[u_+(s_k^{+o})(s_k^{+o})^2/s_k^{+o}(3-1)] = -1/4[u_+(s_k^{+o})s_k^{+o}]$$ (6.20)

$$S_-[\tilde{x}_t/\varsigma_t]|_{s_k^+}^{t\to\tau_k^{-o}} = 1/2[u_-(s_k^{+o})](\tau_k^{-o} - s_k^{+o})^{-1}(s_k^{+o})^2 = 1/2[u_-(s_k^{+o})(s_k^{+o})^2/s_k^{+o}(3-1)] = 1/4[u_-(s_k^{+o})s_k^{+o}],$$

which satisfy

$$S_+[\tilde{x}_t/\varsigma_t]|_{s_k^+}^{t\to\tau_k^{-o}} = -S_-[\tilde{x}_t/\varsigma_t]|_{s_k^+}^{t\to\tau_k^{-o}},$$  (6.21a)

$$S_+[\tilde{x}_t/\varsigma_t]|_{s_k^+}^{t\to\tau_k^{-o}} - S_-[\tilde{x}_t/\varsigma_t]|_{s_k^+}^{t\to\tau_k^{-o}} = -1/2[\bar{u}(s_k^{+o})s_k^{+o}] = \Delta S[\tilde{x}_t/\varsigma_t]|_{s_k^+}^{t\to\tau_k^{-o}}.$$  (6.21b)

Relations

$$4S_+[\tilde{x}_t/\varsigma_t]|_{s_k^+}^{t\to\tau_k^{-o}}/s_k^{+o} = -\bar{u}(s_k^{+o})(s_k^{+o}) = -2\times 1_{s_k^{+o}},$$

satisfy conditions

$$4S_+[\tilde{x}_t/\varsigma_t]|_{s_k^+}^{t\to\tau_k^{-o}}/s_k^{+o} \times 4S_-[\tilde{x}_t/\varsigma_t]|_{s_k^+}^{t\to\tau_k^{-o}}/s_k^{+o} = -\bar{u}(s_k^{+o}) \times \bar{u}(s_k^{+o}) = -(2\times 1_{s_k^{+o}}) \times (2\times 1_{s_k^{+o}}) = -4\times 1_{s_k^{+o}},$$ (6.22a)

$$4S_+[\tilde{x}_t/\varsigma_t]|_{s_k^+}^{t\to\tau_k^{-o}}/s_k^{+o} - 4S_-[\tilde{x}_t/\varsigma_t]|_{s_k^+}^{t\to\tau_k^{-o}}/s_k^{+o} = -\bar{u}(s_k^{+o}) - \bar{u}(s_k^{+o}) = -(2\times 1_{s_k^{+o}}) - (2\times 1_{s_k^{+o}}) = -4\times 1_{s_k^{+o}}.$$ (6.22b)

These entropy expressions at any *current* moment $t$ within $\Delta_t = (t-s_k^{+o})$ do not comply with (4.1A,B).

The same results hold true for the entropy functional increments under functions

$$u_+ = +1_{s_k^{+o}} \bar{u}, u_- = -1_{\tau_k^{-o}} \bar{u}.$$  (6.23)

Indeed. For this functions on $\Delta_t = (t-s_k^{+o})$ we have

$$\Delta S[\tilde{x}_t/\varsigma_t]|_{s_k^+}^{t} = -1/2(u_-(t) - u_+(s_k^{+o}))(t-s_k^{+o})^{-1}(s_k^{+o})^2$$  (6.24)

which for $t \to \tau_k^{-o}$ holds

$$\Delta S[\tilde{x}_t/\varsigma_t]|_{s_k^+}^{t\to\tau_k^{-o}} = -1/2(u_-(\tau_k^{-o}) - u_+(s_k^{+o}))(\tau_k^{-o} - s_k^{+o})^{-1}(s_k^{+o})^2,$$

and satisfies relations

$$S_+[\tilde{x}_t/\varsigma_t]|_{s_k^+}^{t\to\tau_k^{-o}} - S_-[\tilde{x}_t/\varsigma_t]|_{s_k^+}^{t\to\tau_k^{-o}} = \Delta S[\tilde{x}_t/\varsigma_t]|_{s_k^+}^{t\to\tau_k^{-o}},$$  (6.24a)

$$S_+[\tilde{x}_t/\varsigma_t]|_{s_k^+}^{t\to\tau_k^{-o}} = -S_-[\tilde{x}_t/\varsigma_t]|_{s_k^+}^{t\to\tau_k^{-o}}$$  (6.24b)

which determines

$$S_+[\tilde{x}_t/\varsigma_t]|_{s_k^+}^{t\to\tau_k^{-o}} = -1/4(u_-(\tau_k^{-o}) - u_+(s_k^{+o}))(\tau_k^{-o} - s_k^{+o})^{-1}(s_k^{+o})^2$$  (6.25a)

$$S_-[\tilde{x}_t/\varsigma_t]|_{s_k^+}^{t\to\tau_k^{-o}} = 1/4(u_-(\tau_k^{-o}) - u_+(s_k^{+o}))(\tau_k^{-o} - s_k^{+o})^{-1}(s_k^{+o})^2.$$  (6.25b)

We get the entropy expressions through opposite directional discrete functions in (6.25a,b):

$$S_+[\tilde{x}_t/\varsigma_t]|_{s_k^+}^{t\to\tau_k^{-o}} 4(\tau_k^{-o}-s_k^{+o})^{-1}(s_k^{+o})^2 = -(u_-(\tau_k^{-o})-u_+(s_k^{+o})),$$

$$S_-[\tilde{x}_t/\varsigma_t]|_{s_k^+}^{t\to\tau_k^{-o}} 4(\tau_k^{-o}-s_k^{+o})(s_k^{+o})^{-2} = u_-(\tau_k^{-o})-u_+(s_k^{+o}),$$

which satisfy additivity at

$$-2(u_-(\tau_k^{-o})-u_+(s_k^{+o})) = -2(-1_{\tau_k^{-o}}\bar{u}-1_{s_k^{+o}}\bar{u})] = 2\bar{u}[1_{\tau_k^{-o}}+1_{s_k^{+o}}] = 4[1_{\tau_k^{-o}}+1_{s_k^{+o}}]. \tag{6.26}$$

While for each

$$S_+[\tilde{x}_t/\varsigma_t]|_{s_k^+}^{t\to\tau_k^{-o}} 4(\tau_k^{-o}-s_k^{+o})(s_k^{+o})^{-2} = -\bar{u}[-1_{\tau_k^{-o}}-1_{s_k^{+o}}], \tag{6.26a}$$

$$S_-[\tilde{x}_t/\varsigma_t]|_{s_k^+}^{t\to\tau_k^{-o}} 4(\tau_k^{-o}-s_k^{+o})(s_k^{+o})^{-2} = \bar{u}[-1_{\tau_k^{-o}}-1_{s_k^{+o}}] \tag{6.26b}$$

satisfaction of both 4.1A, B:

$$-(u_-(\tau_k^{-o})-u_+(s_k^{+o}))\times(u_-(\tau_k^{-o})-u_+(s_k^{+o})) = -[u_-(\tau_k^{-o})-u_+(s_k^{+o})]^2,$$

requires

$$\bar{u} = -2j, \tag{6.27}$$

when $-[u_-(\tau_k^{-o})-u_+(s_k^{+o})]^2 = (-2j)^2[-1_{\tau_k^{-o}}-1_{s_k^{+o}}]^2.$ \hfill (6.27a)

Simultaneous satisfaction of both 4.1.A, B leads to

$$\Delta S[\tilde{x}_t/\varsigma_t]|_{s_k^+}^{t\to\tau_k^{-o}} 2(\tau_k^{-o}-s_k^{+o})(s_k^{+o})^{-2} = -2\bar{u}[-1_{\tau_k^{-o}}-1_{s_k^{+o}}] = 4j[1_{\tau_k^{-o}}+1_{s_k^{+o}}],$$

$$-(-1_{\tau_k^{-o}}\bar{u}-1_{s_k^{+o}}\bar{u})\times(-1_{\tau_k^{-o}}\bar{u}-1_{s_k^{+o}}\bar{u}) = (-2j)^2(-1_{\tau_k^{-o}}+1_{s_k^{+o}})^2. \tag{6.27b}$$

At $o(t)\to 0$, these admit an instant existence of both $(-1_{\tau_k^{-o}}\bar{u},+1_{s_k^{+o}}\bar{u})$.

Thus, under function (6.25), the entropy expressions are imaginary:

$$S_+[\tilde{x}_t/\varsigma_t]|_{s_k^+}^{t\to\tau_k^{-o}} 4(\tau_k^{-o}-s_k^{+o})(s_k^{+o})^{-2} = -2j[-1_{\tau_k^{-o}}+1_{s_k^{+o}}] = -2j[1_{s_k^{+o}}^{\tau_k^{-o}}], \tag{6.28a}$$

$$S_-[\tilde{x}_t/\varsigma_t]|_{s_k^+}^{t\to\tau_k^{-o}} 4(\tau_k^{-o}-s_k^{+o})(s_k^{+o})^{-2} = 2j[-1_{\tau_k^{-o}}+1_{s_k^{+o}}] = 2j[1_{s_k^{+o}}^{\tau_k^{-o}}], \tag{6.28b}$$

at their multiplicative and additive relations:

$$S_-[\tilde{x}_t/\varsigma_t]|_{s_k^+}^{t\to\tau_k^{-o}} 4(\tau_k^{-o}-s_k^{+o})(s_k^{+o})^{-2} \times S_+[\tilde{x}_t/\varsigma_t]|_{s_k^+}^{t\to\tau_k^{-o}} 4(\tau_k^{-o}-s_k^{+o})(s_k^{+o})^{-2} = 4, \tag{6.29a}$$

$$S_+[\tilde{x}_t/\varsigma_t]|_{s_k^+}^{t\to\tau_k^{-o}} 4(\tau_k^{-o}-s_k^{+o})(s_k^{+o})^{-2} - S_-[\tilde{x}_t/\varsigma_t]|_{s_k^+}^{t\to\tau_k^{-o}} 4(\tau_k^{-o}-s_k^{+o})(s_k^{+o})^{-2} = -j2[1_{\tau_k^{-o}}+1_{s_k^{+o}}] = -j2[1_{s_k^{+o}}^{\tau_k^{-o}}],$$

$$S_+[\tilde{x}_t/\varsigma_t]|_{s_k^+}^{t\to\tau_k^{-o}} - S_-[\tilde{x}_t/\varsigma_t]|_{s_k^+}^{t\to\tau_k^{-o}} = 1/2j[1_{s_k^{+o}}^{\tau_k^{-o}}], \tag{6.29b}$$

Relations (6.26),(6.26a,b) satisfy additivity only at points $\tau_k^{-o}, s_k^{+o}$.

Between these points, within $\Delta_t = (t-s_k^{+o})\to o(t)$, entropy expressions (6.28a, b), (6.29b) are imaginary.

Time direction may go back within this interval until an interaction occurs.

Within this interval, entropy $S_t = S_+ - S_-$ holds relations

$$(S_+ - S_-)^2 = S_+^2 + S_-^2 - 2S_+ \times S_-, S_+ = -1/2 jS_t^+, S_- = 1/2 jS_t^-, S_+^2 = -1/4 S_t^{\pm 2}, S_-^2 = 1/4 S_t^{\pm 2}, \tag{6.30}$$

leading to

$$-2S_+ \times S_- = -2(-1/4)jjS_t^{\pm 2} = -1/2S_t^{\pm 2}, \text{ while } S_+^2 + S_-^2 = -1/2S_t^{\pm 2} \text{ and } (S_+ - S_-)^2 = (-jS_t^+)^2 = -(S_t^+)^2.$$

At fulfillment of 4.1A,B, it follows relations

$$S_{t=\tau}^2 = -1/2 jS_{t\to\tau}, S_{t=\tau}^2 = S_+ \times S_- = -1/4S_{t\to\tau}^{\pm 2}, S_{t\to\tau} = \pm 2 jS_{t=\tau}, \tag{6.30a}$$

from which also follows

$$S_{t \to \tau} = 2j.$$ (6.30b)

These examples concur with *Secs. 6.2, 6.3 and illustrate it. Results show that a window of interaction with environment opens only on the impulse border twice: at begging, between moments $\delta_k^{\tau+}/4$ and $\tau_k^{-o}$, when the entropy flow with energy accesses impulse, and at the end of a gap when an entangle entropy converts to equivalent information.*

### 3.6.4. The relation between the curved time and equivalent space length within an impulse

Let us have a two-dimensional rectangle impulse with wide $p$ measured in time $[\tau]$ unit and high $h$ measured in space length $[l]$ unit, with the rectangle measure

$$M_i = p \times h.$$ (6.31)

The problem: Having a measure of wide part of the impulse $M_p$ to *find* high $h$ at equal measures of both parts:

$$M_p = M_h \text{ and } M_p + M_h = M_i.$$ (6.32)

From that it follows

$$M_h = 1/2 M_i = 1/2 p \times h.$$ (6.33)

Assuming that the impulse has only wide part $1/2 p$, it measure equals $M_p = (1/2 p)^2$.

Then from $M_p = (1/2 p)^2 = M_h = 1/2 p \times h$ it follows

$$h/p = 1/2.$$ (6.34)

Let us find a length unit $[l]$ of the curved time unit $[\tau]$ rotating on angle $\pi/2$ using relations

$$2\pi h [l]/4 = 1/2 p [\tau] \quad (6.35a), \quad [\tau]/[l] = \pi h / p.$$ (6.35)

Substitution (6.31) leads to ratio of the measured units:

$$[\tau]/[l] = \pi / 2.$$ (6.36)

Relation (6.36) sustains orthogonality of these units in time-space coordinate system, but since initial relations (6.32) are linear, ratio (6.36) represents a linear connection of time-space units (6.35) through curving the time unit in the impulse-jumps (6.9). According to Prop.4.3, the impulse' invariant entropy implies the multiplication starting rotation in Secs 3.6.1-2. The microprocess, built in rotation movement, curving the impulse time, adjoins the initial orthogonal axis of time and space coordinates (Fig.1).

The impulses, preserving multiplicative and additive measures (6.31), (6.32), have common ratio of $h/p = 1/2$, whose curving wide part $p = 1/2$ brings universal ratio (6.36), which concurs with Lemma 4.1 (4.2a).

At above assumption, measure $M_h$ does not exist until the impulse-jump curves its only time wide $1/2 p$ at transition of the impulse, measured only in time, to the impulse, measured both in time $1/2 p$ and space coordinate $h$.

According to (6.33), measure $M_h$ emerges only on a half of that impulse' total measure $M_i$.

The transitional impulse could start on border of the virtual impulses $\downarrow\uparrow$, where transition curving time $\delta t_p = 1/2 p$ under impulse-jump during $\delta t_p \to 0$ leads to

$$M_p \to 0 \text{ at } M_h \Rightarrow M_i = p \times h.$$ (6.37)

If a virtual impulse $\downarrow\uparrow$ has equal opposite functions $u_-(t), u_+(t+\Delta)$, at $\bar{u}_+ = \bar{u}_-$, the additive condition for measure (4.2a): $U_a(\Delta) = 0$ is violated, and the impulse holds only multiplicative measure $U_m(\Delta) \neq 0$ in relation (4.2C): $U_m(\Delta) = U_{am}$ which is finite only at $\bar{u}_+ = \bar{u}_- \neq 0$.

If any of $\bar{u}_+ = 0$, or $\bar{u}_- = 0$, both multiplicative $U_m(\Delta) = 0$ and additive $U_a(\Delta) = 0$ disappear.

At $\bar{u}_- \neq 0$, $U_a(\Delta)$ is a finite and positive, specifically, at $\bar{u}_- = 1$ leads to $U_a(\Delta) = 1$ preserving measure $U_{amk} = |U_a|_k$. Existence of the transitional impulse has shown in Secs. 3. 4 (4.19a, b), 3. 6.1, and 3. 6.2.

An impulse-jump at $o[t_o^{\mp}] \to \delta t_p \to 0$ curves a "needle pleat" space at transition to the finite form of the impulse. The Bayes probabilities measure may overcome this transitive gap.

Since entropy (2.1) is proportional to the correlation time interval, whose impulse curvature $K_s = h[l]^{-1}$ is positive, this entropy is positive. The curving needle cut changes the curvature sign converting this entropy to information.

### 3.6.4.1. The curvature of the impulse
An external step-down control carries entropy evaluated by [17]:
$$\delta_{ue}^i = 1/4(u_{io} - u_i), \tag{6.38}$$
where $u_{io} = \ln 2 \cong 0.7 Nat$ is total cutoff entropy of the impulse and $u_i \cong 0.5 Nat$ is its cutting part.

The same entropy-information carries the impulse step-up control, while both cutting controls carry $\delta_{ueo}^i \cong 0.1 Nat$.

That evaluates information wide of each single impulse control's cut which the impulse carries:
$$\delta_{ue}^i \cong 0.05 Nat. \tag{6.38a}$$

Following [17], the starting the step-down part, the step-up part, transferring entropy to create information, and the final cutting part generating information carry the entropy measures accordingly: $\delta_{ue1}^i \cong 0.025 Nat, \delta_{ue2}^i \cong 0.02895, \delta_{ue3}^i \cong 0.01847 Nat$.

These relations allow estimate Euclid's curvature $K_{e1}$ of the impulse step-down part, related to currying entropy $0.25 Nat$ and its increment $\delta K_{e1}$:
$$K_{e1} = (r_{e1})^{-1}, r_{e1} = \sqrt{1 + (0.025/0.25)^2} = \mp 1.0049875, K_{e1} \cong -0.995037, \delta K_{e1} \cong -0.004963. \tag{6.39}$$

The cutting part's curvature estimates relations
$$K_{eo} = (r_{eo})^{-1}, r_{eo} = \mp\sqrt{1 + (0.1/0.5)^2} = 1.0198, K_{eo} \cong -0.98058, \delta K_{eo} \cong -0.01942. \tag{6.39a}$$

The transferred part's curvature estimates relations
$$K_{e2} = (r_{e2})^{-1}, r_{e2} = \sqrt{1 + (0.02895/0.25)^2} \cong 1.0066825, K_{e2} \cong +0.993362, \delta K_{e2} \cong 0.006638 \tag{6.39b}$$
which is opposite to the step-down part.

The final part cutting all impulse entropy estimates curvatures
$$K_{e3} = (r_{e2})^{-1}, r_{e3} = \sqrt{1 + (0.01847/\ln 2)^2} \cong \pm 1.014931928, K_{e3} \cong +0.99261662, \delta K_{e3} \cong -0.00738338.$$
whose sign is same as the step-down part.

Thus, the entropy impulse is curved with three different curvature values (Fig.1a).

These values estimates each impulse' curvatures holding the invariant entropies, which emerge in minimax cutoff impulse carrying entropy $S_{ki} = 0.5$ and a priori probability (6.15) after getting multiple probing impulse $m_p$.

Since the rectangle impulse, cutting time correlation, has measure $M = |1|_M$, the curving impulse, cutting the curving correlation, defines measure

$$r_{iM} = M \times K_{ei}, \qquad (6.40)$$

where not cutting time correlations possess Euclid's curvature $K_{iM} = 1$.

Accordingly, the impulse with both time and space measure $|M_{io}| = \pi$, which could appear in transitional impulse curvature of cutting part $K_{eo}$, defines correlation measures

$$r_{icM} = M_{io} \times K_{eo}. \qquad (6.40a)$$

Ratio $r_{icM} / r_{iM} = \pi / |1| K_{ei} / K_{eio}$ (6.40b) measures increment of the curved impulse correlations at appearance the impulse with emerging space coordinate, relative to the curved correlation of the impulse with only time correlation measure. Counting (6.40b) leads ratio $r_{icM} / r_{iM} = (\pi / |1|) K_{eo} / K_{e3} \cong 3.08$.

Relative increment of correlation $\Delta r_{iM} / r_{iM} = (r_{iM} + r_{icM}) / r_{iM} = 1 + r_{icM} / r_{iM} \square 4$ concurs with (5.12), which in limit: $\lim_{\Delta r(\Delta t), \Delta t \to 0} [\Delta r_{iM} / r_{iM}) = \dot{r}_{icM} / r_{iM}$ brings the equivalent contribution to integral functional IPF (4.21).

Measure $|M_{io}| = |[\tau] \times [l]| = \pi$ satisfies $[\tau] = \pi / \sqrt{2}, [l] = \sqrt{2}$ at $[\tau] / [l] = \pi / 2$. (6.40c)

The shortening of cutting time intervals (Sec.3.5) triples density of each minimax cut impulse (3.4) preserving it invariant curving correlation measure (6.40). Since any impulse with virtual 0-1 Kolmogorov probability preserve its virtual measure (6.40b), the related time virtual correlations is able to create the space that triple its measure. That compresses the impulse rising curvature of the invariant impulse, which increases probability of cutting its time and emerging space coordinate.

After accumulating energy these information curvatures evaluate the impulse gravity.

### 3.6.5. Identifying an edge of the observer's certainty-reality ending uncertainty within the gap

#### 3.6.5.1. Random, quantum and a mean time intervals

Random process is function $x(\omega)$ of collections of random events ($\omega$), as its variable, depending on the process' probability distribution $P[x(\omega)]$ in a random field. Details are in [33a].

A sequence of distributions $P_t = P_i[x_i(\omega_\eta)]$ for each independent index $i = 1, 2, ..., k, ..., n, ...$ holds multi-dimensional probability distribution on trajectories of the random process $\tilde{x}_t = \{x_i(\omega_\eta)\}$, depending on collection of $\omega_\eta$.

The sequence of distributions, in particular, defines a sequence of continuous or discrete time intervals $t_i^*$ that are generally random for random $\omega_\eta$. In a random field, sequence of probability distributions $P_f = P_i[x_i(\omega)]$ for each independent index $i = 1, 2, ..., k, ..., n, ... \to \infty$ of random functions function $x(\omega)$ of collecting random events ($\omega$), holds a sequence of continuous or discrete time intervals $t_i^*$ that are generally random for random $\omega$.

The random field formalizes multiple random interactions where random time intervals also interact and therefore correlate. Each correlation averages $t_i^*$ through mean time interval $t_i$ [13].

*3.6.5.2. Arrow of time in interactive observation*

In quantum mechanics' microprocess, time is reversible until interaction–measurement (observation) affects quantum wave function. Real arrow of time arises in natural macro processes which average multiple microprocesses.

Thus, natural arrow of time ascends during multiple interactions.

*3.6.5.3. The cutting jump enables rotating the microprocess time and creating space interval during observation*

Each observation, processing the interactive impulses, cuts the correlation of random distributions.

The entropies of the cutting correlations curve a virtual impulse and eventually bring information-physical curvature to real impulse (Fig.1a, Sec. 3.6.4). The cutting jump enables rotating time $t_i$ on an orthogonal angle during the curved jump generating the related space interval (Fig.1, Sec3.5.3). Thus, the time and then space intervals emerge in the interacting impulse as a phase interval (3.6.2) whose real or virtual-probing impulses indicate the impulse observation.

Intensifying the time intervals interactions during the observation grows the intensity of entropy per the interval (as entropy density) increasing on each following interval, which identifies entropy forces being necessary to start the jumping cut [17, Sec.II.3.2]. The jump starts the impulse microprocess whose probabilistic wave function encloses the probability field available for the observation.

The jump initiates multiplicative impulse action $\uparrow_{\delta_k^+} \downarrow_{\tau_k^{-o}}$ ((4.13), Sec.3.4) applied to the opposite imaginary conjugated entropies, which rotates the entropies with its inner time interval equal $\tau_k^{-o}$ with enormous angular speed [17, Sec.III, (2.22b)] and high entropy density. This short high density starting impulse identifies the jump.

Interval $\tau_k^{-o}$ determines that impulse wide and the curvature equal to (6.19) forming in the rotation.

This primary negative curvature of the curved impulse (Fig1a) attracts an observing virtual impulse entropy for space creation within the forming space correlation.

The results Secs 3.4, 3.5.4 show that the jumping rotation follows from invariant entropy increment of a discrete impulse preserving its probability measure, which initiates the multiplicative action, rotating the conjugated impulse entropy up to superposition and entanglement. Ultimately, it is a consequence of the observer probabilistic causation which connects uncertain an outer entropy with an impulse inner entropy and information in internal logic.

Comments. When the correlation reaches conditions (3.4.5.6.40b), time interval satisfies (3.4.5.6.40c), the self-observing random process, interacting through the curved impulse Fig1a, starts rotating this time up to its jump beginning random microprocesses. This primary time interactions are the emerging actions of the initial probability field of interacting events. From this field emerge first correlation-time and then space coordinates. Within the field, the emerging initial time has a discrete probability measure satisfying the Kolmogorov law, which interacts through these probabilities.

*3.6.5.4. The interacting curvatures of step-up and step-down actions, and memorizing a bit*

Each impulse (Fig.1a) step-down action has negative curvature (6.19,6.19a) corresponding attraction, step-up reaction has positive curvature (6.19b) corresponding repulsion, the middle part of the impulse having negative curvature transfers the attraction between these parts.

In probing virtual observation [17], the rising Baeys probabilities increase reality of interactions bringing energy.

When an external process interacts with the entropy impulse, it injects energy capturing the entropy of impulse' ending step-up action (Sec.3.4). The inter-action with other (an internal) process generates its impulse' step-down reaction, modeling 0-1 bit (Fig.3A B).

**A virtual impulse (Fig.3A) starts step-down action with probability 0 of its potential cutting part; the impulse middle part has a transitional impulse with transitive logical 0-1; the step-up action changes it to 1-0 holding by the end interacting part 0, which, after the inter-active step-down cut, transforms the impulse entropy to information bit.**

**In Fig. 3B, the impulse Fig. 3A, starting from instance 1 with probability 0, transits at instance 2 during interaction to the interacting impulse with negative curvature $-K_{e1}$ of this impulse step-down action, which is opposite to curvature $+K_{e3}$ of ending the step-up action**

($-K_{e1}$ is analogous to that at beginning the impulse Fig.3A). The opposite curved interaction provides a time–space difference (a barrier) between 0 and 1 actions, necessary for creating the Bit. The interactive impulse' step-down ending state memorizes the Bit when the interactive process provides Landauer's energy with maximal probability (certainty) 1.

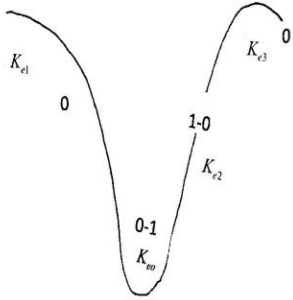

Figure 3A                  Figure 3B

The step-up action of an external (natural) process' curvature $+K_{e3}$ is equivalent of potential entropy $e_o = 0.01847\, Nat$ which carries entropy $\ln 2$ of the impulse total entropy 1 Nat.

The interacting step-down part of internal process impulse' invariant entropy 1 Nat has potential entropy $1 - \ln 2 = e_1$. Actually, this step-down opposite interacting action brings entropy $-0.25\, Nat$ with anti-symmetric impact $-0.025\, Nat$ which carries the impulse wide $\cong -0.05\, Nat$) (Sec. 3.6.4.1) with the total $\cong -0.3\, Nat$ that equivalent to $-e_1$.

Thus, during the impulse interaction, the initial energy-entropy $W_o = k_B \theta_o e_o$ changes to $W_1 = -k_B \theta_1 e_1$, since the interacting parts of the impulses have opposite-positive and negative curvatures accordingly; the first one repulses, the second attracts the energies. The internal process needs minimal entropy $e_{10} = \ln 2$ for erasing the Bit which corresponds Landauer's energy $W = k_B \theta \ln 2$.

If the internal interactive process accepts this Bit by memorizing (through erasure), it should deliver the Landauer energy compensating the difference of these energies-entropy: $W_o - W_1 = W$ in balance form

$$k_B \theta_o e_o + k_B \theta_1 e_1 = k_B \theta \ln 2. \quad (6.40A)$$

Assuming the interactive process supplies the energy $W$ at moment $t_1$ of appearance of the interacting Bit, we get $k_B \theta_1(t_1) = k_B \theta(t_1)$. That brings (6.40a) to form

$$k_B \theta_o 0.01847 + k_B \theta (1 - \ln 2) = k_B \theta \ln 2, \theta_o / \theta = (2 \ln 2 - 1) / 0.01847 = 20.91469199. \quad (6.40B)$$

The opposite curved interaction decreases the ratio of above temperatures on $\ln 2 / 0.0187 - (2 \ln 2 - 1) / 0.01847 = 16.61357983$, with ratio $(2 \ln 2 - 1) / \ln 2 \cong 0.5573$.

External impulse with maximal entropy density $e_{do} = 1 / 0.01847 = 54.14185$ interacting with internal curved impulse transfers minimal entropy density $e_{d1} = \ln 2 / 0.01847 = 37.52827182$.

Ratio of these densities $k_d = e_{do} / e_{d1} = 1.44269041$ equals

$$k_d = 1 / \ln 2. \quad (6.40C)$$

Here the interacting curvature, enclosing entropy density (6.40C), lowers the initial energy and the related temperatures (6.40B) in the above ratio. From that follow

*Conditions creating a bit in interacting curved impulse*

1. The opposite curving impulses in the interactive transition require keeping entropy ratio 1/ln2.
2. The interacting process should possess the Landauer energy by the moment ending the interaction.
3. The interacting impulse hold invariant measure M=[1] of entropy 1 Nat whose the topological metric preserves the impulse curvatures. The last follows from the impulse' max-min mini-max law under its stepdown-stepup actions, which generate the invariant [1] Nat time-space measure with topological metric π (1/2circle) preserving opposite curvatures.

Recent results [27] prove that physical process, which holds the invariant entropy measure for each phase space volume (for example, minimal phase volume $v_{eo} \cong 1.242$ per a process dimension [17, Sec.III.1) characterized by the above topological invariant, satisfies Second Thermodynamic Law.

Energy $W$, which delivers the internal process, will erase the entropy of both attracting and repulsive movements covering energy of the both movements that are ending at the imulse stopping states. The erased impulse total cutoff entropy is memorizes as equivalent information, encoding the impulse Bit in the impulse ending state.

The ending observer's probing logic, which capture entropy increment (4.20), Sec.3.4 moving along negative curvature of its last up impulse and overcoming the gap acquires the equal information (4.20a) that compensates for the movements logical cost. Thus, the attractive logics of an invariant impulse, converting its entropy to information within the impulse, performs function of *logical Demon Maxwell* (DM) in the microprocess. (More details are in [47]).

*Topological transitivity at the curving interactions*

The impulse of the external process holds its 1 Nat transitive entropy until its ending curved part interacts, creating information bit during the interaction.

Theoretically, when a cutting maximum of entropy reaches a minimum at the end of the impulse, the interaction can occur, converting the entropy to information by getting energy from the interactive process.

The invariant' topological transitivity has a duplication point (transitive base) where one dense form changes to its conjugated form during orthogonal transition of hitting time. During the transition, the invariant holds its measure (Fig.3) preserving its total energy, while the densities of these energies are changing.

The topological transition separates (on the transitive base) both primary dense form and its conjugate dense form, while this transition turns the conjugated form to orthogonal.

At the turning moment, a jump of the time curvature switches to a space curvature Sec.3.6. with rising a space waves, Sec.3.6.2. This is what real DM does using for that an energy difference of the forms temperature [28].

Forming transitional impulse with entangled qubits leads to possibility memorizing them as a quantum bit.

That requires first to provide the asymmetry of the entangled qubits, which stats the anti-symmetric impact by the main impulse step-down action ↓ interacting with opposite action ↑ of starting transitional impulse.

That primary anti-symmetric impact $-0.025 \times 2 = 0.05 Nat$ starts curving both main and transitional impulses with curvature $K_{e1} \cong -0.995037$, enclosing $0.025 Nat$, while the starting step-up action of the transitional impulse generates curvature $K_{e2} \cong +0.993362$ enclosing $e_o = 0.01847 Nat$.

Difference $(0.025-0.01847) Nat$ estimates entropy measuring total asymmetry of main impulse $0.0653 Nat = S_{as}$.

The entangled qubits in the transitional impulse evaluates entropy volume 0.0636 Nat [17], which the correlated entanglement can memorize in the equivalent information of two qubits.

That is the information "dimon cost" for the entangled correlation

Starting asymmetric impact brings minimal asymmetry $0.05 Nat$ beginning the transition asymmetry.

The middle part of the main impulse generates curvature $K_{e2} \cong +0.993362$ which encloses entropy $0.02895 Nat$.

Difference $0.02895 - 0.025 = 0.00395 Nat$ adds asymmetry to the starting transitional entropy, while $0.02895 - 0.01845 = 0.0105$ estimates the difference between the final asymmetry of the main impulse and ended asymmetry of transitional impulse.

With starting entropy of the curved transitional impulse $0.05 Nat$, the ending entropy of the transitional impulse asymmetry would be $0.0653 - 0.0105 - 0.00395 = 0.05085 Nat$. Memorizing this asymmetry needs compensation with a source of equivalent energy. It could be supplied by opposite actions of the transitional step-down ↓ and main step up interacting action ↑ ending transitional impulse. That action will create the needed curvature at the end of the main impulse, adding $0.0653 - 0.05085 = 0.01445, 0.01445 - 0.0105 = 0.00395 Nat$ to entropy of transitional impulse curvature sum $0.05085$.

Another part $0.0105$ will bring the difference of entropy' curvature $0.02895 - 0.01845 = 0.0105$ with total $0.0653$.

Thus, $0.05085 Nat = s_{as}$ is entropy of asymmetry of entropy volume $s_{ev} = -0.0636$ of transitional impulse, whereas $0.0653 Nat = S_{as}$ is entropy of asymmetry of the main impulse which generates the same entangled entropy volume that step-action of the main impulse transfer for interacting with external impulse; $s_{as}$ is the information "demon cost" for the entangled correlation, which the curvature of the transitional impulse encloses.

It measures information memorizing two qubits in the impulse measure 1 Nat.

During curved interaction this primary virtual asymmetry compensates the asymmetrical curvature of a real external impulse, and that real asymmetry is memorized through the erasure by the supplied external Landauer's energy.

The ending action of external impulse creates classical bit with probability

$P_k = \exp-(0.0636^2) = 0.99596321$.

Since the entanglement in the transitional impulse creates entropy volume creates, potential memorizing pair of qubits has the same probability.

Therefore, both memorizing classical bit and pair of quits occur in probabilistic process with high probability but less than 1, so it happens and completes not always.

The question is how to memorize entropy enclosed in the correlated entanglement, which naturally holds this entropy and therefore has the same probability.

If transitional impulse, created during interaction, has such high probability, then its curvature holds the needed asymmetry, and it should be preserved for multiple encoding with the indentified difference of the locations of both entangled qubits.

Information, as the memorized qubits, can be produced through interaction, which generates the qubits within a material - devices (a conductor-transmitter) that preserve curvature of the transitional impulse in a Black Box, by analogy with [45]. At such invariant interaction, the multiple connected conductors memorize the qubits' code.

The needed memory of the transitional curved impulse encloses entropy $0.05085 Nat$.

*The time intervals of the curved interaction*

If the external space action curves the internal interactive part, the joint interactive time-space measures the last interactive impact. But if the internal curvature of $K_{e1} \cong -0.995037$ enclosing $-0.025 Nat$ presents only by moment $t_{o1}$ before an interaction, then interacting time-space interval measures the difference of these intervals $|t_{o1}| - t_o \cong 0.0250 - 0.01847 = 0.00653 Nat$.

For that case, the internal curved inter-action attracts the energy of external interactive action until internal process energy $W$ by moment $t_1$. If No part presents at $t_o$ and Yes part arises by $t_1$, then the internal impulse spends $1 - \ln 2 + \ln 2 = 1$ Nat on creating and memorizing a bit.

If the external process hits the internal process having energy $W$, the inter-action of this process brings that energy by moment $t_1$ as the reaction, which carries $W$ to hold the bit. The same energy will erase the bit and memorize it according to (6.40B-6.40B). The internal impulse spends ~1 Nat on creating and memorizing a bit while its curving part holds $(1-\ln 2) \cong 0.3\,Nat$, since the curved topology of interacting impulses decreases the needed energy ratio to (6.40C).

Thus, time interval $t_o - t_1$ creates the bit and performs the DM function.

Multiple interactions generate *a code* of the interacting process at the following *conditions*:
1. Each impulse holds an invariant probability –entropy measure, satisfying the Bit conditions.
2. The impulse interactive process, which delivers such code, must is a part of a real physical process that keeps this invariant entropy-energy measure. That process memorizes the bit and creates information process of multiple encoded bits, which build the process information dynamic structure.

For example, a water, cooling interacting drops of a hot oils in the found ratio of temperatures, enables spending an external hot energy on its chemical components to encode other chemical structures, or the water kinetic energy will carry the accepting multiple drops' bits as an arising information dynamic flow.

3. Building the multiple Bits code requires increasing the impulse information density in three times with each following impulse acting on the interacting process (Sec 3.4). Such physical process generating the code should supply it with the needed energy. To create a code of the bits, each interactive impulse, produced a Bit, should follow three impulses measure π, i.e. frequency of interactive impulse should be f=1/3 π=~0.1061.

The interval 3 π gives opportunity to joint three bits' impulses in a triplet as elementary macro unit and combats the noise, redundancies from both internal and external processes.

Multiplication mass M on curvature $K_{e2}$ of the impulse equals to relative density Nat/ Bit=1.44 which determine M=1.44/$K_{e2}$. At $K_{e2}$=0.993362, we get a relative mass M=1.452335645.

The opposite curved interaction lovers potential energy, compared to other interactions for generating a bit.

The multiple curving interactions create topological bits code, which sequentially forms moving spiral structure [16]. Therefore, the curving interaction dynamically encodes bits in natural process developing information structure Fig. 2 of the interacting information process. The rotation increases density of information bit through growing impulse' curvature.

*Relative information observer*

Ratio of the impulse space and time units
$$h_k / o_k = c_k$$
defines the impulse linear speed $c_k$.

Using the invariant impulse measure, this speed determines
$$c_k = |1|_M / (o_k)^2.$$

More Bits concentrating in impulse leads to $o_k \to 0$ and to $c_k \to \infty$, which is limited by the speed of light.

The persisting increase of information density grows the linear speed of the natural encoding, which associates with a rise of the impulse curvature.

The curvature encloses the information density and enfolds the related information mass [17].

The information observer progressively increases both its linear speed and the speed of natural encoding combined with growing curvature of its information geometry.

The IPF integrates this density in observer' geometrical structure [16] whose rotating speed grows with increasing the linear speed.

Considering any current information observer with speed $c_o$ relative to maximal at $c_k > c_o$ leads to a wider impulse' time interval of observer' $c_o$ for getting the invariant information compared to that for observer $c_k$.

The IPF integrates less total information for observer $c_o$, if both of them start the movement instantaneously.

Assuming each observer total time movement, memorizing the natural encoding information, determines its life span, indicates that for observer $c_o$ is less than for the observer $c_k$ which naturally encodes more information and its density.

At $c_o / c_k \to 1$ both observer approach the maximal encoding.

The discussed approach is an *information versions* of Einstein's theory of relativity applied to moving information observer.

### 3.6.5.5. Evaluation an edge of reality through Plank's fine structural constant of uncertainty $h_\alpha^o$ within the gap

According to M. Born [29]: *If alpha ( $h_\alpha^o$ ) were bigger than it really is, we should not be able to distinguish matter from ether [the vacuum, nothingness], and our task to disentangle the natural laws would be hopelessly difficult. The fact however that alpha has just its value 1/137 is certainly no chance but itself a law of nature. It is clear that the explanation of this number must be the central problem of natural philosophy.*

During the observation, each prior finite correlation increases following posterior correlation through Bayes probability connecting them until it reaches the uncertain gap, defined through the fine structural constant of uncertainty $h_\alpha^o$ [17] which includes the Plank constant equivalent of energy.

This universal parameter constrains the considered observed probes logic by the gap end located on edge of reality. Within the gap, the simultaneous entangled quantum conjugated entropies $S_{\mp a}^* = 2h_\alpha^o$ can emerge with probability $p_{\pm a} = \exp(-2h_\alpha^o) = 0.9855507502$ limited by minimal uncertainty measure $h_\alpha^o = 1/137$ (Sec.3.6.3).

The coupling entropy units confine an entangling *qubit as a part of potential Bit*.

Plank constant $h^o$ of two dimensional phase-space volume $(h^o)^3 = v_p$ not determines a prior uncertain part of last probing impulse on the gap, and its volume extends in the curving cutoff correlation, bringing a space dimension at the end of rotation. Analyzing the volume extension we estimate an access to the uncertain gap.

The Plank phase volume spreads over number $N$ of interacting probing states, proportional to impulse action $A$:

$$N = A / 2\pi h^o, \qquad (6.41)$$

where $N$ limits the dimension of available Hilbert space. Unitary transformations $U_+$ or $U_-$ in Hilbert space, caused by interactive displacement operator $D_\pm$ in phase space, orthogonal rotate the shifts of state $\delta_x^\pm$ and momentum $\delta_p^\pm$:

$$D_\pm = U_\pm = \exp j(\delta_x^\pm + \delta_p^\pm)/\hbar. \qquad (6.41a)$$

Impulse "displacement" $a$ - difference between the shifts caused by interaction, holds orthogonality and depends on $A$:

$$a = (h^o)^2 / A \text{ or } a = h^o / 2\pi N \qquad (6.42)$$

From that follows volume

$$v_p = a 2\pi N (h^o)^2 \qquad (6.42a)$$

which creates "spotty structure on scales corresponding to sub-Planck"[30].

Displacements root $\delta \sim \sqrt{a}$ shift the state to make it orthogonal i.e., distinguishable from the unshifted.

Time scale, developing under rotation $t_r = \Lambda^{-1} \ln A/\hbar$, depends on action $A$, Lyapunov's exponent $\Lambda$ and standard Plank constant $\hbar$ [30].

If we assume that the spotty structure changes the physical structure parameter from $h_\alpha^o = 1/137$ to its up-to-date value $h_\alpha^{o*} = 1/137.036$, then, at other equal conditions, it changes the Plank constant in the inverse ratio

$$h_*^o / h^o = 137.036/137 = 1.000262774 \qquad (6.43)$$

which corresponds changing the spotty Plank volumes:

$$v_{p*}^o / v_p^o = (h_*^o / h^o)^3 = 1.000788529 \qquad (6.43a)$$

For the interacting relative phase volume $v_{eo} \cong 1.242$ the above ratio reaches

$$h_*^{o1} / h_*^o = \sqrt[3]{v_{oe}} = \sqrt[3]{1.272} \cong 1.0831 = 148.449686/137.036, \qquad (6.43b)$$

Substituting the changed Plank constant (6.43) to (6.42) brings the relative displacement

$$a^* / a = h_*^o / h^o \times N / N_*, a^* / a = 1.000262774 N / N_* . \qquad (6.44)$$

And for (6.43b) it follows

$$a_*^{o1} / a_*^o = h_*^{o1} / h_*^o \times N_* / N_*^1, a_*^{o1} / a_*^o = 1.0831 N_* / N_*^1 . \qquad (6.44a)$$

To get previous $a^* / a = 1$ through only changing $N_* / N$ it requires the change in above ratio.

For the estimated in [17] number of virtual impulses, needed to reach standard Plank verge: $N = m_o = 8800$, increases to $N_* \cong 8828$. Or the exact $N_* / N = 1$ relation (6.44) requires increment of $\Delta a^* / a = 0.000262774$.

The relative displacement (6.44a) requires $N_*^1 \cong 9562$ probing impulses. At $N = m_o = 8800$ and standard Plank constant

$$\hbar = \frac{h}{2\pi} = 1.054\,571\,800(13) \times 10^{-34} \text{ J·s} = 6.582\,119\,514(40) \times 10^{-16} \text{ eV·s}.$$

The displacement $a = h^o / 2\pi N = \hbar / N$ is in inverse proportion of $N$:

$$a = 1/N \times \hbar = 1.13636 \times 10^{-4} \times \hbar, \text{ or relative } a/\hbar = 1.13636 \times 10^{-4}. \qquad (6.45)$$

At $N_* \cong 8828, a^*/\hbar = 1.13276 \times 10^{-4}$ and at $N_*^1 \cong 9562, a^1/\hbar = 1.04581 \times 10^{-4}$ \qquad (6.45a)

At $h_\alpha^{o*} = 1/137.036$, $p_{\pm a} = \exp(-2h_\alpha^{o*}) \cong 0.985511281$. \qquad (6.46)

And for $h_*^{o1} = 1/148.449686$, $p_{\pm a} = \exp(-2h_\alpha^{o*1}) \cong 0.9866617771$. \qquad (6.46a)

Probability (6.46a) identifies physical structural parameter $h_\alpha^{o1}$ which counts sub-Plank spot, resulting from interactive probing probability impulses during the observation.

On a path from uncertainty to certainty, the increasing number of probing impulses $N$ allows the observer closer approaches the gap of reality through decreasing uncertain displacement (6.45a) of the sub-Planck spots.

After entropy volume of the $N+$ probe increases to overcome uncertain volume (6.42a), the entropy reaches the edge of certainty-reality with increasing probability (6.46a).

Since a Bit is created at probability (6.46a) approaching 1 with the number of each interactions $N_* \cong 8828$, each impulse observation can create the Bit with frequency

$$F_{im} = 1/8828 = 10^{-4} \times 1.13276. \qquad (6.47)$$

Moreover, because each bit creation needs a final interaction of the impulses with opposite curvatures (Sec.3.6.5.4), such interaction needs $N = 8800$, which evaluates probability (6.46), and the frequency of appearance that impulse is

$$F_{imo} = 1/8800 = 10^{-4} \times 1.13636 \qquad (6.47a)$$

### 3.7. Information dynamic processes determined by EF and IPF functionals
#### 3.7.1. *The EF-IPF connection.*

Since the IPF functional integrates finite information and converges with the initial entropy functional, which had expressed through the additive functional, the EF covers both cutoff information contributions and entropy increments between them.

The IPF at $n \to \infty$ integrates unlimited discrete sequence of the EF cutoff fractions.

The integration of the discrete fractions and solving a classical variation problem for the IPF to find continuous extreme dynamics presents a *difficult mathematical task*.

Initial entropy functional (I) presents potential information functional of the Markov process until the applied impulse control, carrying the cutoff increments, transforms it to physical IPF. The IPF maximal limit (5.1) approaching EF at $o(t \to T) \to 0, n \to \infty$ avoids the direct access to Markov random process.

Extreme of this integral provides dynamic process $x(t)$, which minimizes the distance (IV) and dynamically approximates movement $\tilde{x}_t$ to $\varsigma_t$ evaluating transition to Feller kernel. Process $x(t)$ carries information collected by the maximal IPF at $n \to \infty$, as the IPF information dynamic macroprocess [26], when each interval $\Delta_t = (t - s) \to o(t)$.

Increment of the EF at the end of interval $o_m \to 0$ approaches zero satisfying:

$$\lim_{t_m = T} \Delta S_m[\tilde{x}_t(\tau_m \to t_m))] \to 0. \tag{7.1}$$

The IPF extracts the finite amount of integral information on all cutoff intervals, approaching initial $S[\tilde{x}_t / \varsigma_t]$ (I) before its cutting. The sequential cuts on $(T - s)$ lose the process correlations and the states functional connections, which transforms initial random process to a limited sequence of independent states.

In the limit, the IPF (with probability (V)-within each cutoff) extracts a deterministic process, which approaches the EF extremal trajectories. Thus, information, collected from the diffusion process by the IPF, approaches its source, measured by the EF, when intervals become infinite small $\Delta_t \to o(t)$.

However without impulse action releasing information from the EF it is unfeasible for an observer.

Impulse composes Yes-No actions establishing Bit and all communications (Secs.2.3).

**3.7. 2.*Estimation of extremal process.***

Mathematical expectations of Ito's Eqs:

$$E[a] = \dot{\overline{x}}(t) = E[c\tilde{x}(t)] = cE[\tilde{x}(t)] = c\overline{x}(t) \tag{7.2}$$

approximates a regular differential Eqs

$$\dot{\overline{x}}(t) = c\overline{x}(t), \tag{7.3}$$

whose common solution averages the random movement by dynamic *macroprocess* $\overline{x}(t)$:

$$\overline{x}(t) = \overline{x}(s) \exp ct, \overline{x}(s) = E[\tilde{x}(s)]. \tag{7.3a}$$

Within discrete $o(t) = \delta_o$, the opposite controls $u_+, u_-$, satisfying relation

$$c^2 = |u_+ u_-| = c_+ c_- = \overline{u}^2, c_+ = u_+, c_- = u_-, \ |u_+ u_-| = \overline{u}^2$$

are imaginable, presenting an opposite discrete complex:

$$u_+ = j\overline{u}, u_- = -j\overline{u}. \tag{7.4}$$

Conditions 4.1A, B are fulfilled at

$$\overline{u} = -2j. \tag{7.4a}$$

when

$$u_+u_- = j\overline{u}(-j\overline{u}) = \overline{u}^2, \ -j\overline{u} - (+j\overline{u}) = -2j\overline{u}, \ \overline{u}^2 = -2j\overline{u}, \overline{u} = -2j. \tag{7.4b}$$

The controls are real when
$$u_+ = j(-2j) = 2, u_- = -j(-2j) = -2. \tag{7.5}$$

Relations (7.3), (7.4), satisfy two differential equations
$$\dot{x}_+(t) = j\overline{u}x_+(t), \dot{x}_-(t) = -j\overline{u}x_-(t) \tag{7.6}$$

describing microprocess ($x_+(t)$, $x_-(t)$) under controls (7.4,7.5) on time interval $\Delta_t = t - s, \Delta_t \to o(t)$.

Solutions of (7.6) takes forms
$$\ln x_+(t) = Cu_+t, \ln x_-(t) = Cu_-t, x_+(t) = C\exp(j\overline{u}t), x_-(t) = C\exp(-j\overline{u}t), C = x_-(s^{+o}) = x_+(s^{+o}), \tag{7.7a}$$
$$x_+(t) = x_+(s^{+o})(\cos\overline{u}t + j\sin\overline{u}t), x_-(t) = x_-(s^{+o})(\cos(\overline{u}t - j\sin\overline{u}t), \tag{7.7}$$

where moment $t = t^e$ of reaching minimal entropy functional identifies Eqs (2.12, 2.12b).
Correlation function for microprocess (7.7):
$$r(x_+(t), x_-(t)) = r_s = x_+(s^{+o}) \times x_-(s^{+o}), Cos^2(\overline{u}t) + Sin^2(\overline{u}t) = 1 \tag{7.7b}$$

depends on interaction at moment $\delta^{+o}/4, \delta^{+o}/2$ on an interactive edge of the impulse.
During this fixed correlation, the conjugated anti-symmetric entropies (6.14a) interact producing entropy flow (6.16).
Applying formula [13, p.27] for the correlation between moments $\delta^{+o}/4, \delta^{+o}/2$ (Sec.3.2.4) leads to

$$r_s = \sqrt{(\delta^{+o}/4)(/\delta^{+o}/2)}, \ r_s = \sqrt{0.5}. \tag{7.7c}$$

During this correlation acts impulse $[1^{\delta_k^{\tau+}/2}_{\delta_k^{\tau+}/4}]\overline{u}_k, \overline{u}_k = |\pm 1|_k$ which includes both opposite controls instantly.

If real control, cutting the entropy influx, does not cover it, the states' correlation will not dissolve, and the state, carrying both opposite controls, will hold during the correlation. It corresponds the states' entanglement in microprocess, which may exists between the entropy gap (5.26a,b) balancing entropy (5.29) at the anti-symmetric actions.

Solutions (7.7) describe microprocess on $o(t) \to o(\tau_n^{-o})$ compared to impulse macroprocess (7.3a) averaging (7.7).

The microprocess becomes an inner part of the dynamics process, minimizing distance (IV), when its time intervals satisfy optimal time between the impulse cutoff information at

$$\tau_k^{-o}/\tau_k^{+o} = 3, \delta_k = \tau_k^{+o} - \tau_k^{-o} = 2\tau_k^{-o}, \tau_k^{-o} = 3\delta_k/2. \tag{7.8}$$

It implies that imaginary time interval (5.5b) triples the cutoff discrete intervals:
$$\Delta_k = 3\delta_k, \tag{7.8a}$$

while microprocess (7.7) locates within these cutoff discrete intervals.
To find dynamic process $x(t)$, we consider solution of (7.3) under real control (7.5) starting at moment $t = t^e$:
$$x_\pm(t^e) = x(s^+)\exp(u_\pm t^e), \ t^e = s_k^{+o}b_k(t)/b_k(s_k^{+o}), \ u_\pm = \pm 2 \tag{7.9}$$

approximating an extreme of the entropy functional within each $\Delta_t = t - s$ following from (2.10).
The solutions in form
$$dx(t)/x(t) = cdt, \ln x(t) = ct, t = s_k^{+o}b_k(t)/b_k(s_k^{+o}), \ln x(t) = cs_k^{+o}b_k(t)/b_k(s_k^{+o}) \tag{7.10}$$

starting on time $t = t^e$, integrate on minimal time distance $\Delta_t = t - s$ the process
$$x(t) = \exp[cs_k^{+o}b_k(t)/b_k(s_k^{+o})], x(s_k^{+o}) = \exp(cs_k^{+o}), c = \ln(x(s_k^{+o}))/s_k^{+o}, x(s_k^{+o}) = \overline{x}(s_k^{+o}),$$

$$x(t) = \exp[\ln(x(s_k^{+o}))b_k(t)/b_k(s_k^{+o})] = \exp[\ln(x(s_k^{+o}))t/s)], \ln x(t) = \ln(x(s_k^{+o}))t/s, \quad (7.10a)$$

which at $t \to T$ approaches

$$\ln x(T) = \lim_{t \to T}[\ln(x(s_k^{+o}))T/s], x(T) \to x(s_k^{+o}))T/s. \quad (7.10b)$$

Process $x(t)$ (7.3a), (7.10) is the extremal solution of *macroprocess* $\bar{x}(t)$, which averages solution of Ito Eq. under the optimal controls' multiple cutoffs of the EF for $n$-dimensional Markov process within interval $\Delta_t = (t-s) \to o(t)$.

Process $x(t)$ carries the EF increments, while the information dynamic macroprocess collects the maximal IPF contributions at $o(t) \to 0$, $n \to \infty$, and on each $\Delta_t \to o(t)$.

Information, collected from the diffusion process by the IPF, approaches the EF entropy functional.

Finding process $x(t)$ which the EF generates requires solution of the EF variation problem.

### 3.8. The solution of variation problem for the entropy functional

Applying the variation principle to the entropy functional, we consider an integral functional

$$S = \int_s^T L(t, x, \dot{x}) dt = S[x_t], \quad (8.1)$$

which minimizes the entropy functional (I) of the diffusion process in the form

$$\min_{u_t \in KC(\Delta, U)} \tilde{S}[\tilde{x}_t(u)] = S[x_t], Q \in KC(\Delta, R^n). \quad (8.1a)$$

Specifically, for integral (1.2), it leads to optimal solution of variation problem

$$\text{extr } S[\tilde{x}_t / \varsigma_t] = \underset{c^2(t)}{\text{extr}} 1/2 \int_s^T c^2(t) A(t,s) dt, c^2(t) = \dot{x}(t). \quad (8.1b)$$

<u>Proposition 8.1.</u>

1. An *extremal solution* of variation problem (8.1a, 8.1) for entropy functional (I), brings the following equations of extremals for vector $x$ and conjugate vector $X$ accordingly:

$$\dot{x} = a^u, a^u = a(u,t,x) (t,x) \in Q, \quad (8.2)$$

$$\dot{X} = -\partial P/\partial x - \partial V/\partial x, \quad (8.3)$$

Where a function

$$P = (a^u)^T \frac{\partial S}{\partial x} + b^T \frac{\partial^2 S}{\partial x^2}, \quad (8.4)$$

is a potential of functional (I) depending on function of action $S(t,x)$ on extremals (8.2, 8.3); $V(t,x)$ is integrant of additive functional (Ia), which defines the probability function (II).

<u>Proof.</u> Using the Jacobi-Hamilton (JH) equations for function of action $S = S(t,x)$, defined on the extremals $x_t = x(t), (t,x) \in Q$ of functional (8.1), leads to

$$-\frac{\partial S}{\partial t} = H, H = \dot{x}^T X - L, \quad (8.5)$$

where $X$ is a conjugate vector for $x$ and $H$ is a Hamiltonian for this functional.

(All derivations here and below have vector form).

From (8.1a) it follows

$$\frac{\partial S}{\partial t} = \frac{\partial \tilde{S}}{\partial t}, \frac{\partial \tilde{S}}{\partial x} = \frac{\partial S}{\partial x}, \tag{8.6}$$

where for the JH we have

$$\frac{\partial S}{\partial x} = X, -\frac{\partial S}{\partial t} = H. \tag{8.6a}$$

The Kolmogorov Eq. for functional (I) on diffusion process allows joining it with Eq. (8.6a) in the form

$$-\frac{\partial \tilde{S}}{\partial t} = (a^u)^T X + b\frac{\partial X}{\partial x} + 1/2 a^u (2b)^{-1} a^u = -\frac{\partial S}{\partial t} = H, \tag{8.7}$$

where dynamic Hamiltonian $H = V + P$ includes differential function $V = d\varphi/ds$ of additive functional (Ia) and potential function (8.4) which, at satisfaction (8.7), imposes the constraint on transforming the diffusion process to its extremals:

$$P(t,x) = (a^u)^T X + b^T \frac{\partial X}{\partial x}. \tag{8.8}$$

Applying Hamilton equations $\frac{\partial H}{\partial X} = \dot{x}$ and $\frac{\partial H}{\partial x} = -\dot{X}$ to (8.7) we get the extremals for vector $x$ and $X$ in the forms (8.2) and (8.3) accordingly. •

More details in [26].

<u>Proposition 8.2.</u> A *minimal solution* of variation problem (8.1a,8.1) for the entropy functional (I) brings the following equations of extremals for $x$ and $X$ accordingly:

$$\dot{x} = 2bX_o, \tag{8.9}$$

satisfying condition

$$\min_{x(t)} P = P[x(\tau)] = 0. \tag{8.10}$$

Condition (8.10) is a dynamic constraint, which is imposed on the solutions (8.2), (8.3) at some set of the functional's field $Q \in KC(\Delta, R^n)$, where the following relations hold:

$$Q^o \subset Q, Q^o = R^n \times \Delta^o, \Delta^o = [0,\tau], \tau = \{\tau_k\}, k = 1,...,m \tag{8.11}$$

for cutting process $x(t)_{t=\tau} = x(\tau)$ at $\tau$-localities.

Hamiltonian $H_o = -\frac{\partial S_o}{\partial t}$ \hfill (8.12)

defines function of action $S_o(t,x)$, which on extremals Eq.(8.9) satisfies condition

$$\min(-\partial \tilde{S}/\partial t) = -\partial \tilde{S}_o/\partial t. \tag{8.13}$$

Hamiltonian (8.12) and Eq. (8.9) determine a second order differential Eq. of extremals:

$$d^2 x/dt^2 = dx/dt[\dot{b}b^{-1} - 2H_o]. \tag{8.14}$$

<u>Proof.</u> Using (8.4) and (8.6), we find the equation for Lagrangian in (8.1) in the form

$$L = -b\frac{\partial X}{\partial x} - 1/2\dot{x}^T (2b)^{-1} \dot{x}. \tag{8.15}$$

On extremals (8.2, 8.3), both functions drift and diffusion in (I) are nonrandom.

After substitution the extremal Eqs to (8.1), the integral functional $\tilde{S}$ on the extremals holds:

$$\tilde{S}[x(t)] = \int_s^T 1/2(a^u)^T (2b)^{-1} a^u dt, \tag{8.15a}$$

which should satisfy variation conditions (8.1a), or
$$\tilde{S}[x(t)] = S_o[x(t)], \tag{8.15b}$$
where both integrals are determined on the same extremals.

From (8.15), (8.15a,b) it follows
$$L_o = 1/2(a^u)^T (2b)^{-1} a^u, \text{ or } L_o = \dot{x}^T (2b)^{-1} \dot{x}. \tag{8.16}$$

Both expressions for Lagrangian (8.15) and (8.16) coincide on the extremals, where potential (8.7) satisfies condition (8.10):
$$P_o = P[x(t)] = (a^u)^T (2b)^{-1} a^u + b^T \frac{\partial X_o}{\partial x} = 0, \tag{8.17}$$

while Hamiltonian (8.12), and function of action $S_o(t,x)$ satisfies (8.13).

From (8.15b) it also follows
$$E\{\tilde{S}[x(t)]\} = \tilde{S}[x(t)] = S_o[x(t)]. \tag{8.17a}$$

Applying Lagrangian (8.16) to Lagrange's equation
$$\frac{\partial L_o}{\partial \dot{x}} = X_o, \tag{8.17b}$$

leads to equations for vector
$$X_o = (2b)^{-1} \dot{x} \tag{8.17c}$$

and extremals (8.8).

Both Lagrangian and Hamiltonian here are the *information forms* of JH solution for the EF.
Lagrangian (8.16) satisfies the maximum principle for functional (8.1,8.1a) from which also follows (8.17a).
Functional (8.1) reaches its minimum on extremals (8.8), while it is maximal on extremals (8.2,8.3) of (8.6).
Hamiltonian (8.7), at satisfaction of (8.17), reaches minimum:
$$\min H = \min[V + P] = 1/2(a^u)^T (2b)^{-1} a^u = H_o \tag{8.18}$$

from which it follows (8.10) at
$$\min_{x(t)} P = P[x(\tau)] = 0. \tag{8.19}$$

Function $(-\partial \tilde{S}(t,x)/\partial t) = H$ in (8.6) on extremals (8.2,8.3) reaches a *maximum* when constraint (8.10) is not imposed.

Both the minimum and maximum are conditional with respect to the constraint imposition.

Variation conditions (8.18), imposing constraint (8.10), selects Hamiltonian
$$H_o = -\frac{\partial S_o}{\partial t} = 1/2(a^u)^T (2b)^{-1} a^u \tag{8.20}$$

on the extremals (8.2,8.3) at discrete moments $(\tau_k)$ (8.11).

The variation principle identifies two Hamiltonians: $H$-satisfying (8.6) with function of action $S(t,x)$, and $H_o$ (8.20), whose function action $S_o(t,x)$ reaches absolute minimum at moments $(\tau_k)$ (8.11) of imposing constraint $P_o = P_o[x(\tau)]$.

Substituting (8.2) and (8.17b) in both (8.16) and (8.20), leads to Lagrangian and Hamiltonian on the extremals:
$$L_o(x, X_o) = 1/2 \dot{x}^T X_o = H_o. \tag{8.21}$$

Using $\dot{X}_o = -\partial H_o / \partial x$ brings $\dot{X}_o = -\partial H_o / \partial x = -1/2 \dot{x}^T \partial X_o / \partial x$,
and from constraint (8.10) it follows

$\partial X_o / \partial x = -b^{-1} \dot{x}^T X_o$, and $\partial H_o / \partial x = 1/2 \dot{x}^T b^{-1} \dot{x}^T X_o = 2 H_o X_o$, (8.22)

which after substituting (8.17b) leads to extremals (8.9).

From the Eq. for conjugate vector (8.3), Eqs. (8.7), (8.8) and (8.17c) follows the constraint (8.10) in form

$$\frac{\partial X_o}{\partial x} = -2 X_o X_o^T \quad (8.23)$$

Differentiation of (8.9) leads to a second order differential Eqs on the extremals:

$$\ddot{x} = 2b\dot{X}_o + 2\dot{b}X_o, \quad (8.24)$$

which after substituting (8.22) leads to (8.14). • This solution simplifies proof of Theorem 3.1[26].

### 3.8a. *About the initial conditions for the entropy functional and its extremals*

1. According to definition (II), the *initial conditions for the EF* determines ratio of a primary a priori- a posteriory probabilities $p(o_s^p) = \frac{P_{s,x}^a}{P_{s,x}^p}(o_s^p)$ beginning the probabilistic observation.

Start of the observation evaluates $p(o_s^p) \cong 1.65 \times 10^{-4}$ [17] and $\Delta s_{ap}(o_s^p) = -\ln p(o_s^p) = 0.5 \times 10^{-4}$ with minimal posterior probability $P_{poo} \approx 1 \times 10^{-4}$. That gives estimation of an average initial entropy of the observation:

$S(o_s^p) = [-\ln p(o_s^p) \times P_{poo}] \cong 0.5 \times 10^{-8} Nat$. (8.25)

Based on physical coupling parameter $h_\alpha^o = 1/137$, physical observation theoretically starts with entropy

(6.17): $S(o_{rs}^p) = 2/137 \cong 0.0146 Nat$, (8.26)

while entropy of the first real 'half-impulse' probing action starts at moment $t^{oe}$:

$S_{ko}(t^{oe}) = 0.358834 Nat$ (8.27)

with a priori-a posteriory probabilities $P_{ako} = 0.601, P_{pko} = 0.86$.

In this approach, involving no material entities, physical process begins with the real probing action, and the physical coupling may start with minimizing this entropy at beginning the information process.

At real cut of posteriory probability $P_{po} \to 1$, ratio of their priory-a posteriory probabilities $P_{ao} / P_{po} \cong 0.8437$ determines minimal entropy shift between interacting probabilities $P_{ao} \to P_{po}$ during real cut: $\Delta s_{apo} = -\ln(0.8437) \cong 0.117 Nat$, which after averaging at $P_{po} = 1$ leads to

$S(o_r^p) = 0.117 Nat$. (8.28)

Minimal entropy cost on covering the gap during it conversion to information is $s_{ev} \cong 0.0636 Nat$ [17].

Theoretical start of observation (8.26) is a potential until an Observer gets information from that entropy.

If we *define* the launch of real Information Observer by minimal converting entropy (8.28), then opening real observation *defines* the entropy of first real 'half-impulse' probing actions (8.27), which could be multiple for a multi-dimensional process. Since the potential observing physical process with structural entities is possible (virtual) with entropy (8.26), this observation we call *virtual* until real observation generating information Observer starts and confirms it.

The term 'virtual' associates with physically possibility, until this physical objectivity becomes information-physical reality for the information Observer.

Specifically, if start of virtual observation associates with entropy binding primary a priori-a posteriori probability (8.25), then Virtual Observer identifies the entropy of appearing potential structures (8.26).

Here, we have *identified* both beginning of virtual and physical observations and the virtual and real information Observers, based on the actual quantitative parameters, which are independent on each particular Observer.

But it may impose specific limitations after the Observer forms [17, Sec.2.5.2].

2. The initial conditions for the EF *extremals* determine function $x_\pm(t^e) = x(s^+)\exp(u_\pm t^e)$ which at moment $t^e = s_k^{+o} b_k(t^e)/b_k(s_k^{+o})$ starts virtual or real observations, depending on required minimal entropy of related observations (8.25-8.28). It brings functions

$$x_\pm(t^e) = x(s^+)\exp(\pm 2t^e) \quad (8.29a), \qquad t^e = s_k^{+o} b_k(t^e)/b_k(s_k^{+o}), \tag{8.29b}$$

where (8.29b), at known dispersion function $b_k(t^e, s_k^{+o})$, identifies dependency $t^e = t^e(s_k^{+o}, b_k)$, while (8.29a) identifies initial conditions for the EF extreme conjugated process:

$$x_\pm(t^e) = x(s^+)\exp(\pm 2t^e(s_k^{+o}, b_k)). \tag{8.29}$$

Applying the EF solutions (6.10, 6.10a) at opposite relative time $t_-^* = -t_+^*$ lead to the entropy functions
$S_+(t_+^*) = [\exp(-t_+^*)(\cos(t_+^*) - j\sin(t_+^*))]$, $S_-(t_-^*) = [\exp(-t_+^*)(\cos(-t_+^*) + j\sin(-t_+^*))]$ at

$$S_\pm(t_\pm^*) = 1/2 S_+(t_+^*) \times S_-(t_-^*) = 1/2[\exp(-2t_+^*)(\cos^2(t_+^*) + \sin^2(t_+^*) - 2\sin^2(t_+^*))] =$$
$$1/2[\exp(-2t_+^*)((+1 - 2(1/2 - \cos(2t_+^*))))] = 1/2\exp(-2t_+^*)\cos(2t_+^*) \tag{8.30}$$

This interactive entropy $S_\pm(t_\pm)$ becomes minimal interactive threshold (8.28) at $t_+^* = t_*^e$, which starts the information Observer. From (8.28) it follows:

$$S(t_+^* = t_*^e) = 1/2\exp(-2t_*^e)\cos(2t_*^e) = 0.117 \tag{8.31}$$

with relative time $t_*^e = \pm \pi/2t^e$.

Solution (8.31) will bring real $t^e$ which, after substitution in (8.29b), determines $s_k^{+o} = s_k^{+o}(t^e, b_k(t^e, s_k^{+o}))$ if dispersion functions $b_k(t^e, s_k^{+o})$ are known. Substituting relative $s_{k*}^{+o} = s_k^{+o}(t_*^e, b_k)$ to

$$S(s_{k*}^{+o}) = 1/2\exp(-2s_{k*}^{+o})\cos(2s_{k*}^{+o}) \tag{8.32}$$

allows finding unknown initial posteriory entropy $S_\pm(s_k^{+o})$ starting virtual Observer at $s_k^{+o} = s_{k*}^{+o}/(\pi/2)$.

To find the moment of time starting virtual Observer at maximal uncertainty measure (8.25), when dispersion functions unknown, only joint pre-requirements (8.32) and (8.25) are available.

Applying (7.7) for initial conditions holds $\ln x(s_k^{+o}) = x(s_k^{+o}) u_\pm s_k^{+o}$ which after integration leads to

$1/2[\ln x(s_k^{+o})]^2 = u_\pm (s_k^{+o})^2$, $\ln x(s_k^{+o}) = \sqrt{2u_\pm}(s_k^{+o})$ and to

$x_\pm(s_k^{+o}) = \exp(\pm\sqrt{2u_\pm})(s_k^{+o})$, $x_+(s_k^{+o}) = \exp(\pm\sqrt{2\times 2})s_k^{+o}$, $x_-(s_k^{+o}) = \exp(\pm j\sqrt{2\times 2})s_k^{+o}$, $u_+ = 2, u_- = -2$.

It brings both real and complex initial conditions for starting extremal processes in virtual observer:

$$x_+^i(s_k^{+o}) = \exp(\pm 2 s_k^{+o}), x_-^i(s_k^{+o}) = \exp(\pm 2 j s_k^{+o}) = \cos(2s_k^{+o}) \pm j\sin(2s_k^{+o}), \tag{8.33}$$

where the first corresponds virtual trajectory before an impulse generates complex microprocess – the second one.

Finally the trajectories of extreme processes (8.29a) by moment $t_i^e$ of $i$ dimension takes form

$$x_{\pm}^{i}(t_{i}^{e}) = x_{\pm}^{i}(s_{k}^{+o})[Cos(2s_{k}^{+o}) \pm jSin(2s_{k}^{+o})]\exp(\mp 2t_{i}^{e}). \tag{8.34}$$

Let us numerically validate the results (8.31-8.34). Solution of (8.31):

$\ln 1/2 - 2t_{*}^{e} + \ln[Cos(2t_{*}^{e})] = \ln 0.117, -0.693 + 2.1456 = 2t_{*}^{e} - \ln[Cos(2t_{*}^{e})], 1.4526 = 2t_{*}^{e} + \ln[Cos(2t_{*}^{e})]$

leads to $2t_{*}^{e} \approx 1.45, t^{e} = 1.45/\pi \approx 0.46$ - as one of possible answer.

Applying condition (8.26) to (8.32): $S(s_{k*}^{+o}) = 1/2\exp(-2s_{k*}^{+o})Cos(2s_{k*}^{+o}) = 2/137$ (8.32a) leads to solution $-0.693 + 4.22683 = 2s_{k*}^{+o} - \ln[Cos(2s_{k*}^{+o})], 3.534 = 2s_{k*}^{+o} - \ln[Cos(2s_{k*}^{+o})]$ with a result $s_{k*}^{+o} \approx 1.767, s_{k}^{+o} \approx 1.12$.

Applying condition of beginning virtual observation (8.25) at relative time $o_{s*}^{p}$ to

$$S(o_{s*}^{p}) = 1/2\exp(2o_{s*}^{p})Cos(2o_{s*}^{p}) = 0.5 \times 10^{-8} \tag{8.34a}$$

leads to $-0.693 + 12.8 = 2o_{s*}^{p} - \ln[Cos(2o_{s*}^{p})]$ with solutions $2o_{s*}^{p} \approx 12$ and $o_{s}^{p} \approx 3.85$.

Applying condition of starting information observation (8.27) with (8.30) at relative time $t_{*}^{oe}$, leads to solution

$2t_{*}^{oe} \approx 0.33, t^{oe} = 2t_{*}^{oe}/\pi \approx 0.1$.

These times are counting from the real Observer after overcoming the threshold (8.28). That means that $o_{s}^{p} \approx 3.85$ evaluates time interval of virtual observation, while virtual observer starts on time interval $s_{k}^{+o} \approx 1.12$, and real observer starts on $t^{e} \approx 0.46$ whereas observing the first real 'half-impulse' probing action takes *part* of this time: $t^{oe} \approx 0.1$;

States $x_{+}^{i}(s_{k}^{+o}) = \exp(\pm 2 \times 1.12)$ hold probability $P_{ako} = 0.601$ and have multiple correlations $r_{\pm}^{x}(s_{k}^{+}) = x_{+}^{i}(s_{k}^{+o})x_{-}^{i}(s_{k}^{+o})$, starting virtual observer (8.32, 8.32a). Conjugates processes (8.34) interact through correlation

$$r_{\pm}^{x}(t_{i}^{e}) = x_{+}^{i}(t_{i}^{e}) \times x_{-}^{i}(t_{i}^{e}) = r_{\pm}^{x}(s_{k}^{+})[Cos(2s_{k}^{+o})^{2} + jSin(2s_{k}^{+o})^{2}]\exp(-2t_{i}^{e})\exp(+2t_{i}^{e}) = 1 \tag{8.34b}$$

reaching information observer' threshold (8.31) with relative probability $P_{ao}/P_{po} \cong 0.8437$ and two conjugated entropies

$$S_{\pm}(t^{e}) = 1/2\exp(\pm \pi/2 \times t^{e})Cos(\pm \pi/2 \times t^{e}), \tag{8.35}$$

following from (8.31), which process the entangle movement.

The starting extemal process (8.34) evaluate two pairs of real states for the conjugated process:

$$\begin{aligned}x_{\pm}^{r}(t_{i}^{e}) &= 9.39 \times 0.999 \times 0.3985 = 3.738, x_{\mp}^{r}(t_{i}^{e}) = 0.1064 \times 0.999 \times 2.509 = 0.2666 \\ x_{\mp}^{r1}(t_{i}^{e}) &= 0.1064 \times 0.999 \times 0.3985 = 0.042, x_{\pm}^{r1}(t_{i}^{e}) = 9.39 \times 0.999 \times 2.509 = 23.536\end{aligned}. \tag{8.35a}$$

Imaginary initial conditions evaluate also four options:

$$x_{+}^{im1}(t^{e}) = \exp(2 \times 1.12)[\pm jSin(2 \times 1.12)] \times \exp(-0.92) = j3.064 \times 0.039 \times 0.3985 \cong \pm j0.0475 \tag{8.35b}$$

$$x_{-}^{im2}(t^{e}) = \exp(-2 \times 1.12)[\mp jSin(2 \times 1.12)] \times \exp(0.92) = \mp j0.323 \times 0.039 \times 2.509 \cong \mp j0.0316. \tag{8.35c}$$

**3.8b.** *Applying equation of extremals $\dot{x} = a^{u}$ to a dynamic model's traditional form* [31]:

$$\dot{x} = Ax + u, u = Av, \dot{x} = A(x + v), \tag{8.36}$$

where $v$ is a control reduced to the state vector $x$, allows finding optimal control $v$ that solves the initial variation problem (VP) and identifies matrix $A$ under this control's action.

Proposition 8.3.

The reduced control is formed by a feedback function of macrostates $x(\tau) = \{x(\tau_{k})\}, k = 1,...,m$ in form:

$$v(\tau) = -2x(\tau). \tag{8.36a}$$

Or it is applied to a speed of the macroprocess:
or applied to macroprocess' speed:

$$u(\tau) = -2Ax(\tau) = -2\dot{x}(\tau), \tag{8.36b}$$

at the localities of moments $\tau = (\tau_k)$ (8.11), when matrix $A$ determines equations

$$A(\tau) = -b(\tau)r_v^{-1}(\tau), r_v = E[(x+v)(x+v)^T], b = 1/2\dot{r}, r = E[\tilde{x}\tilde{x}^T] \tag{8.37}$$

and $A$ identifies the correlation function or its derivative, or directly dispersion matrix $b$ from (2.1):

$$|A(\tau)| = b(\tau)(2\int_{\tau-o}^{\tau} b(t)dt)^{-1}, \tau - o = (\tau_k - o), k = 1..., m. \qquad \bullet \tag{8.37a}$$

<u>Proof.</u> Using Eq. for conjugate vector (8.3) allows writing constraint (8.10) in the form

$$\frac{\partial X}{\partial x}(\tau) = -2XX^T(\tau), \tag{8.38}$$

where for model (8.36) it leads to

$$X = (2b)^{-1}A(x+v), X^T = (x+v)^T A^T (2b)^{-1}, \frac{\partial X}{\partial x} = (2b)^{-1}A, b \neq 0, \tag{8.38a}$$

and (8.38a) acquires form

$$(2b)^{-1}A = -2E[(2b)^{-1}A(x+v)(x+v)^T A^T (2b)^{-1}], \tag{8.38b}$$

from which, at a nonrandom $A$ and $E[b] = b$, the identification equations (8.37) follows strait.
Completion of both (8.38a,b) performs the control's action, which is found using (8.38b) in form

$$A(\tau)E[(x(\tau) + v(\tau))(x(\tau) + v(\tau))^T] = -E[\dot{x}(\tau)x(\tau)^T], \text{ at } \dot{r} = 2E[\dot{x}(\tau)x(\tau)^T].$$

This relation after substituting (8.36) leads to

$$A(\tau)E[(x(\tau) + v(\tau))(x(\tau) + v(\tau))^T] = -A(\tau)E[(x(\tau) + v(\tau))x(\tau)^T] \text{ and then to}$$

$$E[(x(\tau) + v(\tau))(x(\tau) + v(\tau))^T + (x(\tau) + v(\tau))x(\tau)^T] = 0,$$

which is satisfied at applying the control (8.36a).
Since $x(\tau)$ is a discrete set of states, satisfying (8.11), (8.13), the control has a discrete form.
Each stepwise control (8.36a), with its inverse value of doubling controlled state $x(\tau)$, applies to both (8.38a,b), implements (8.38) which, following from variation conditions (8.1a), fulfills this condition.
This control, applied to the additive functional (IIa), imposes constraint (8.8, 8.10) which limits transformation segments of random process to the process extremals. By applying the control step-down and step-up actions to satisfy conditions (8.7) and (83.10), the control sequentially starts and terminates the constraint, while extracting the cutoff hidden information on the $x(\tau)$-localities.
Performing the transformation, this control initiates the identification of matrix $A(\tau)$ (8.37, 8.37a) during its time interval, solving simultaneously the identification problem [32]. •
Obtaining this control here *simplifies* some results of Theorems 4.1 [26].
<u>Corollary 8.3.</u> Control that turns the constraint on creates the Hamilton dynamic model with complex conjugated eigenvalues of matrix $A$. After the constraint's termination, the control transforms this matrix to its *real* form (on the diffusion process' boundary point [24]), which identifies diffusion matrix in (8.37a). Thus, within each extremal segment, the information dynamics is reversible; irreversibility rises at each constraint termination between the segments.

The IMD Lagrangian integrates both the impulses and constraint information on its time space-intervals. •

Proposition 8.4.

Let us consider controllable dynamics of a closed system, described by operator $A^v(t,\tau)$ with eigenfunctions $\lambda_i^v(t_i,\tau_k)_{i,k=1}^{n,m}$, whose matrix equation:

$$\dot{x}(t) = A^v x(t), \qquad (8.39)$$

includes feedback control (8.36a). The drift vector for both models (8.36) and (8.39) has same form:

$$a^u(\tau, x(\tau,t)) = A(\tau,t)(x(\tau,t) + v(\tau)); A(\tau)(x(\tau) + v(\tau)) = A^v(\tau)x(\tau) \qquad (8.39a)$$

Then the followings hold true:

(1)-Matrix $A(t,\tau)$ under control $v(\tau_k^o) = -2x(\tau_k^o)$, applied during time interval $t_k = \tau_k^1 - \tau_k^o$, in form $A(t_k, \tau_k^1)$ depends on initial matrix $A(\tau_k^o)$ at the moment $\tau_k^o$ according to Eq

$$A(t_k, \tau_k^1) = A(\tau_k^o)\exp(A(\tau_k^o)t_k)[2 - \exp(A(\tau_k^o)t_k)]^{-1}. \qquad (8.39b)$$

(2)- The identification Eq.(8.37) at $\tau_k^1 = \tau$ holds

$$A^v(\tau) = -A(\tau) = 1/2 b(\tau) r_v^{-1}(\tau), b(\tau) = 1/2\dot{r}_v(\tau), \qquad (8.39c)$$

whose covariation function $r_v(\tau_k^o)$, starting at the moment $\tau_k^o$, by the end of this time interval, acquires form

$$r_v(\tau_k^1) = [2 - \exp(A(\tau_k^o)t_k)]r(\tau_k^o)[2 - \exp(A^T(\tau_k^o)t_k)]. \qquad (8.39d)$$

(3a)-At the moment $\tau_k^o + o$ following $\tau_k^o$ at applying control $v(\tau_k^o) = -2x(\tau_k^o)$, the controllable matrix gets form

$$A^v(\tau_k^1)_{t_k \to 0} = A^v(\tau_k^o + o) = -A(\tau_k^o), \qquad (8.39e)$$

changing its sign from the initial matrix.

(3b)- When this control, applied at the moment $\tau_k^1$, ends the dynamic process on extremals in the following moment at $x(\tau_k^1 + o) \to 0$, function $a^u = A^v x(t)$ in (8.36) turns to

$$a^u(x(\tau_k^1 + o)) \to 0; \qquad (8.39f)$$

which brings (8.39a) to its dynamic form $a^u = A(\tau_k^1 + o)v(\tau_k^1 + o) \to 0$ that requires turning the control off.

(3c)- At fulfilment of (8.39f), Ito stochastic Eq. [21] includes only its diffusion component, which identifies dynamic matrix $A(\tau_k^1 + o)$, being transformed in following moment $\tau_{k+1}^1 : A(\tau_k^1 + o) \to A(\tau_{k+1}^1)$ via correlation matrix $r(\tau_{k+1}^1)$ using relation

$$A(\tau_{k+1}^1) = 1/2\dot{r}(\tau_{k+1}^1)r^{-1}(\tau_{k+1}^1) \text{ at } r(\tau_{k+1}^1) = E[\tilde{x}(t)\tilde{x}(t + \tau_{k+1}^1)^T] = r^v(\tau_{k+1}^1)_{v(\tau_k^1 + o) \to 0}.$$

(3d)-Dispersion matrix (in (8.37)) on extremals with covariation matrix (8.39d) acquires forms

$$\partial r_v(\tau_k^1)/\partial t_k = -A(\tau_k^o)\exp(A(\tau_k^o)t_k)r(\tau_k^o)[2 - \exp(A^T(\tau_k^o)t_k)] + [2 - \exp(A(\tau_k^o)t_k)]r(\tau_k^o)[-A^T(\tau_k^o)\exp(A^T(\tau_k^o)t_k)] \quad (8.40)$$

which for symmetric matrix $A(\tau_k^o)$ holds relations

$$\partial r_v(\tau_k^1)/\partial t_k \big|_{t_k = \tau_k^1} = -2A(\tau_k^o)\exp(A(\tau_k^o)\tau_k^1)r(\tau_k^o), b(\tau_k^1) = -A(\tau_k^o)\exp(A(\tau_k^o)t_k)r(\tau_k^o),$$

$$b(\tau_k^1 = \tau_k^o) = -A(\tau_k^o)\exp(A(\tau_k^o)0_k)r(\tau_k^o) = -A(\tau_k^o)r(\tau_k^o), \ b(\tau_k^1)/b(\tau_k^o) = \exp(A(\tau_k^o)\tau_k^1)A(\tau_k^o)^{-1}. \qquad (8.40a)$$

Ratio in (8.40a) for a single dimension, at $A(\tau_k^o) = \alpha_1(\tau_{k1}^o)$, leads to

$$b(\tau_{k1}^1)/b(\tau_{k1}^o) = \exp\alpha_1(\tau_{k1}^o)/\alpha_1(\tau_{k1}^o) \qquad (8.40b)$$

which after applying relation (2. 12) in form $b(\tau_{k1}^1)/b(\tau_{k1}^o) = \tau_k^1/\tau_{k1}^o$ leads to

$$\tau_k^1 = \tau_{k1}^o \exp\alpha_1(\tau_{k1}^o)/\alpha_1(\tau_{k1}^o), \qquad (8.40c)$$

connecting interval $t_k = \tau_k^1 - \tau_k^o$ with eigenvalue at $\tau_{k1}^1 = \tau_{k1}^o[\alpha_1(\tau_{k1}^o)]$, which measures information speed on the interval.

(3e)- Equation for conjugated vector (8.17) on each extremal segments follows from relation:

$$X_o(t_k) = 2b(t_k)^{-1}\dot{x}(t_k) = 2A(\tau_k^o)\exp(A(\tau_k^o)t_k)x(\tau_k^o)A^T(\tau_k^o)\exp(A^T(\tau_k^o)t_k). \qquad (8.41)$$

(3f)- Entropy increment $\Delta S_{io}$ on optimal trajectory at

$$E[\frac{\partial\tilde{S}}{\partial t}(\tau)] = 1/4Tr[A(\tau)] = H(\tau), A(\tau) = -1/2\sum_{i=1}^{n}\dot{r}_i(\tau)r_i^{-1}(\tau), (r_i) = r, \qquad (8.41a)$$

measured on the cutting localities, determines the time interval of the nearest segments according to formula:

$$\Delta S_{io} = I_{x_t}^p = -1/8\int_S^T Tr[\dot{r}r^{-1}]dt = -1/8Tr[\ln(r(T)/\ln r(s)], (s = \tau_o, \tau_1, ..., \tau_n = T). \qquad \bullet \quad (8.42)$$

<u>Proof</u> (1). Control $v(\tau_k^o) = -2x(\tau_k^o)$, imposing the constraint at $\tau_k^o$ on both (8.36) and (8.39) and terminating it at $\tau_k^1$ on time interval $t_k = \tau_k^1 - \tau_k^o$, brings solutions of (8.36) by the end of this interval:

$$x(\tau_k^1) = x(\tau_k^o)[2 - \exp(A(\tau_k^o)t_k)]. \qquad (8.43)$$

Substituting this solution to the right side of $\dot{x}(\tau_k^1) = A^v(\tau_k^1)x(\tau_k^1)$ and to its derivative on the left side leads to

$$-x(\tau_k^o)(A(\tau_k^o)t_k)\exp(A(\tau_k^o)t_k) = A^v(\tau_k^o)x(\tau_k^o)[2 - \exp(A(\tau_k^o)t_k)]],$$

or to connection of both matrixes $A^v(\tau_k^1)$ and $A(\tau_k^1)$ (at the interval end) with the matrix $A(\tau_k^o)$ (at the interval beginning):

$$A^v(t_k, \tau_k^1) = -A(\tau_k^o)\exp(A(\tau_k^o)t_k)[2 - \exp(A(\tau_k^o)t_k)], \qquad (8.44)$$

and to $A^v(\tau_k^1) = -A(\tau_k^1)$ by moment $\tau_k^1$, from which folows (8.39e). Other proof parts are straight forward. •

The initial conditional probability measure (II) determines the probability measure along the extremal trajectory:
$$p[x(t)] = p[x(s)]\exp(-S[x(t)]) \qquad (8.44a)$$
where starting probability
$$p[x(s)] = \exp(-S[x(s)]) \qquad (8.44b)$$
follows from formula (8.32) and numerical values (8.34a),(8.35,8.35a,b).

*Finding the invariant relations.*

Using (8.36b) in form $u(\tau) = -2Ax(\tau) = -2\dot{x}(\tau)$, and $c^2 = |u_+u_-| = c_+c_- = \bar{u}^2, c_+ = u_+, c_- = u_-$ leads to

$$c^2 = \dot{x}(\tau) = -2A^v x(\tau), \ln x(\tau)/\ln x(s) = A^v(\tau - s) = u_\pm(\tau - s),$$
$$A^v(\tau - s) = u_\pm(\tau - s) = inv, c^2 = \dot{x}(\tau) = a^u = inv \qquad (8.45)$$

where $(\tau - s)$ is equivalent of interval $t_k = \tau_k^1 - \tau_k^o$ between the discrete moments (8.11) at imposing constraint (8.10,8.38) on each impulse invariant interval. As result of constraint (8.10), it follows
$$A^v(\tau - s) = u_\pm(\tau - s) = inv = \mathbf{a}_o, \qquad (8.45a)$$
where invariant $\mathbf{a}_o = \mathbf{a}_o(\gamma_k)$ [33] depends on ratio of imaginary to real eigenvalues of operator (8.44):

$\gamma_k = \beta_{ko}/\alpha_{ko}$. From (8.45a), in particular, it follows numerical value for real

$$A^v = [\pm 2(\tau - s)]_{n \times n}. \qquad (8.45b)$$

That concurs with Eqs (8.33) and with the impulse invariant information in Corollaries 4.2.
For optimal model with information invariant $\mathbf{a}_o(\gamma_k \to 0.5) = \ln 2$, it leads to $(\tau - s) = \ln 2/2 \cong 0.346$ which evaluates $\delta_k = \tau_k^{+o} - \tau_k^{-o} \cong 0.35$ in [17].

Invariant $\mathbf{a}_o = \ln 2 \cong 0.7$ measures information generating at each impulse cut (Secs. 3.3,3.4).
Correlation matrix (8.39d), measured by the `optimal model's invariant, takes form

$$r_v(\tau_k^1) = [2 - \exp \mathbf{a}_o)]r(\tau_k^o)[2 - \exp(\mathbf{a}_o)] = r(\tau_k^o) \times [1.5^2], \qquad (8.46)$$

where vector $x(\tau_k^1) = x(\tau_k^o) \times [1.5]$ measures its extreme relation (8.43). Conjugate vector

$$X_o(\tau_k^1) = 2A(\tau_k^o)\exp(\mathbf{a}_o)x(\tau_k^o)A^T(\tau_k^o)\exp(\mathbf{a}_o) \qquad (8.46a)$$

for a single dimension holds

$$X_{o1}(\tau_k^1) = 2\alpha_1(\tau_{k1}^o)^2 \exp(2\mathbf{a}_o)x(\tau_{k1}^o), \qquad (8.46b)$$

or at $\alpha_1(\tau_{k1}^o) = 2\mathbf{a}_o/t_k$, $t_k \to 0$, $X_{o1}(\tau_k^1) = (2\mathbf{a}_o/t_k)^2 \exp(2\mathbf{a}_o)x(\tau_{k1}^o) \to \infty$.

It means at decreasing time intervals $t_k$ between the impulse' generated information invariants $\mathbf{a}_o$, information force grows infinitely. Whereas, the force grows in square function of time intervals, which involve potential overrunning the impulse with a minimal $t_k$. It leads to possibility of pulling together the action and its result for the minimal time impulse.
The EF-IPF *estimate* invariant's measure $\mathbf{a}_o(\gamma_k)$, counting both segment's and inter-segment's increments:

$$\tilde{S}_{\tau m}^i = \sum_{k=1}^m (\mathbf{a}_o(\gamma_k) + \mathbf{a}_o^2(\gamma_k)), \tilde{S}_\tau = \sum_{i=1}^n \tilde{S}_{\tau m}^i, \qquad (8.47)$$

where $m$ is number of the segments, $n$ is the model dimension (assuming each segment has a single $\tau_k$-locality). However, to *predict* each $\tau_k$- locality, where information should be measured, only invariant $\mathbf{a}_o(\gamma_k)$ needs. Sum of process's invariants

$$\tilde{S}_{\tau m}^{io} = \sum_{k=1}^m \mathbf{a}_o(\gamma_k), \quad \tilde{S}_\tau^o = \sum_{i=1}^n \tilde{S}_{\tau m}^{io} \qquad (8.48)$$

estimates EF entropy with maximal process' probability (8.14a) expressed through $\mathbf{a}_o = \mathbf{a}_o(\gamma_k)$.
This entropy allows encoding the *random process* using Shannon's formula for an average optimal code-word length:

$$l_c \geq \tilde{S}_\tau^o / \ln D, \qquad (8.49)$$

where $D$ is the number of letters of the code's alphabet, which encodes $\tilde{S}_\tau^o$ (8.48).
An elementary code-word to encode the optimal process' segment is

$$l_{cs} \geq \mathbf{a}_o(\gamma_k)/\log_2 D_o, \qquad (8.50)$$

where $D_o$ is extremal segment's code alphabet, which implements the invariant macrostates connections.
At $\mathbf{a}_o(\gamma_k \to 0.5) \cong 0.7$, $D_o = 2$, it follows $l_{cs} \geq 1$, or (8.50) encodes a bit per the encoding letter.

Values of $x_{\pm}^r(t^e), x_{-}^{im1}(t^e), x_{-}^{im2}(t^e)$ (8.35, 8.35a, b,c) start Hamiltonian process (Sec.3.8) and correlation (8.37),(8.34a) which identify the initial segment's state $x(\tau_{k1o}^o) = x(\tau_k^o)$ (in (8.43)) and the eigenvalues of matrix (8.39) in process (Fig.2) forming information units. The connection to each following segment on the extremal trajectory determines the segment state $x(\tau_{k1}^o)$ (8.43). Each $x(\tau_{k1o}^o), x(\tau_{k1}^o)$ enfolds the impulse microprocess and starts the macroprocess segment.

Moment $t^e$ identifies dispersion and correlation on the trajectories that determine operator of information speed $A(t, \tau_k)$, Hamiltonian, and both EF-IPF on the optimal trajectory segments. Optimal control (8.36a) starts with each segments initial states. The details of information micro-macrodynamics (IMD), based on the *invariant* $\mathbf{a}_o(\gamma_k)$ *description*, are in [17,26,3], where the dynamics' scale parameter $\gamma_{k,k+1}^{\alpha} = \alpha_k / \alpha_{k+1}$ depends on frequency spectrum of observations detecting through the identified $\gamma_k$.

The observer self-scales observation initiates its time-space IMD which builds distributed information network [16,33]. *Above equations finalize both math applications of main general results and validate them numerically.*

### 3.9. Analysis of the observer path functional and logic in multi-dimensional observing process. Discussion

Initial Kolmogorov probability distribution $P[x(\omega)]$ and the following EF (V) represent all $n$-dimensional Markovian model of the observing process.

Probability of each process dimension are local for each its random ensemble being a part of whole process ensemble. Total multi-dimensional probability integrates conditional Kolmogorov's–Bayes probabilities (IVa).

That is why the EF presents a potential information functional of the Markov process until the applied impulse control, carrying the cutoff contributions (3. 2.3.5), transforms it to the informational path functional (3.2.5.1).

Markov random process is a source of each information contribution, whose entropy increment of cutting random states delivers information, hidden between these states' correlation. The finite restriction on the cutting function determines the discrete impulse's step-up and step-down actions between impulse cutoff $\delta_k = \tau_k^{+o} - \tau_k^{-o}$, which supply entropy hidden between impulses on $\tau_k^{-o}$, cut it and transfers to $\tau_k^{+o}$ where the cutting entropy produces the equivalent physical information and memorizes it. Each interval $\Delta_t \to o(t)$ following next cutoff, delivers new hidden process' information.

Information is a physical entity, which distinguishes from entropy that is observer's virtual-imaginable.

In the multi-dimensional diffusion process, the step-wise controls, acting on the process all dimensions, sequentially stops and starts the process, evaluating the multiple functional information. Impulses delta-function $\delta u_t$ or discrete $\delta u_{\tau_k}$ implement transitional transformations (II), initiating the Feller kernels along the process and extracting total kernel information for $n$-dimensional process with $m$ cuts off.

The maximal sum measures the interstates information connections held by the process along the trajectories during its real time $(T - s)$, which are hidden by the random process correlating states.

The EF functional information measure on trajectories is not covered by traditional Shannon entropy measure.

The dissolved element of the functional's correlation matrix at the cutoff moments provides independence of the cutting off fractions, leading to orthogonality of the correlation matrix for these cut off fractions.

Intervals between the impulses do not generate information, being imaginary-potential for getting information, since no real double controls are applying within these intervals. The minimized increments of entropy functional between the cutoffs allow prediction each following cutoff with maximal conditional probability.

A sequence of the functional a priori-posteriori probabilities provides Bayesian entropy measuring a *probabilistic causallity*, which is transforming to physical casualty in information macrodynamics. Sum of information contributions, extracted from the EF, approaches its theoretical measure (I) which evaluates the upper limit of the sum.

Since the sum of additive fractions of the EF on the finite time intervals is less than EF, which is defined by the additive functional, the additive principle for a process' information, measured by the EF is *violated*.

Each $k$-cutoff "kills" its process dimension after moment $\tau_k^{+o}$, creating process that balances killing at the same rate [34]. Then $k = n$, and condition (3.5.1) requires infinite process dimensions, continuing the processes balance creation.

The EF measure, taken along the process trajectories during time $(T-s)$, limits maximum of total process information, extracting its hidden cutoff information (during the same time), and brings more information than Shannon traditional information measure for multiple states of the process.

The limited time integration and the last cutting correlation bind the EF last cut.

Maximum of the process cutoff information, extracting its total hidden information, approaches the EF information measure. Or total physical information, collected by IPF in the infinite dimensional Markov diffusion process, is finite. Since rising process dimension up to $n \to \infty$ increases number of the dimensional kernels, information of the cutting off kernels grows, and when the EF transforms to the IPF kernels, the IPF finally measures all kernels finite information.

Total process uncertainty-entropy, measured by the EF is also finite on the finite time of its transition to total process certainty-information, measured by the IPF. The EF integrates this time (3.2.1.2) along the trajectories.

The IPF formally defines the distributed actions of multi-dimensional delta-function on the EF via the multi-dimensional additive functional (Ia), which leads to analytical solution and representation by Furies series. The delta-impulse, generating spectrum of multiple frequencies, is source of experimental probabilities in formal theory [1].

Since entropy requires a direction of time course–arrow of time, cutting entropy memorizes the cutting time interval which freezes the probability of events with related dynamics of information micro- macroprocess.

Total multi-dimensional probability for the antisymmetric entangled local space entropies rotating on angle $\varphi_{\mp}^2 = \mp \pi/4$ (Se.3.6.1) integrates its local conditional Kolmogorov's –Bayes probabilities (IVa) in final probability consistent with the Aspect-Bell's tests [35]. The correlated values of the entropies starting on these angles at (3.6.4) emerges as a process probabilistic logic created through the observer s probes-observations.

The Bell's experimental test, based on a partial representation of the process probabilities, violates Bell's inequality [36], while the multi-dimensional process' interactions cover all probabilities assigning the probability spaces to experimental contexts.

During the multi-dimensional process, sequence of dimensional impulses cut the entangled space entropy (analogously to "a quantum collapse"), and each cutting entropy produce information which transfers this logic to real information logic.

Within the gap, each entangle entropy holds probability ½. By overcoming the gap, the sequential cuts increase total process probability up to 1, which do not require the Bell test, and leads to the process reality.

Within the gap may proceed the sequential logics of qubits which can build a quantum computation.

In natural interactive process, both the observer logic and information can emerge spontaneously. That concurs with [37],[38]. Elementary unit of information created during the cut-off interactive impulse contains $S_{e\delta t} \cong 0.75 Nat$ (Eq.3.4).

In [17], the time interval of creation a Bit during transition trough entropy-information gap evaluates

$$\delta_{to} \cong 0.4 \times 10^{-15} \sec. \tag{9.1}$$

The information analog of Plank constant $\hat{h}$, at maximal frequency of energy spectrum of information wave in its absolute temperature, evaluates maximal information speed of the observing process:

$$c_{mi} = \hat{h}^{-1} \cong (0.536 \times 10^{-15})^{-1} Nat/\sec \cong 1.86567 \times 10^{15} Nat/\sec, \tag{9.2}$$

which, at real time (9.1) of the gap transmission, estimates this time-entropy equivalent

$$S_{e\delta to} \cong 1.86567 \times 10^{15} \times 0.4 \times 10^{-15} \cong 0.746268 Nat .\tag{9.3}$$

That brings $S_{e\delta t} \cong S_{e\delta to}$ which confirms that energy for conversion this entropy delivers the real time course during the gap transitive movement. Applying (9.2) for information Bit $I_B = 1/\ln 2 \cong 1.442695 Nat$ leads to information equivalent of transmission time interval $\delta_t \to \delta_{tI} = \ln 2 Nat$ which holds minimal time interval of the Bit transmission

$$\delta t_m = (\ln 2)^{-1}/1.86567 \times 10^{15} = 0.7732852 \times 10^{-15} \sec .\tag{9.4}$$

The cutting off random process brings the persistent Bit which sequentially and automatically converts entropy to information, holding the cutoff information of random process, which connects the Bits sequences in the IPF.
Each bit on different IN' hierarchical level encloses distinct information density which depends on the curving geometry. (By analogy with bits in different computer languages at micro and macro levels encoding with various speeds).
The IPF maximum, integrating unlimited number of Bits' units with finite distances, limits the total information carrying by the process' Bits. The limited total time of collecting the cutoff information (at $n \to \infty$) decreases the time interval of each cutoff, which increases the inclusive quantity of information extracted on this interval [42].
At the same cutting off information for each impulse, density of this information, related to the impulse interval, grows, which integrates sum of all previous cutoffs.
This process starts with growing memory of cutting process correlation which after each cut binds the impulse code sequence and memorizes both the code time followed by its length, as a source of logical complexity.
The limitation on the cutting discrete function determines two process' classes for cutting EF functional: microprocess with entropy increment on each time interval $o(t)$, as a carrier of information contribution to each $\delta_k$ driving the integration, and macroprocess over real time $(T-s)$ describing the total process of transformation EF-IPF. The microprocess specifics within each $o(t)$ are: an imaginary time compared to real time of real microprocess on information cutoff $\delta_k$; two opposite sources of entropy-information as an interactive reaction from random process at transition $\delta_k^{\tau+}$ to $\tau_k^{-o}$, which carry two symmetrical conjugated imaginary entropy increments until the interactive capturing brings their dissimilarity (the asymmetry break between controls on $\delta_k^{\tau+}$ evaluates relations (3.5.25), (3.5.27)).

The entropy increments correlate at $\delta_k^{\tau+}$-locality, before the cutting action on $\tau_k^{-o}$ transforms the adjoin increments to real information. The cutting action dissolves the correlation and generates the impulse information contribution by moment $\tau_k^{+o}$; the entropy cut memorizes the cutting information, while a gap within $\delta_k^{\tau+}$ delivers external influx of entropy (3.5.27), covered by real step-wise action, which carries energy for the cut. Microprocess may exist within Markov kernel or beyond, and even prior interactions, independently on cutting randomness.
Microprocess with imaginary time between the impulses belongs to random process, whose cutoff transfers it to information microprocess within each impulse $\delta_k$. Time course on $\Delta_k$ is a source of the entropy increment between impulses, which moves the nearest impulses closer. The relative intervals (5.4) has a measure of information density (5.3).
Between the impulse No and Yes action emerges a transitional impulse which transforms an interim time to the following space interval that ends holding the cutting information.
*This transition depends on the gap separating the micro- and macroprocess.*
Superimposing interaction, measured through the multiplication of conjugated entropy and probability functions (Sec.2.2.6), brings the observable values of emerging space coordinates during the microprocess.

Within the observing probability field, the emerging initial time has a discrete probability measure satisfying the Kolmogorov law and interacting through these probabilities.

Basic orthogonal relation between real time and space coordinates follows from Minkowski metric with imaginary time.

The conjugated and probabilistic dynamics of the impulse' microprocess is different from Physics in Quantum Mechanics. Shifting $\delta_k^{\tau+}$ in real time course $\Delta_k$ moves to automatically convert its entropy to information, working as Maxwell's Demon, which enables compensate for the transitive gap (Sec.3.5) from the time course entropy to information.

The macroprocess, integrating the imaginary entropy between impulses with an imaginary microprocesses and the cutoff information of real impulses, builds information process of the collected entropies converted to physical information process. The EF extremal trajectories (Sec. 3.6.) describe the macroprocess starting with the microprocess trajectories, which at $n \to \infty$ are extremals of both EF and IPF.

The EF imaginary entropy, measured by logarithmic conditional probability of the observing random process, is distinct from Boltzmann physical entropy satisfying Second Thermodynamic Law.

The EF-IPF transformations provide the Information Path from Randomness and Uncertainty to Information, Thermodynamics, and Intelligence of Observer [43]. This includes a virtual observer, acting with imaginary control and integrating imaginary entropy increments in EF, which is mathematically established in Secs.3.2.2.5-2.6 and numerically verified in Secs.3.2.2.8a,b.

It is impossible to reach reality in quantum world without overcoming uncertainty-information gap located on the edge of reality (Sec.3.6.4-5).

Actual information observer acts with real control cutting impulse's information, which integrates the IPF.

During the process' time-space observation, the virtual and actual Observer rises, created on a Path from uncertainty to certainty-information without any priory physical law.

The moment of creation information we associate with arising a conscience of the elementary observer (Sec.3.6.5.4).

Multiple interactive impulse actions on the Path coverts the integrated maxmin entropy to information in real time-space, transforming the information to physical laws.

The information processes correctly describe all worlds' natural process because they logically chose and connect all observations in certain–most probable sequence of events, which include hidden information between interacting events.

Primary virtual observer through probing impulses sequentially increases the observing correlations, reducing entropy fractions, and integrates them in entropy functional.

The chosen fractions create the information observer whose path functional integrates the fractions in information process, describing a factual events-series as physical processes.

The known paradox between truth: yes(no) and lie: no truth-yes(no) or lie: no(yes) solves each particular observer by sending probes on its path to requested answer from the imaginary probes up to real information.

That mathematically leads to imaginary impulse, where the time rotation on angle $\pi/2$ creating a space, corresponds its multiplying on imaginary symbol $j$. Virtual observer, located inside each sending probe, rotates with its imaginary time enclosing logic of this paradox. The brain processing with nuclear spins [39] has such possibility.

According to Feynman [40], a physical law describes the most probable events of observation process, which implies applying a variation principle (VP) for finding the law. However, a common form of arising physical law formulates only informational VP whose solution brings maximal probability on the VP extremals-information processes (Sec.3. 2.2.8).

**II. The emerging self-organization of observer information in evolution dynamics toward intelligence**

Assuming an observable random field of probable events conceals a randomly distributed energy**,** we show how the field's interactive impulses start self-connection, developing in interactive virtual observer, generate information bit of information observer, which self-organizes observation in evolving information dynamics, creating cooperative networks, and finally an intelligence of self-evolving observer.

## 1. Stages and levels of the emerging observer self-organization of information and the evolution dynamics

I. Observable process.

In a probability field of interacting events, an infinite sequence of independent events satisfying Kolmogorov 0-1 law, affect a Markov diffusion process' probabilities, distributed in this field.
The Markov transitional probabilities change the process a priori-a posteriori Bayes probabilities, the probability density of random No-Yes impulses 0-1 or 1-0. That links the Kolmogorov's 0-1 probabilities, the Markov process' Bayes probabilities, and the Markov No-Yes impulses in common Markov diffusion process.
These abstract objective probabilities quantify the probabilistic link measures.
The random interactive actions may randomly shift each impulse 0-1 to a following 0-1 or to 1-0, which connects them in a correlation through the Markov process drift and diffusion.
The probabilistic impulse' Yes-No actions represent an act of a virtual observation where each observation measures a probability of potential events.
The arising correlation reduces the conditional entropy measures connecting the probabilistic observations in a virtual observing process with No-Yes actions. That defines first level of this stage.
This correlation connects the Bayesian a priori-a posteriori probabilities in a temporal memory that does not store virtual connection, but renews, where any other virtual events (actions) are observed.
Memorizing this action indicates start of observation with following No-Yes impulse at level two.
The starting observation limits minimal entropy of virtual impulse, which depends on minimal increment of process correlation overcoming a maximal finite uncertainty at level three.

II. The impulse' max-min self-action.

Each impulse' opposite No-Yes interactive actions (0-1) carries a virtual impulse which potentially cuts off the random process correlation (at first level) whose conditional (Bayes) entropy decreases with growing the cutting correlations.
If a preceding No action cuts a maximum of the cutting entropy (and a minimal probability), then following Yes action gains the maximal entropy reduction-its minimum (with its maximal probability) during the impulse cutoff (at level two).
The impulse' maximal cutting No action minimizes absolute entropy that conveys Yes action (raising its probability).
That leads to a maxmin of conditional (relational) entropy between the impulse actions, transferring the probabilities.
The maxmin rules the impulse observations (at level three).
The virtual impulse, transferring the Bayes probabilistic observation, virtually probes an observable random process, processing the maxmin observation.
The probing impulse, consisting of step-down No and step-up Yes actions, preserves probability measure of these maxmin actions along the observation process.
This sequence of interacting impulses, transforming opposite No-Yes actions, increases each following Bayesian posterior probability and decreases the relative entropy reducing the entropy along the observation.
The observation under random probing impulses with opposite Yes-No probability events reveals hidden correlation [71], which connects the process' Bayesian probabilities increasing each posterior correlation (at level three).
The maxmin self-defines the variation principle which *formalizes* description of the observing evolution path (level four).

III. Virtual Observer

If the observing process is self–supporting through automatic renewal these virtual probing actions, it calls a Virtual Observer, which acts until these actions resume, up to the emergence of real Information Observer (if it appears).

Such virtual observer belongs to a self–observing process, whose Yes action virtually starts next impulse No action, etc. Both process and observer are temporal, ending with stopping the observation.

Starting the virtual self-observation limits a *threshold* of the impulse's connection on first level on this stage.

The virtual observations may not link up to the real ones, but are preceding them.

The sequentially reduced relational entropy conveys probabilistic causality along the process with temporal memory collecting correlations, which interactive impulse innately cuts. The cutting entropy defines second level of this stage.

IV. Emerging the observer time

The correlation holds entrance of a *time interval* of the impulse–observation (at level one).

The time interval, connecting the probability-entropy in correlation, measures an uncertainty distance between nearest current observations (at level two).

The measuring, beginning from the starting observation, identifies an interval from the start, which is also virtual, disappearing with each new connection that identifies a next interval temporally memorized in that correlation connection. The difference of the probabilities actions temporary holds memory of the correlation, as a virtual measure of an *adjacent distance* between the impulses' No-Yes actions.

It indicates a probabilistic accuracy of measuring correlation in a *time interval's unit* (at level three).

The impulses of the observable random process hold virtually observing random time intervals.

V. Emergence of the impulse space interval and space-time geometry in an Observer structure.

With growing correlations, the intensity of entropy per the interval (as entropy density) increases on each following interval, indicating a shift between virtual actions, measured in a time interval's unit measure $|1|_M$. (Level one).

The shift merges the impulse border interactive actions , which generate an interactive jump of a high entropy density.

The jump cutting action ↓ of growing density *curves* an emerging *½ time units* of the border impulse' time interval.

The jump, curving that time, initiates a *displacement* starting rotation the impulse opposite Yes-No actions. (Level two).

This originates the curved space shifts, quantified by the impulse $\bar{u}_k$ invariant probability (1 or 0) measure $p[\bar{u}_k]$ on *two shifting space units* (as a counterpart to the curved time, Fig.1a). (Level three).

The displacement within the impulse $\bar{u}_k$ changes the impulse primary time to a discrete space form while *preserving* its measure $M[\bar{u}_k] = |1/2 \times 2| \xrightarrow{p[\bar{u}_k]} |1|_M$ in the emerging time-space coordinate system.

The measure is conserved in following time-space correlated movement transferring a minimax.

The virtual observer, being displaced from the initial virtual process, sends the discrete time-space impulses as virtual probes to *self-test* the preservation of Kolmogorov probability measure of the observer process with probes' frequencies defined by the probabilities (Level four). The Observer *self-supporting* probes increase frequencies checking a growing probability. Such test checks this probability via symmetry condition indicating the probability correctness and identifying time-space of the virtual observer' location. (Level five). The memory temporary holds a difference of the starting space-time correlation as accuracy of its closeness, which determines a time-space shape of the observer (Level six).

The evolving shape gradually confines the running rotating movement, which *self-supports,* by developing both the shape and Observer. The virtual Observer *self-develops* its space-time virtual geometrical structure during virtual observation, which gains its real form with sequential transforming the integrated entropy to equivalent information. (Level six).

VI. The microprocess.

The microprocess emerges inside Markov diffusion process, which, therefore, should possess the Markovian additive and multiplicative properties. Satisfaction the following conditions indicates the levels of the emerging microprocess below. 1.

At growing Bayes a posteriori probability along observations, neighbor actions $\downarrow\uparrow$ with probabilities 0 and 1 may merge generating interactive jump between them.

The jump minimizes the time interval, which identifies a virtual radius of rotation displacing and curving these actions.

The curving jump initiates extreme gradient entropy, which brings the extreme discrete displacement that rotates the opposite actions' anti-symmetric entropy increments in a starting microprocess.

Because time and space of the displacement emerge with the jump, the microprocesses opposite entropies evolve together with the emerging time-space.

Action $\downarrow$ starts entropy on a displacement verge beginning the jumping impulse $\downarrow\uparrow$. At a minimal displacement size, emerges action $\uparrow$ with opposite entropy increments, which develop that impulse time-space volume.

Since that displacement rises between the actions with real probabilities 0 or 1, the displacement has no real probabilities.

The process within the displacement has studied as a sub-markov process Markov path, loops [44], connected with Schrodinger bridge [45] which is unique Markov process in the class of reciprocal processes introduced by Bernstein [46].

2. When the sub-markov process gets negative entropy measure $S^*_{\mp a} = -2$ through jumping action $\downarrow$, with relative probability $p_{a\pm} = \exp(-2) = 0.1353$, it initiates the microprocess with the minimal time on the verge of the displacement.

3. The developing anti-symmetric entropy increments connect the observation in correlation connections, increasing with growing the classical prior probabilities of multiple random actions $\downarrow$ and the poster probabilities of the random action $\uparrow$. Both of them are virtual, whose manifold decreases with growing the probability measure.

At the maximal probabilities, only a pair of the Markov additive entropies increments with axiomatic symmetric probabilities (which contains symmetrical-exchangeable states) advances in the correlated superposition of both actions $\downarrow\uparrow$. These random actions, belonging the Markov process, measures a multiplicative probability.

At satisfaction of the symmetry condition, the actions axiomatic probabilities transform to the microprocess 'quantum' probability with pairs of conjugated entropies and their correlated movements.

4. The conjugated entropies increments rotating on angle $\pi/4$ raise the space interval that holds virtual transitive action $\uparrow$ within the microprocess, which initiates the correlated entanglement.

Maximal correlation adjoins the conjugated symmetric entropies, uniting them in a running pair entanglement.

The conjugated entropies, rotating the space interval on angle $-\pi/4$, transforms the transitive action $\uparrow$ to action $\downarrow$ that settles a *transitional impulse* $\uparrow\downarrow$ which finalizes the entanglement at total angle $\pi/2 = \pi/4 - (-\pi/4)$.

The transitional impulse, holding actions $\uparrow\downarrow$ opposite to the primary jumping impulse $\downarrow\uparrow$, intends to generate an inner conjugated entanglement, involved, for example in left and rights rotations ($\mp$).

The transitional impulse, interacting with the opposite correlated entropies $\mp$, reverses it on $\pm$.

Since the correlated entropies are virtual, transition action within this impulse $\uparrow\downarrow$ is also virtual and its interaction with the forming correlating entanglement is reversible.

5. Such interaction logically erases (cuts) each previous directional rotating the entangle entropy increment of the entropy volume. This erasure emits minimal energy $e_l$ of quanta [47].

The transitional impulse absorbs this emission inside the virtual impulse, which logically memorizes the entangle units making their mirror copy. Such operation performs function of logical Maxwell Demon.

6. The observing correlation starts its integral entropy measure along all impulses discrete time-space intervals.

Multiple impulses initiate a manifold of virtual Observers with random space-time volume and shape in a collective probabilistic movement. In such movement, random impulses, proceeding between a temporary fixed random No-Yes actions, are multiple, whose manifold decreases with growing the probability measure. That also decreases the manifold of the virtual observers with the multiple probabilities, which include the microprocess. At a maximal probability, only a pair of anti-symmetric entropy observers advances in a superposition, correlation and entanglement.

The running pair entanglement confines an entropy volume of the pair superposition, which encloses the condensed correlating entropies of the microprocess.

The states of cutting correlations hold hidden process' inner connections [41, 41a], which actually initiates growing correlations up to running the superposition, entanglement, and conversion of the cutting entropy in information.

Both entangled pair' anti-symmetric entropy fractions appear simultaneously with the starting space interval.

The correlation, binging this couple with maximal probability, is extremely tangible.

A pair of correlated conjugated entropies of the virtual impulse are not separable with no real action between them.

The entangled entropy increments with their volumes, captured in rotation, adjoin the entropy volumes in a stable entanglement, when the conjugated entropies reach equalization and anti-symmetric correlations cohere. The stable entanglement minimizes quantum uncertainty of the ongoing virtual impulses and increases their Bayesian probability.

7. The entangle logic is memorized, when the rotating step-up action $\uparrow$ of the transitional impulse moves to transfer the entangle entropy volume to ending step-up action $\uparrow$ of the jump, which follows real step-down action $\downarrow$.

*The entanglement starts before the space is formed and ends with beginning the space during reversible relative time interval* of the impulse $\delta t_{\pm}^{ko}/\tau_k \cong 0.03125$  $0.015625\pi$ part of $\pi$ with time interval $\tau = 1nat$. The last interactive action kills and finally memorizes the joint entangle unithe engled ts in the action ending state as the information Bit. The killing is the irreversible erasure encoding the Bit, which requires energy.

The energy, capturing between these interacting actions $\uparrow \downarrow$, carries the last action.

That action, cutting the entropy volume, initiates an irreversible process, satisfying the Landauer principle and compensating for the cost of Maxwell Demon. Such a process preempts memorizing any bits and forming triplets.

The memorized bit freezes the energy spent on the erasure for its creation as the bit equivalent ln2.

8. The microprocess is different from that in quantum mechanics (QM), because it arises inside the evolving probing impulse under No–Yes virtual and final real actions.

The microprocess' superposing rotating anti-symmetric entropy increments, which the cutting EF defines, have additive time–space complex amplitudes correlated in time–space entanglement that do not carry and bind energy, just connects the entropy in joint correlation.

These complex amplitudes models elementary interaction with no physics, while real cut brings the physical bit. Whereas the QM probabilistic particles carry analogous conjugated probability amplitudes correlated in time-space entanglement.

Theoretically, a pure probability predictability challenges Kolmogorov's probability measure at quantum mechanics' entanglement when both additivity and symmetry of probability for mutual exchangeable events vanish. It happens in this Observer probabilistic approach with appearance of information in the microprocess by cutting the entangle entropy. Memorizing the process information also cuts the microprocess in the evolving observation.

Conclusively, the impulse step-down cut $\downarrow$, extracting each Bit hidden position, erases it at a cost of the cutting real time interval, which encloses the energy of the natural interactive process. The impulse step-up' $\uparrow$ stopping state at the end of the impulse's time interval memorizes (encodes) the information Bit of the impulse.

Each impulse encoding merges its memory with the time of encoding, which minimizes that time.

Forming transitional impulse with entangled qubits leads to possibility memorizing them as quantum bit.

The needed memory of the transitional curved impulse encloses entropy $0.05085 Nat$ (Sec.I. 3.5.4).

9. The logical operations with information units achieve the approach a goal: integrating the discrete information hidden in the cutting correlations in the creating structure of Information Observer.

The relational entropy conveys probabilistic causality with temporal memory of correlations, while the cutoff memorizes certain information causality during the objective probability observations.

The self-observing virtual observer self-generates elementary Bit self-participating in building self-holding geometry and logic of its prehistory, and predicting evolving dynamics without any physical law.

The gap between entropy and information.

As maximal a priori probability approaches $P_a \to 1$, both the entropy volume and rotating moment grow. Still, between the maximal a priori probability of virtual process and a posteriory probability of real process $P_p = 1$ is a small microprocess' gap, associated with time-space probabilistic transitive movement, separating entropy and its information (at $P_a < 1$ throughout $P_p \to 1$).

It implies a distinction of statistical possibilities with the entropies of virtually from the information-certainty of reality. The gap holds a hidden real locality which impulse cuts within the hidden correlation. The rotating momentum, growing with increased volume, intensifies the time–space volume transition over the gap, acquiring physical property near the gap end, and, when last posterior probability $P_p = 1$ overcomes last prior virtual probability, the momentum curves a physical cut of the transferred entropy volume.

It is impossible to reach a reality in quantum world without overcoming the gap between entropy-uncertainty and information-certainty, which is located on edge of reality. Within the gap, the simultaneous entangled quantum conjugated entropies $S^*_{\mp a} = 2h^o_\alpha$, at probabilities $p_{\pm a} = \exp(-2h^o_\alpha) \to 1$, confine the entangling qubits, limited by minimal uncertainty measure $h^o_\alpha = 1/137$ -structural parameter of energy, which includes the Plank constant's equivalent of energy. Injection of the energy corresponds

$p_{\pm a} = \exp(-2h^o_\alpha) = 0.9855507502$ (3)

which starts both the erasure of entropy and memory, while actual killing with probability $P_k = 0.99596321$ (3a) ends erasure and memory. A gap to reality evaluates $1 - P_k \cong 0.004$ .(3b)

Both memorizing and encoding the classical and qubits have high probability but less than 1 not reaching absolute reality.

Information process

This process emerges from observing process of virtual observer (Level 1) continuing as microprocess of conjugated entropy flows within an interacting impulse (Level 2).

Information arises from multiple random interactive impulses when some of them erase-cut other providing Landauer's energy on first level of this stage. An asymmetrical inter-action erasing the impulse becomes bit of information as a phenomenon of interactions defining second level of the stage (Level 3).

The impulse cutoff correlation sequentially converts the cutting entropy to information that memorizes the probes logic in Bit, participating in next probe-conversions which generate interactive information process at Level 3.

The origin of information, thus, associates with the impulse ability of both cut and stipulate observing process, generating information under the cut, whose memory is the impulse' cutting time interval.

Each impulse' cutting information 1 Nat, deducting from its bit $S = \ln 2 \cong 0.7 Nat$, creates a *free* information $f_s \cong 0.3 Nat$ which enables connecting a multiple bits through *information attraction* at Level 4.

Multiple cuts of the more probable posterior correlations in the interactive multi-dimensional observation is a source of $f_s$ that composes multiple Bits with memories of the collecting impulse' cutting time intervals, freezing the observing events dynamics in information processes. (Level 4)

Information attraction of multiple bits through growing their $f_s$ originates time-*cooperation* in the information process at Level 5. The integration of cutting Bits time intervals along the observing time course converts it to the *Information*

*Observer* inner time course that is opposite to the observable time course in which the virtual process entropy increases (Level 6). The difference between each previous impulse' minimal and the following impulse' maximal information, which is random but predictable by the EF between these moments, can model "mutation" in evolving information process at EF-IPF measure estimating at Level 7.

The emerging macroprocess composes basic triplet units
The rotating movement (Fig.4) connects the microprocess imaginary entropy and information Bits *in a macroprocess,* where the free information binds the diverse Bits in *collective* information time-space macro trajectories (Figs.2,5) (Level1).

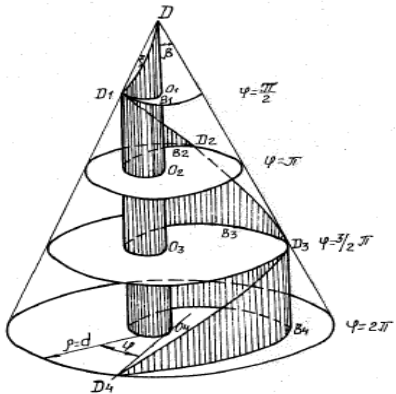

**Figure 4. Forming a space-time spiral trajectory with radius** $\rho = b\sin(\varphi \sin \beta)$ **on the conic surface at the points D, D1, D2, D3, D4 with the spatial discrete interval DD1=** $\mu$**, which corresponds to the angle** $\varphi = \pi k /2$**,** $k=1,2,...$ **of the radius vector's** $\rho(\varphi,\mu)$ **projection of on the cone's base (O1, O2, O3, O4) with the vertex angle** $\beta = \psi^o$**.**

*The observing information moves the macroprocess, whose rotation depends on forming the entropy gradient.* Minimum three rotating Bits join in triplet (UP) unit measuring macrorocess information $\mathbf{a}_{io}(\gamma_i)$ (at Level 2), whose size limits the unit' starting maximal and ending minimal information speeds, attracting new UP by its free information $\mathbf{a}(\gamma_i)$. Parameter $\gamma_i = \beta_{toi}/\alpha_{tio}$ connects the entangled entropy's frequency of rotating volume $f_{io}$ at $\beta_{toi} = c_{ev} \to f_{io}$ - holding imaginary UP part, and the ending value of speed of transition of rotating entropy volume $\alpha_{tio}$ holding the UP real part. Since each killing has a frequency of creating the bit, it implies frequency $f_{io}$ will periodical appears as a bleaching signal from the microprocess. The microprocess within impulse is reversible, keeping transitional measure $\pi$, until the impulse cutting action transforms to information bit composing the irreversible macroprocess with frequency $f_{io} = \pi/3$.

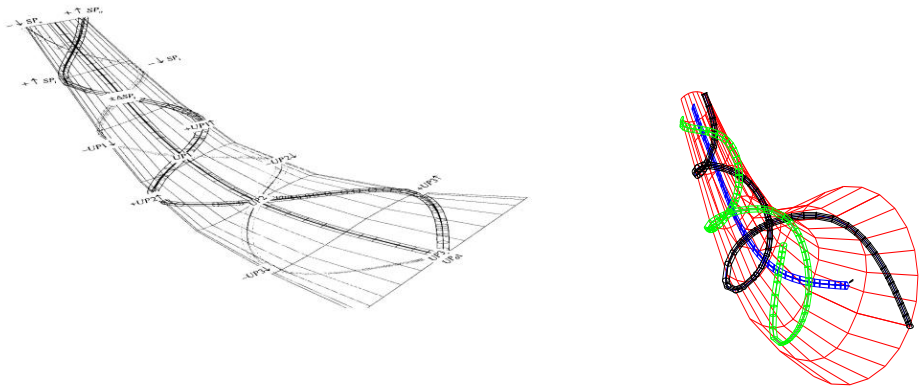

**Fig.5.** Time-space opposite directional-complimentary conjugated trajectories $+\uparrow SP_o$ and $-\downarrow SP_o$, of Hamiltonian macroprocess (Sec.I.3.8), forming the spirals located on conic surfaces. Trajectory on the spirals bridges $\pm\square SP_i$ binds the contributions of process information macro unit $\pm UP_i$ through the impulse joint No-Yes actions, which model a line of switching inter-actions (the middle line between the spirals). Two opposite space helixes and middle curve are on the right.

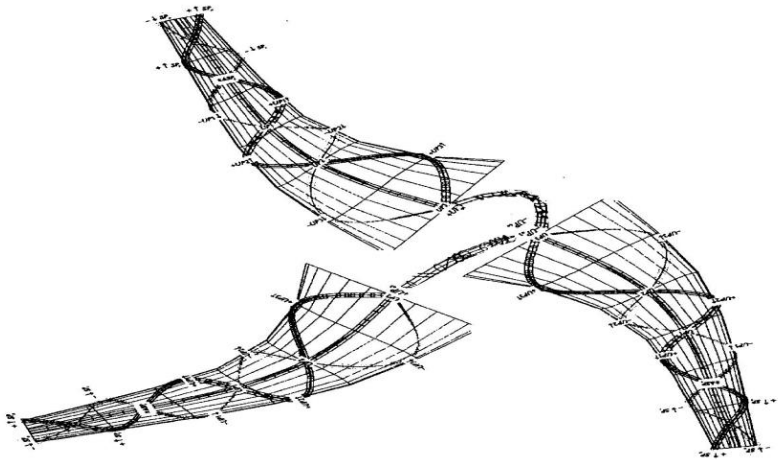

**Fig.6.** Assembling three formed units $\pm UP_{i0}$ at a higher triplet's level connecting the units' equal speeds (Fig.2).

At forming UP, the information speeds of two cooperating bits, selected automatically during the attracting minimax movement, should coincide with the rotating speed of third bit (at level 3), which joins the two cutoff Bits and third Bit that delivers information for the next cutting Bit (Figs.3, 5) and binds it in the third Bit, providing local stability of UP. This attracting process forms a rotating loop of harmonized speeds-information frequencies analogous to the Efimoff scenario **[47]** of resonance frequencies which gives a rise of many three units systems, (Fig.6).

The loop includes Borromean knot and ring, which was early proposed in Borromean Universal three-body relation.

Forming UP depends on the bit's *fitness* for the triple cooperation in the UP, which alters at changing the bit's moving speed.

The macroprocess free information integrates the Bits in an information path functional (IPF), which encloses the Bits joint in the UP time–space geometry in the process' information structure.

The mutation between the bits brings a variety of the bits with different moving speeds which changes its fitness for particular UP. The variety of three bits' finesses could form different $UP_i$, which the minimax selects at level 4, or none if they do not fit its triple self-connection ending level 4.

Emerging Information networks

1. Assembling $UP_i$ in cooperative units $UP_{oi}$ structuring information network (IN) During the attractive macro-movement, the multiple $UP_i$ triples adjoin in $UP_{oi}$ triplets cooperating in time-space hierarchical network (IN) (Fig.7), Level 1.

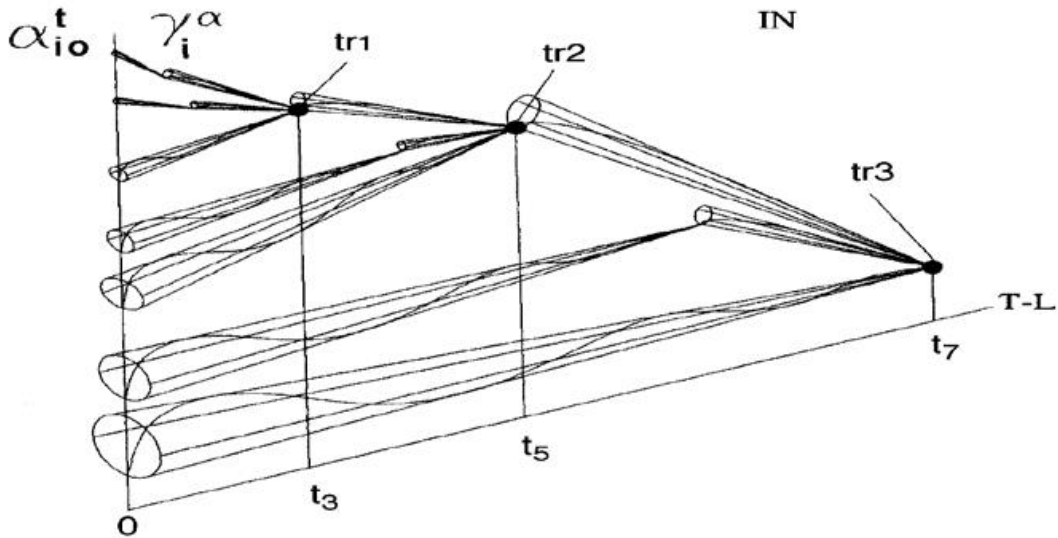

**Fig. 7. The IN information geometrical structure of hierarchy of the spiral space-time dynamics (Figs.5,6) of triplet nodes (tr1, tr2, tr3, ..); $\{\alpha_{io}^t\}$ is a ranged string of the initial eigenvalues, cooperating on $(t_1, t_2, t_3)$ locations of T-L time-space.**

Particularly, information speeds of primary triplets' $UP1, UP2$ connect them to $UP3$ building the triplet's $UP_{o1}$ knot by the information spent on their attractive movement. The knot, attracting free information, forms a rotating loop which attracts next forming $UP_{o2}$ and then $UP_{o3}$ possibly from different process dimensions, when the triple fits the cooperative minimax conditions analogous to the UP(Level 2).

Free information of cooperating $UP_{oi}$ forms hierarchical IN information structure of nested knots-nodes. (Level 3). Each $UP_{oi}$ has *unique position* in the IN hierarchy, which defines exact location of such information logical structures.

The position depends on each unit' information measure $\mathbf{a}_{iuo}(\gamma_{iu})$ parameter $\gamma_{iu}$.

The IN node hierarchical level classifies *quality* of assembled information, while the currently ending IN node integrates information enfolding all IN's levels. (Level 4)

New information for the IN delivers the requested node information's interactive impulse, through impact on the cutoff memorized entropy of observing data events.(Level 5).

Appearing new quality of information currently builds the IN temporary hierarchy, whose high level enfolds information logic that requests new information for the running observer's IN, which extends the IN hierarchy and logic. (Level 6).

Ending IN triplet integrates these cooperative qualities holding information, frequency, space-time location, which evaluate quality of all IN. (Level 7).

In the IN hierarchy where the node quality is descending with growing the node hierarchical level, the ending triplet-node has higher quality, compared to other nodes. The IN nested structure harmonizes its nodes quality which the ending node enfolds.

Variety of distinctive $UP_k$ can make multiple different networks $IN_{oj}$, which harmonizes its nodes' specific cooperative qualities. The ending nodes enfold particular quality information and its time space positions, depending on the value of each nested node parameter $\mathbf{a}_{iuo}(\gamma_{iu})$.

2. Assembling $UP_{oi}$ in cooperative units structures a higher level $1_{oi}$ of the information networks.

The attracting minimax movement assembles each three of ending of the network $IN_{oj}$ nodes $UP_{oi}$ in new formed $1_{oi}$ unit forming new loops on higher structural organization level, which connects the cooperating speeds of each triple (Fig.6).

Particularly, the attractive motion of rotating triples units $(+UP_{o1}, -UP_{o2}, +UP_{o3})$, taken from opposite (conjugated) ending nodes of the macroprocess networks, can cooperate them in triplet $-1_o(+UP_{o1}, -UP_{o2}, +UP_{o3})$ and then cooperate the opposite triplet $+2_o(-UP_{o5}, +UP_{o6}, -UP_{o7})$ adjoining both to third rotating triplet. Triplet unit $-1_o$ enfolds qualities of the above three networks though a loop of resonance frequencies, which depend on the nodes location and information values.

Thus, this unit's resonance frequency joins the qualities of the IN ending nodes in new quality which enfolds the adjoin qualities at next level. Since the resonance frequency is only a part of spectrum of these network nodes frequencies, it's less than the highest of them. The highest of these frequencies form new higher information level for $-1_o$ compared to any of the three, whose frequencies corresponds a lesser quality. The resonance frequency encloses higher information quality from all three.

As a result information qualities of each of triplet units $-1_o, +2_o, -3_o$ will grow, while they joining in rotating cooperative circles forms new triplet unit $1_{o3}$ which encloses the fitting from the above three.

Multiple triplets $1_{oj}, j = 3,5,7,...$ sequentially cooperate new network $IN_1$ which harmonizes its nodes higher qualities and enfolds in its ending node the highest of these qualities.

3. Each harmonized IN form a *domain* of observer with its specific qualities and high density of information enclosing all IN node densities.

Sequential built triplet knots, memorizing only current $1_{oj}, j = 3,5,7,...,$ while the previous units information have erased, automatically implement the IPF minimax, minimizing total time of building each composite information unit.

The minimax leads to sequential decreasing ending information speed of each node, and therefore to decreasing starting information speed on next cooperative unit.

It restricts spectrum of information frequencies for each self-built IN, decreasing the spectrum with growing the IN structural organization level.

Each self-built IN has limited number of cooperating triplets and the IN nodes, which is restricted by condition of ability for last IN cooperation, whose violation leads to the IN instability with rising chaotic movement.

Building each high level cooperative unit hardens the requirements to the fitness for the variety of primary bits, units UP, $UP_i$, $-UP_{o1}$, etc, which sequentially decreases their variety.

Any considered INs should satisfy invariant relations for ratios of starting information speeds $\gamma_1^\alpha = \alpha_{io}/\alpha_{i+1o}$ and $\gamma_2^\alpha = \alpha_{i+1o}/\alpha_{i+2o}$ connected by dynamic invariant $\mathbf{a}(\gamma)$ that binds the ending eigenvalues of triplet's segments:

$$\gamma_1^\alpha = \frac{\exp(\mathbf{a}(\gamma)\gamma_2^\alpha) - 0.5\exp(\mathbf{a}(\gamma))}{\exp(\mathbf{a}(\gamma)\gamma_2^\alpha/\gamma_1^\alpha) - 0.5\exp(\mathbf{a}(\gamma))}, \gamma_2^\alpha = 1 + \frac{\gamma_1^\alpha - 1}{\gamma_1^\alpha - 2\mathbf{a}(\gamma)(\gamma_1^\alpha - 1)}, \quad (3b)$$

where invariants $\mathbf{a}_{io}$ and $\gamma_i$ connect Eqs [16]:

$$2\sin(\gamma_i \mathbf{a}_{io}) + \gamma_i \cos(\gamma_i \mathbf{a}_{io}) - \gamma_i \exp(\mathbf{a}_{io}) = 0. \quad (3c)$$

Self-control of the multiple IN cooperative domains, integrating quality of the domain information

1. The main information mechanism, governing both each IN cooperation, the domain, and multiple IN domains cooperation, is space–time spiral rotation.

The basic parameters of this mechanism are invariants determined by the initial minimax depending on $\gamma_i^\alpha$ for each IN except the vertex angle $\beta = \psi^o$ of conic rotation (Fig.4), which is changing at growing the quality domain level.

That depends on each resonance frequency which determines quality of the domain information.

Primary radius of rotation determines emerging displacement distance $d_a = r_{e1}$ from which follows initial angle $\beta_o$ of the rotation trajectory of cone Fig.4 from relation

$$r_{e1} = \rho = b\sin(\varphi \sin\beta_o) \text{ at } \varphi = \pi k/2, k=1, b=1/4. \quad (4)$$

According to [51], the vertex angle can be changed discretely according to formula

$$\sin\psi_i^o = (2k)^{-1}, k=1, 2,.. \quad (4a)$$

where at $k=1$, $\psi_i^o = \pi/6$, and at following $k=2$, $\psi_{k=2}^o = \pi/2$, $\psi_{k=3}^o = 2\pi/3$. (4b)

Here $k$ is the number of current domain level; total numbers of $k$ domains determine the numbers of changing resonance frequencies, while the specific frequencies are not required, and they are different for diversity of observers.

Since the maxmin leads to growing a regular observer quality of the domain information, it implies grows of domain quantity. Switch to the following domain increases the angle of cone rotation, extending the growing IN domain.

Transferring the angle of the mechanism rotation according to the decreasing the domain frequency moves the quality of domain information on a higher information level. Whereas, running each existing domain rotation continues its functioning for the current observer.

The forming IN hierarchical structure of the nested domains requests new information by the impulses-bits, sending each request down along the hierarchy to the bit's window.

The highest level of the IN domain send the request for maximal density of needed information since the increase of information density corresponds increasing quality of information to be enfolded in the current higher observer IN domain.

2. Each requested domain information performs an impulse with different information density of the requested IN.

The information equivalent of the impulse wide $\delta_{ue}^i \cong 0.05 Nat$ limits its size and the extension minimal time interval $\delta_{te} \approx 1.6 \times 10^{-14}$ sec and the speed between the nearest impulses on time interval $\Delta_t$:

$$c_{ika} \cong 0.0516 \times 10^{14} Nat/\sec. \tag{5a}$$

That requires shortening time interval of inner communication for transporting the impulse information within its wide. It also limits potential speed of information attraction restraining the new information to obtain.

3. Thus, each Observer *owns the time of inner communication,* depending on the requested information density, and *time scale,* depending on *density* of accumulated (bound) information in each IN.

Each self-organizing information triplet is a macrounit of specific self-forming information time-space cooperative distributed network enables self-scaling, self-renovation, and adaptive self-organization. Such unit of the IN higher level cooperates the three INs lower levels ending units in its starting triplet.

Each IN has invariant parameter of its triplet cooperation $\gamma_i^\alpha \cong 3.45$. \hfill (5b)

The microprocess real speed $\alpha_{io}$ determines starting information after killing the entropy volume.

Within the reversible microprocess ($+t = -t$), ratio of current imaginary to real component of entropy is

$$S_-(t)/S_+(t) = \beta_{it}/\alpha_{it} = [1 + jtg(-t)]/[1 - jtg(+t)] = 1. \tag{5c}$$

It means while the entangled conjugated entropies are equal, the frequency of each bleaching signal stays invariant.
Information speed of starting a single real information (after killing the entropy volume):

$$\alpha_{io} = c_{iv} \cong 2.4143 \times 0.596 \times 10^{15} Nat/\sec \cong 1.44 \times 10^{15} Nat/\sec, \tag{5}$$

determines invariant .

Killing minimal entropy volume creates a bit.

In forming a first IN triplet, this speed determines the first IN eigenvalue, which attracting the second and third bit's eigenvalues (as real information speeds) creates first triplet invariant structure. The triplet parameter $\gamma_{iko}^\alpha(\gamma_{io}) \cong 3.45 - 3.8$ determines frequency $f_{ik} = 1/3(\gamma_{iko}^\alpha) \cong 1.15$ of appearance new triplet, which is closed to optimal $\pi/3 \cong 1.05$.

4. Any quality domain may request needed information increasing its IN information quality.

The requested information from any lower level of the IN domains with less information density requires longer communication time to the observer external bit's for obtaining a needed information with less density.

The question is: At satisfaction the above objective limitations on getting maximal observer quality domain with highest information density, could any particular observer chooses specific needed domain information, or creates it and then builds and develops?

<u>Acquisition of the IN current information via an observer' feedback with observation</u>
The IN node' interaction with the observable information spectrum delivers the needed information frequency when interactive impact of requested information $\mathbf{a}(\gamma_i)$ on the delivered $\mathbf{a}_\tau(t-s)$ forms an impulse function whose step-control carrying attracting information $\mathbf{a}(\gamma_i)$ initiates the observer dynamic process to acquire the delivering information $\mathbf{a}_\tau(t)$.

Assume the IN current $i$ node with speed $\alpha_{i\tau}$ requests an external information frequency, which will bring speed $\alpha_{k\tau}$ attracting the IN's $k$ node and spends on the request information of attraction $\mathbf{a}(\gamma_i)$.

Incoming information $\mathbf{a}_\tau(t-s)$ is delivered during time interval $\Delta_t = t-s$ by sequence of impulse control functions, which on each time interval $t_k$, $k = 0,1,2,....,m$ brings invariant information $\mathbf{a}_o(\gamma_i)$ as a part of $\mathbf{a}_\tau(t-s)$, where $\gamma_i$ is a priory unknown.

An interactive impact of requested information $\mathbf{a}(\gamma_i)$ on delivered $\mathbf{a}_\tau(t-s)$ evaluates Riemann-Stieltjes integral $I_s = \int_{-\infty}^{\infty} f(t-s)dg(s)$ applied to an information function $f(t-s) \to \mathbf{a}_\tau(t-s)$ [33] in the form:

$$I_s = \int_{-\infty}^{\infty} \mathbf{a}_\tau(t-s)\mathbf{a}(\gamma_i)\delta(s)ds = \mathbf{a}_\tau(t)\mathbf{a}(\gamma_i). \qquad (6)$$

Solution (6) is found [33] for step-function $dg(s) \to \mathbf{a}(\gamma)du(s)$, which holds derivation $du(s)$ forming impulse function $du(s) = \delta(s)ds$.

The interactive impulse impact of the requested node information on the external information measures information:

$$\mathbf{a}(\gamma_i)\mathbf{a}_\tau(t_{k+1}) = \mathbf{a}_o^2(\gamma_i) \qquad (7)$$

which *binds* information $I_{ik} = \mathbf{a}(\gamma_i)\mathbf{a}_\tau(t_{k+1})$ according to (6).

The binding impulse memorizes the cutting entropy of information microprocess of external spectrum prompting encoding information of data. The impulse may directly bind the incoming information, as well as information from other observer's INs.

Applying (7) allows finding delivering information $\mathbf{a}_\tau(t_{k+1})$ requested by $\mathbf{a}(\gamma_i)$ with the IN known $\gamma_i$.

For example, at $\gamma_i = 0.3, \mathbf{a}_o(\gamma_i) = 0.743688, \mathbf{a}(\gamma_i) = 0.239661, \mathbf{a}_o^2(\gamma_i) = 0.553$ we get $\mathbf{a}_\tau(t_{k+1}) \cong 2.3$ which for number of impulses $k=3$ during $\Delta_t \cong 3\delta_{ek}^t$ ($\delta_{ek}^t$ is time duration of each $k$) brings $\mathbf{a}_{ok}(\gamma_k) = 0.76924$ that according to balance Eq.

$$(\mathbf{a}_{io}(\gamma_{io}))^2 + \mathbf{a}_i(\gamma_{io}) \cong \mathbf{a}_{io}(\gamma_{io}) = \ln 2 \qquad (8)$$

determines new $\gamma_k \cong 0.05$. That allows finding the new IN $\gamma_{ik}^\alpha \cong 4.81$, which identifies the requested frequency spectrum $f_{ik}$, defined by ratio of information speeds $\gamma_{ik}^\alpha = \alpha_{io}/\alpha_{ko} = (f_{ik})^{-1}$.

This ratio increases compared to that at $\gamma_i = 0.3$.

Thus, known $\alpha_{io}$ determines $\alpha_{ko}$ and time interval $t_{k+1} = \tau_{k+1}^1 - (\tau_{k+1}^o + o_{ko})$ starting at the moment $\tau_{k+1}^o + o_{ko}$ and ending at the moment $\tau_{k+1}^1$ of turning off the $k$ control.

The interval of acquisition of the delivered information continues during $k=3$ step-wise controls, which build new IN triplet during the same time.

For this example, at $\gamma_k \cong 0.05$, Eq (8) brings the following results:

$$\delta_{ek}^t = \mathbf{a}_k(\gamma_k)/\alpha_{ko}, \alpha_{ko} = \gamma_{ik}^\alpha \alpha_{io}, \gamma_{ik}^\alpha \cong 4.81, \mathbf{a}_k(\gamma_k) \cong 0.2564, \alpha_{io} \cong |-2.57|, \delta_{ek}^t \cong 0.1\sec, \Delta_t \cong 0.3\sec \quad (9)$$

and spectrum ratio $f_{ik} \cong 0.2$.

The IN control, delivering new information that needs particular node, runs the IN feedback.
The requested information emanates from the network until it satisfies the impulse requirements, building by the triple co-operations, and the feedback interactive binding will bring growing quality information in evolution of extending that quality.
Each IN new information acquisition runs also interaction with other IN accessible spectrum within observer, which enables self-creation new information both internally and externally. How?
The feedback binds the requested IN node's high density quality of cooperation with some external information. This delivers new information which generally increases the observer information quality by both growing current network nodes and/or developing another IN with other $\gamma_{ik}^\alpha$ that enables synchronizing additional information spectrum. These correspond to growing the observer knowledge collected by both increasing level of the IN connected domain quality and extension of the domain INs through the feedback communication with the observing information.
But what does initiate a particular subjective observer to prioritize the specific needed information, create and develop explicit domain having advantage with the others?

Selection information *and structuring a multiple selective observer*
1. Forming an information dynamic cooperative requires rising cooperative information force between the potential cooperating triplets:

$$X_{ik}^{Im} = -\frac{\delta I_{ik}^m}{\delta l_{ik}^\alpha} = \mathbf{a}_{oi}(\gamma)(\gamma_{ik}^m - 1)$$

where current IN' triplet $m_i$, currying information $\delta I_{ik}^m = \mathbf{a}_i + \mathbf{a}_{oi}^2 \cong \mathbf{a}_{oi}$, attracts $m_k$ triplet, depending on $\mathbf{a}_{oi}(\gamma)$ with IN invariant parameter $\gamma_{ik}^m$, and on relative distance

$(t_i^m - t_k^m)/t_i^m = (l_i^m - l_k^m)/l_i^m = \partial l_{ik}^{m*}$.

That cooperative force' measures the attracting potential directly in Nats (bits).
2. The information potential, relative to information of first IN triplet, determines relative cooperative force between these triplets:

$$X_{1k}^{Im1} = (\gamma_{1k}^m - 1). \tag{10a}$$

The required relative cooperative information force of the first and second triplets:

$$X_{12}^\alpha \geq [\gamma_{12}^\alpha - 1], \tag{10b}$$

at limited values $\gamma_{12}^\alpha \to (4.48 - 3.45)$, restricts the related cooperative forces by inequality

$$X_{12}^\alpha \geq (3.48 - 2.45). \tag{10c}$$

The quantity of information, needed to provide this information force, is

$I_{12}(X_{12}^\alpha) = X_{12}^\alpha \mathbf{a}_o(\gamma_{12}^\alpha)$,

where $\mathbf{a}_o(\gamma_{12}^\alpha)$ is invariant, evaluating quantity of information concentrated in a selected triplet by $\mathbf{a}_o(\gamma_{12}^\alpha) \cong 1 bit$ at $\gamma_{12}^\alpha = \gamma_{1o}^\alpha$, From that it follows

$$I_{12}(X_{12}^\alpha) \geq (3.48 - 2.45) bits.  \quad (10)$$

3. The invariant's quantities $\mathbf{a}_o(\gamma_{io} \to 0)$, $\mathbf{a}(\gamma_{io} \to 0)$ provide *maximal* cooperative force $X_{12}^{am} \cong 3.48$. Minimal quantity of information, needed to form a very first triplet, estimates dynamic invariants

$$I_{o1} \cong 0.75 Nats \cong 1 bits, \quad (11a)$$

Therefore, total information, needed to start adjoining next triplet to the IN, estimates

$$I_{o12} = (I_{o1} + I_{12}(X_{12}^\alpha)) \geq (4.48 - 3.45) bit. \quad (11b)$$

which supports the node cooperation [16] and initiates the IN feedback below. This information equals or exceeds information of the IN current node needed for sequential cooperation the next triplet. Minimal triplets' node force $X_{1m}^\alpha = 2.45$ depends on the ratio of starting information speeds of the nearest nodes, which determines the force scale factor $\gamma_{m1}^\alpha = 3.45$ satisfying the minimax.

The observer, satisfying both minimal information $I_{o1}$ and admissible $I_{12}(X_{12}^\alpha)$ delivering total information $(4.48 - 3.45) bit$, we call a *minimal* selective observer, which includes the control's carried free information for the triplet's node.

These limitations are the observer boundaries of *admissible* information spectrum, which also apply to multi-dimensional selective observer.

4. At satisfaction of cooperative condition (11b), each following observed information speed, enables creating next IN's level of triplet's hierarchy, delivers the required density of the information spectrum. This leads to two conditions: *necessary*-for creating a triplet with required information density, and *sufficient* -for a cooperative force, needed to adjoin this triplet with an observer's IN. Both these conditions should be satisfied through the observer's ability to *select* information of growing density.

At satisfaction the necessary condition, each next information units joins a sequence of triplet's information structures forming the IN, which progressively increases information bound in each following triplet-at satisfaction the sufficient condition, and the IN ending triplet's node conserves all IN information. When acquisition of information brings new $\gamma_{ik}^\alpha \cong 4.81$, that parameter defines cooperative force $X_{1k}^\alpha = 3.48$, which enables transferring to next cooperative stage.

The node *location* in the IN spatial-temporal hierarchy determines *quality* of the information bound in IN node, which depends on the node enclosed information density.

5. *Extending the IN requires quantity and quality of information, which could deliver an external observer, satisfying the requested information emanating from the current IN ending node.*

Let us evaluate the interactive information impact of an external observer, carrying own information, on the IN requested information to form potential new triplet of the current observer.

Such request carries the control information $\mathbf{a}_m \sim 0.25 Nat$ with information speed $\alpha_m^t$ determined by the requested IN node, which encloses information density

$$\gamma_{m+1}^\alpha = (\gamma_{12}^\alpha)^{m+1} = (\gamma_{m=1}^\alpha)(\gamma_{12}^\alpha)^m \times \alpha_{1o}, \quad (12)$$

**71**

where $\alpha_{1o}$ is information speed on a segment of the IN initial triplet, whose ratio to its third segment speed is $\gamma^{\alpha}_{o13} = \alpha_{1o}/\alpha_{3o} \cong 3.45 = \gamma^{\alpha}_{m=1}$, at ratio $\gamma^{\alpha}_{12} = (\gamma^{\alpha}_{m=1})$ which supposedly equals for all $m+1$ triplets; and $\gamma^{\alpha}_{m+1}$ is scale factor for $m+1$-the triplets of requesting IN's node.

The requested density $\gamma^{\alpha}_{m+1}$ requires relative information frequency

$$f_{1m+1} = (\gamma^{\alpha}_{m+1})^{-1}. \tag{13}$$

Thus, the decreasing frequency with tendency of growing information quality in evolving IN delivers information forces which enable automatically overcome each following threshold and transfer to next stage of the quality.

Minimal IN with single triplet node and potential speed of attraction (5) requests its information density with speed

$$c^{\alpha}_{m+1} = (3.45)^2 \times 0.1444 \times 10^{14} \approx 11.9 \times 0.1444 \times 10^{14} \approx 1.7187 \times 10^{14} \, Nat/\sec. \tag{14}$$

To adjoin the requested information in the IN, the requesting control should carry information $I_{o12}$ (11a) with speed (14) which requires time interval for transporting this information:

$$t_m \cong 3.45 \times 0.7 \, Nat/3.564 \times 10^{14} \, Nat/\sec = 0.6776 \times 10^{-14} \sec. \tag{15}$$

This relates to interval of time communication within the observer inner processes, which is less than the impulse wide's time $\delta_{te} \approx 1.6 \times 10^{-14}$ sec that carries the requested information binding $I_{ik}$ (11b).

The decrease ratio $\delta_{te}/t_m \cong 2.36$ corresponds to increase the initial speed of attraction $0.1444 \times 10^{14} \, Nat/\sec$ in $\sim 12$ times. The increased impulse information density decreases the impulse time wide $\delta_{te}$ to

$$\delta_{tm} = \delta_{te}/12 \cong 0.1333 \times 10^{-14} \sec, \tag{16}$$

which is less than the communication time (15) in ~ 5 times.

Impulse with time wide $\delta_{tm}$ (16) transfers the requested information to the observing external process where it interacts via the probing impulses. This information should deliver the step-down cut of external impulse that requires quantity information $0.25 \, Nat$ -the same as the information which carries the requested control.

Communication time (15) to get the needed actual frequency-density (13) from the observing external process also determines the frequency of the probing impulses.

That requires to increase the initial impulse's attracting information density $i_{od} = 0.25 \, Nat/\delta_{te} = 0.25 \, Nat/1.6 \times 10^{-14} \sec = 0.15 \times 10^{14} \, Nat/\sec$ (17)

in ~ 12 times up to

$$i_{md} \cong 1.8 \times 10^{14} \, Nat/\sec. \tag{18}$$

The observer time' increase follows from preserving information invariant

$$\mathbf{a}_{om} = \alpha_{1o}\tau_{1o} = \alpha_{m+1}\tau_{m+1} \tag{18a}$$

along the IN nodes at

$$(\gamma^{\alpha}_{12})^{m+1} = \alpha_{1o}/\alpha_{m+1} = M_{\tau} = \tau_{m+1}/\tau_{1o}, \tag{19}$$

where for $\tau_{1o} = t_{io}, \tau_{m+1} = t_{ko}$ in (18a), the above densities and the time scale (19) evaluates ratio $M_{\tau} = t_{md}/t_{od} = i_{md}/i_{od} = 11.9 \approx 12$.

The increase of information density corresponds increasing quality of information to be enfolded in the current IN.
Thus, each Observer *owns the time of inner communication,* depending on the requested information, and *time scale,* depending on *density* of accumulated (bound) information.

If new node formation requires $k$ cutting information units, time interval of such cuts $\delta^t_{eik}$ will depend of the IN time scale, increasing proportionally: $\delta^t_{eik} = \gamma^\alpha_{m=k}\delta^t_{eio}, \gamma^\alpha_{m=k} = (\gamma^\alpha_{12})^k$, which, for $k=10$ increases the $\delta^t_{eio} \cong 0.2 \times 10^{-15}$ sec in $\cong 2.57 \times 10^{17}$ times up to $\delta^t_{eik} \cong 51.4$ sec. That allows memorizing a "movie" of moving $k$ information units during the time-space dynamic of the entangling units. Thus, free information, carrying attracting information force $X^{\alpha m}_{12} \cong 3.48$ with quantity of the force information $I_{o12}$ (11) delivers related quality to the forming IN by cutting external information with density (12). The cutting information enables forming new IN triplet with $\mathbf{a}_o(\gamma^\alpha_{12}) \cong 1bit \cong \ln 2 Nat$, which should be attached to the current IN. The impulse, carrying this triplet, interacts with existing IN node by impact, which provides information $0.25 Nat$ - the same as at the interaction with an observer external process.

The relative information effect of impact estimates ratio $\ln 2 / 0.25 \approx 3$. Taking into account the attracting information $0.231 Nat$ carrying with the triplet Bit, total increase brings $(3\ln 2 + 0.231) Nat \cong 3.573 bit$ which compensates for $I_{12}(X^\alpha_{12}) \geq (3.48 - 2.45) bits$ requested by the current IN triplet node.

This is minimal threshold for building elementary triplet (10), which enables attract and deliver the requested information to IN node (10a) that the selective observer can select with the needed frequency of the probing impulses.

The triplets are elementary selective objective observers, which unable getting the requested IN level quality of information that ultimately evaluates the IN distinctive cooperative function.

A subjective observer, in addition to the objective observer, *selects* the observing process to acquire needed information according to its *optimal criterion* for *growing the quality.*

The *necessary* and *sufficient conditions* for adjoining information units in the observer's IN sequentially increase quality of the enfolded information.

The identified information threshold separates subjective and objective observers.

Subjective observer enfolds the concurrent information in a temporary build IN's high level logic that requests new information for the running observer's IN and then attaches it to the running IN, currently building its hierarchy.

6. *The limited the IN time-scale, speed of cooperation, and dimension*

Minimal admissible time interval of impulse acting on observable process is limited by $\delta^o_{t\min} \cong \mathbf{a}_{io}\hat{h} \approx 0.391143 \times 10^{-15}$ sec.

Minimal wide of internal impulse $\delta_{te} \approx 1.6 \times 10^{-14}$ sec limits ratio $\delta_{te} / \delta^o_{t\min} \cong 41$ which evaluates the limited IN time scale $(\gamma^\alpha_{12})^{m+1} = M_{\tau m} = \tau_{1o}/\tau_{m+1}$ and scale ratio $\gamma^\alpha_{m+1} = (\gamma^\alpha_{12})^{m+1} = 3.45^{m+1} = 41$. That, for $m+1 = 3$ triplets brings $M = 41.06$ which for minimal IN two bound triplets enfolding $n = 5$ process dimension brings the IN *geometrical* bound scale factor $\sqrt{(\gamma^\alpha_{m1})^n}$ which for this IN holds $\sqrt{(\gamma^\alpha_{m1})^5} \cong 22.1$.

In Physics, three particles' bound stable resonance has been recently observed [48, 49] with predicted scale factor $\cong 22.7$.

If an observer enables condense the external information in a decreased wide of its impulse then the number of the IN enclosed triplet grows. This number limits a cooperative speed of IN's last triplet $m$ node, whose ratio to initial triple node information speed evaluates invariant

$$C_{oc} \cong 1/2 \mathbf{a}_{io}(\gamma) \mathbf{a}_i^{-1}(\gamma)(\gamma_{i=m}^\alpha - 1)(\gamma_{i=m}^\alpha)^m . \qquad (20)$$

Since each pair cooperation requires $1/2\ln 2 Nat$ of information, applied during time of cooperation $\delta_{te}$, the maximal speed of cooperation is

$$c_{oc} = 0.35 Nat / \delta_{te} = 0.21875 \times 10^{14} Nat/\sec, \qquad (20a)$$

whichis closed to maximal potential speed (14):

$$c_{oa} \cong 0.1444 \times 10^{14} Nat/\sec \cong 20 \times 10^{13} bit/\sec . \qquad (20b)$$

The ending node cooperative speed $c_{ico} = 1 bit/\sec \cong 0.7 Nat/\sec$ leads to ratio $c_{iam}/c_{oa} = C_{icm} = 0.2062857 \times 10^{14}$. Applying to (20) the optimal minimax invariants: $C_{ocm} \cong 1/2\ln 2/0.33\ln 2(2.45)(3.45)^m = 1.515 \times 2.45(3.45)^m = 3.77(3.45)^m$, at $C_{icm} = C_{ocm}$ determines related maximal $m_o \cong 23.6, n_o \approx 48$.

Maximal cooperative speed $c_{am} \cong 10^6 bit/s$ and a single neuron' low speed $c_{hco} \approx 10 bit/s$ leads to $C_{oc} \cong 10^5$ and to $m = 7.3 \cong 7, n = 14$. \qquad (21)

At such $m$, a complete IN of human being's information logic can build the IN sections, each with $m \cong 7$ levels, which cooperate in a triplet of future IN, composing all observed integral information (Figs.6,7).

At $c_{hco} \approx 10 bit/s$, it requests starting information frequency $B_f \cong 10^6 bit/s \approx 10^{-3} Gbit/s$.

A single IN's maximal information level for human being IN approximates $m_M \cong 7$, and a minimal selective subjective observer is limited by $4 > m_{M1} > 3$ levels.

The minimal observer with a single triplet can build the minimal IN with two triplets $m_{Min} = 2$ by adding one more triplet. The observer is able building multiple information Networks, when each three ending nodes of maximal admissible $m_M$ can form a triplet structure with enfolds all three local INs, increasing the encoded information in $3m_M$ and then multiply it on $m_M : 3m_M m_M$ by building new IN starting with this triplet.

That process allows progressively increase both quantity and quality of total encoded information in $(3m_M m_M) \times (3m_M m_M) \times ....... = (3m_M m_M)^{m_M} = N_m$ times of the initial IN's node information

$$\mathbf{a}_{1o} = \ln 2 : I_m = \ln 2(3m_M m_M)^{m_M} \qquad (22)$$

with maximal density $I_m^d = \ln 2(\gamma_{12}^\alpha)^{N_m}$ . \qquad (23a)

and time scale $M_\tau = (\gamma_{12}^\alpha)^{N_m}$ . \qquad (23b)

This a huge quantity and quality of information is limited by maximal $m_M$.

Maximal information available from an external random information process is also limited at infinite dimension of the process [42].

*7. Information conditions for self-structuring a multiple selective observer*

Satisfaction of sufficient condition (11) in multiple IN's interactions determines a multiple selective observer, whose attracting cooperative information force grows from $X_{12}^\alpha \geq [(\gamma_{12}^\alpha) - 1]$ to

$$X_{12}^{\alpha N_m} \to (\gamma_{12}^\alpha [\gamma_m])^{N_m} \text{ at } (\gamma_{12}^\alpha [\gamma_m]) \to (\gamma_{12}^\alpha [\gamma_m])^{N_m}, \qquad (24)$$

accumulating maximal information

$$I_{12}(X_{12}^{\alpha N_m}) = X_{12}^{\alpha N_m} \mathbf{a}_o[(\gamma_{12}^{\alpha})^{N_m}], \quad \gamma_m \to 0. \tag{24a}$$

Multiple communications of numerous observers send a message-demand, as *quality messenger* (qmess), enfolding the sender IN's cooperative force (10a-c), which requires access to other IN observers [43].

This allows the observer-sender generates a collective IN's logic of the multiple observers.

Each observer's IN memorizes its ending node information, while total multi-levels hierarchical IN memorizes information of the whole hierarchy. The observer, requesting maximal quality information (by its intentional information (10)), generates probing impulses, which select the needed density-frequency's real information among imaginary information of virtual probes. That brings to the observer as sender all current IN logical information; while the IN information dynamics enable renovate the existing IN in a process of exchanging the requested information with environment, and rebuild the IN by encoding and re-memorizing the recent information. Since the whole multiple IN information is *limited* as well as a total time of the IN existence, the possibility of the IN self-replication arises.

<u>Observer's IN self-replication and conditions of self-generation the observer new information quality.</u>

Each IN node's maximal admissible $m_M$-th level ends with a single dimensional process, which in attempt to attach new triplet, over constrained $\gamma_{k1} \to 1$, loses ability to enfold new attracting information. Such IN stops satisfying the minimax information law, which leads to its instability. That violation of the constrain is *natural intention* to grow for each IN.

Specifically, after completion of the IN cooperation, the control of last IN ending node initiates one dimensional process $x_n(t_n) = x_n(t_{n-1})(2 - \exp(\alpha_{n-1}^t t_{n-1}))$, which at $t_{n-1} = \ln 2 / \alpha_{n-1}^t$ approaches final state $x_n(t_n) = x_n(T) = 0$ with potential infinite relative phase speed

$$\dot{x}_n / x_n(t_n) = \alpha_n^t = -\alpha_{n-1}^t \exp(\alpha_{n-1}^t t_n)(2 - \exp(\alpha_{n-1}^t t_n))^{-1} \to \infty.$$

Since the model cannot reach zero final state $x_n(t_n) = 0$ with $\dot{x}_n(t_n) = 0$, a periodical process arises as result of alternating the macro movements with the opposite values of each *two* relative phase speeds $\dot{x}_{n+k-1} / x_{n+k-1}(t_{n+k-1}) = \alpha_{n+k-1}^t, \dot{x}_{n+k} / x_{n+k}(t_{n+k}) = -\alpha_{n+k}^t$.

That leads to instable fluctuations of these speeds at each $t = (t_{n+k-1}, t_{n+k})$ starting the alternations with $\dot{x}_n / x_n(t_n) = \alpha_n^t$, $k = 1, 2, ..$ at $\gamma \geq 1$.

The instable fluctuations in three-dimensional process involves the oscillation interactions of three ending node other IN's approaching $\gamma \geq 1$, which generate frequency spectrum of model eigenvalues $\lambda_i^*(t_{n+k})$ in each space dimension $i = 1, 2, 3$.

Formal analysis of this instability associates with nonlinear fluctuations, which can be represented [52] by a superposition of linear fluctuations with the frequency spectrum ($f_1, ..., f_m$) proportional to imaginary components of the spectrum eigenvalues ($\beta_1^*, ..., \beta_m^*$), where $f_1 = f_{\min}$ and $f_m = f_{\max}$ are the minimal and maximal frequencies of the spectrum accordingly.

In our model, the oscillations under the interactive control generate imaginary eigenvalues $\beta_i^*(t)$:

$$\text{Im}\,\lambda_{n+k}^i(t) = \lambda_{n+k-1}^i[2 - \exp(\lambda_{n+k-1}^i t)]^{-1}$$

at each $t = (t_{n+k-1}, t_{n+k})$ for these $i$ dimensions. This leads to relation

$$\operatorname{Im}\lambda_n^i(t_{n+k}) = j\beta_n^i(t_{n+k}) = -j\beta_{n+k-1}^i \frac{\cos(\beta_{n+k-1}^i t) - j\sin(\beta_{n+k-1}^i t)}{2 - \cos(\beta_{n+k-1}^i t) + j\sin(\beta_{n+k-1}^i t)}, \quad (25)$$

**at** $\beta_i^* = \beta_{n+k}^i, \beta_i^* \neq 0 \pm \pi k$, where that $\beta_n^i(t_{n+k})$ includes real component

$$\alpha_n^i(t_{n+k}) = -\beta_{n+k-1}^i \frac{2\sin(\beta_{n+k-1}^i t)}{(2 - \cos(\beta_{n+k-1}^i t))^2 + \sin^2(\beta_{n+k-1}^i t)}, \quad (25a)$$

at $\alpha_i^* = \alpha_i^*(t_{n+k}) \neq 0$, with the related parameter of dynamics

$$\gamma_i^* = \frac{\beta_n^i(t_{n+k})}{\alpha_n^i(t_{n+k})} = \frac{2\cos(\beta_{n+k-1}^i t) - 1}{2\sin(\beta_{n+k-1}^i t)}. \quad (26)$$

At $\gamma = 1$ it corresponds $(\beta_{n+k-1}^i t) \approx 0.423 rad (24.267^o) = 0.134645\pi$.

These fluctuations may couple the nearest dimensions by an interactive double cooperation through overcoming *minimal elementary* uncertainty UR, *separating the model's dimensions' measure via invariant* $h_\alpha^o$, which may border the IN maximal stable level $m_M$.

Suppose, the $k-$th interaction, needed for creation a single element in the double cooperation, conceals information $s_c(\gamma) = \mathbf{a}_{ok}^2(\gamma)$, which should compensate for increment of minimal uncertainty of invariant $h_\alpha^o \mathbf{a}_o(\gamma = 0) = 0.76805/137 = 0.0056$ Nat by information equals $\delta \mathbf{a}_k = \mathbf{a}_o(\gamma = 0) - \mathbf{a}_o(\gamma^*)$.

This invariant evaluates the UR information *border* by the increment of the segment's entropy concentrated in UR, while equality $h_\alpha^o \mathbf{a}_o(\gamma = 0) = \mathbf{a}_{ok}^2(\gamma)$ evaluates minimal interactive increment

$\delta \mathbf{a}_k = \mathbf{a}_{ok}^2(\gamma) = 0.0056$ (26a) with minimal information $\mathbf{a}_{ok}(\gamma) = 0.074833148$, (26b)

needed for a single interaction in each dimension.

Each $k-$th interaction changes initial $\gamma \geq 1$ on $-\Delta\gamma = -\gamma^*$ bringing minimal information increment in each dimension (26b). Minimal information attraction enables cooperate a couple needs three these increments, generating information

$3\mathbf{a}_{ok}(\Delta\gamma) = 0.2244499443 \cong 0.23 = \mathbf{a}_k$. (26c)

Information attraction $\mathbf{a}_k$, generated in each dimensional interaction, can cooperate that interactive information in invariant $\mathbf{a}_{ok}(\gamma) \cong 0.7$ which binds three dimensions in single bit through total nine interactions.

The question is: how the interactive fluctuations enable creating a triplet which self-replicates new IN? Dynamic invariant $\mathbf{a}(\gamma) = \mathbf{a}_k$ of information attraction determines ratios of starting information speeds $\gamma_1^\alpha = \alpha_{io}/\alpha_{i+1o}$ and $\gamma_2^\alpha = \alpha_{i+1o}/\alpha_{i+2o}$ needed to satisfy invariant relations (3b).

To create new triplet's IN with ratio $\gamma_1^\alpha = \alpha_{io}/\alpha_{i+1o}$, relation (26) requires such ratio $\frac{\beta_i^*(t_{n+k})}{\beta_{n-1,o}(t_{n-1,o})} = l_{n-1}^m$ which deliver imaginary invariant $(\beta_{n+k-1}^i t) \to (\pi/3 \pm \pi k), k = 1, 2, \ldots$ at each $k$ with ending information frequency $\beta_{lo}(t_o) = \beta_i^*(t_{n+k})$ that would generate needed $\alpha_n^i(t_{n+k}) = \alpha_{lo}^m(t_o)$.

In case $\Delta\gamma \to 0 \to \gamma^*$, it can be achieved in (26) at $2\cos(\beta_{n+k-1}^i t) \to 1$, or at
$$(\beta_{n+k-1}^i t) \to (\pi/3 \pm \pi k), k = 1, 2, \ldots, \text{ with } \alpha_n^i(t_{n+k}) = \alpha_{lo}^m(t_o) = \lambda_{lo}^m \cong -0.577\beta_{n+k-1}^i \quad (27)$$

That determines maximal *frequency's ratio*
$$l_{n-1}^m = \beta_{n+k-1}^i / \beta_{n-1,k=o}^i \quad (27a)$$

which at $\gamma = 1$, $\beta_{n-1,o}(t_{n-1o}) = \alpha_{n-1,o}(t_{n-1,o})$ and $\beta_{n+k-1}^i = \alpha_{l,k=3}^i / 0.577$ holds
$l_{n-1}^m = \alpha_{l,k=3}^i / 0.577 / \alpha_{n-1,k=o}^i$, $\alpha_{l,k=3}^i / \alpha_{n-1k=o}^i = \gamma_1^\alpha$. It identifies (27a) connecting the triplet ratio by invariant
$$l_{n-1}^m = \gamma_1^\alpha / 0.577, \quad (28)$$

generated by the initial $(n-1)$-dimensional spectrum with an imaginary eigenvalue $\beta_{n-1,o}(t_{n-1,o})$ by the end of the interactive movement. Invariant (27a), (28) leads to $\gamma_1^\alpha = \alpha_{l,k=3}^i / \alpha_{n-1,k=o}^i = 3.89$ and to ratio of initial frequencies $l_{n-1}^{m=1} \cong 6.74$. Next nearest $\alpha_{l+1,k=6}^i / \alpha_{lo,k=3}^i = \gamma_2^\alpha$ needs increasing first ratio in $l_{n-1}^{m=2} \cong 3.8$, and the following $\alpha_{l+2,k=9}^i / \alpha_{l+1,k=6}^i = \gamma_2^\alpha$ needs $l_{n-1}^{m=3} \cong 3.8$. The multiplied ratios
$$l_{n-1}^{m=1-3} = l_{n-1}^{m=1} \times l_{n-1}^{m=2} \times l_{n-1}^{m=3} \cong 97.3256 \quad (28a)$$

need to build new triplet, which at $\gamma = 1$ not ends with segment, while $l_{n-1}^{m=1-3}$ identifies maximal ratio of spectrum frequencies generated by the instable fluctuations.

Each three information $3\mathbf{a}_{ok}(\Delta\gamma)$ binds pair of nearest spectrum frequencies, starting with pair $\beta_{n-1,o}^i, \beta_{n,k=1}^i$, which sequentially grows with each $k$ interactive information (26a-c), intensifying the increase of frequency.

First three pairs bind three dimensions in single Bit $\mathbf{a}_{ok}(\gamma) \cong 0.7$ through nine interactions of frequencies requiring ratio $l_{n-1}^{m=1} \cong 6.74$, next pair ratio grows in ~2.24 times, each next in 1.26 times.

Thus, a natural source to produce the very first triplet is nonlinear fluctuation of an initial dynamics, involving, as minimum, three such native dynamics dimensions that enclose some memorized information by analogy with ending IN node, which at $\mathbf{a}_o(\gamma = 1) = 0.58767, \mathbf{a}(\gamma = 1) = 0.29$ brings $\gamma_1^\alpha \cong 2.95$ and needs ratio $l_{n-1}^{m=1} \cong 5.1126$.

Natural source of maximal speed frequency is light wavelength whose time interval $t_{lo} \approx 1.33 \times 10^{-15}$ sec determines maximal frequency $f_{max} \cong 0.7518 \times 10^{15} \text{ sec}^{-1}$ that at each interaction brings information
$$h_\alpha^o \mathbf{a}_o(\gamma = 0) = 0.005 \text{ Nat} \quad (28b)$$

changing the initial frequency.

The required frequency ratio $l_{n-1}^{m=1-3}$ identifies minimal frequency $f_{\min} = 0.7724586 \times 10^{13} \sec^{-1}$.

The triplet information invariant allows finding the equivalent energy invariant for creating such triplet. Invariants $h_\alpha^o \cong 1/137$ coincides with the Fine Structural constant in Physics:

$$\alpha^o = 2\pi \frac{e^2}{4\pi\varepsilon^o hc}, \tag{29}$$

where $e$ is the electron charge magnitude's constant, $\varepsilon^o$ is the permittivity of free space constant, $c$ is the speed of light, $h$ is the Plank constant.

The equality $h^o \cong \alpha^o$ between the model's and physical constants allows evaluate the model's structural parameter energy through energy of the Plank constant and other constants in (29):

$$h = \frac{e^2}{2\varepsilon^o ch^o} = C_h \alpha_h, \alpha_h = (h^o)^{-1} = inv, \; \frac{e^2}{2\varepsilon^o c} = C_h = h/(h^o)^{-1} = 9.0831 \times 10^{32} J \cdot \sec \tag{29a}$$

where $C_h$ is the energy's constant (in [J.s]), which transforms the invariant $\alpha_h$ to $h$.

In this information approach, (29a) evaluates energy that conceals the IN bordered the stable level $m_M$. The triplet creation needs nine such interacting increments, which evaluate the triplet energy's equivalent $e_{tr} = 8.1748 \times 10^{-29} J.\sec$. (29b)

Invariant conditions (26), (29a) enable the model's *cyclic* renovation, initiated by the two mutual attractive processes, which do not consolidate by the moment of starting the interactive fluctuation. After the model disintegration, the process can renew itself with the state integration and the transformation of the imaginary into the real information during the dissipative fluctuations bringing energy for a triplet.

Initial interactive process may belong to different IN macromodels (as "parents") generating new IN macrosystem (as a "daughter"), at end of the "parents" process and beginning of the "daughters".

The macrosystem, which is able to continue its life process by renewing the cycle, has to transfer its coding life program into the new generated macrosystems and provide their secured mutual functioning.

A direct source is a joint information of three different IN nodes $\mathbf{a}_{o1}(\gamma_1), \mathbf{a}_{o2}(\gamma_2), \mathbf{a}_{o3}(\gamma_3)$ enable initiate attracting information with three information speeds where one has opposite sign of the two; while the information values cooperating an initial triplet will satisfy the above invariant relations.

Creating a triplet with specific parameters depends on the starting conditions initiating the needed attracting information. To achieve information balance, satisfying the VP and the invariants, each elementary $\mathbf{a}_{oi}(\gamma_i)$ searches for partners for the needed consumption information. A double cooperation conceals information $s_c(\gamma) = \mathbf{a}_o^2(\gamma)$, while a triple cooperation conceals information $s_{cm}(\gamma) = 2\mathbf{a}_o^2(\gamma)$ and it could produce a less free information, while both of them depend on $\gamma$. With more triplets, cooperating in IN, the cooperative information grows, spending free information on joining each following triplet.

Minimal relative invariant $h_\alpha^o = 0.00729927 \cong 1/137$ evaluates a maximal *increment* of the model's dimensions $m_M \cong 14$ (29c), and the quantity of the hidden invariant information (28b) produces an elementary triple code, enclosed into the hyperbolic structure Fig.8, with its cellular geometry.

This hidden *a non-removable uncertainly also enfolds a potential DSS information code.*

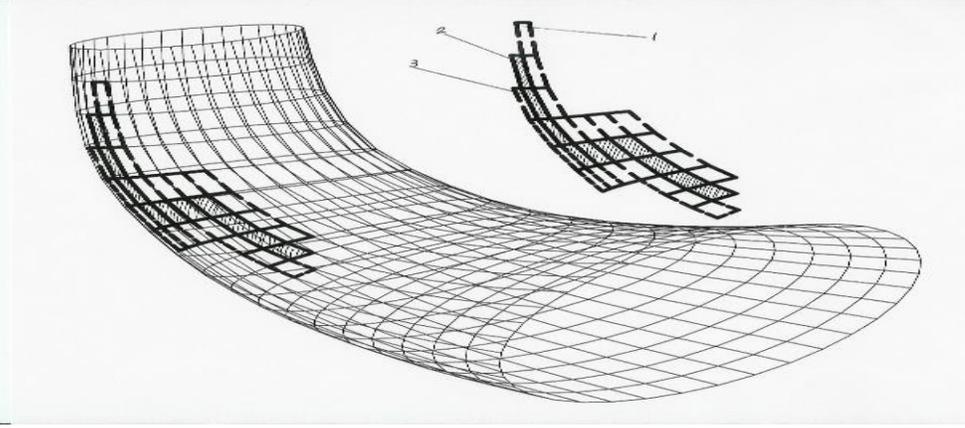

Fig. 8. Structure of the observer bcellular geometry, formed by the cells of the triplet's code, with a portion of the surface cells (1-2-3), illustrating the space formation. This structure geometry integrates information contributions modelling in the Figs.6,7.

Results (26-29a) impose important *restrictions* on both maximal frequency generating new starting IN and maximal IN dimension which limits a single IN, and whose ending node may initiate this frequency. For example, $\gamma = 1$ corresponds to

$(\beta^i_{n+k-1} t) \approx 0.423 rad (24.267^o)$, with $\beta^*_i(t_{n+k}) \cong -0.6 \beta^i_{n+k-1}$. (30)

Here $\beta^*_i(t_{n+k}) \cong \alpha^m_{lo}(t_o)$ determines maximal frequency $\omega^*_m$ of fluctuation by the end of optimal movement: $\alpha^m_{lo}(t_o) \cong -0.6\beta^i_{n+k-1}$, where $\alpha^m_{lo}(t_o) = \alpha_{1o}(t_o)(\gamma^\alpha_{12})^{N_m}$.

Starting speed $\alpha_{1o}(t_o) = 0.002 \sec^{-1}$, $m = 14$, determines $\alpha_{14o}(t_o) = 0.00414 \sec^{-1}$, $\beta_{14} \cong 0.0069 \sec^{-1}$. (30a)

The new macromovement starts with that initial frequency. This newborn macromodel might continue the consolidation process of its eigenvalues, satisfying the considering restrictions on invariants and cooperative dynamics up to ending the consolidations and arising the periodical movements.

This leads to cyclic *micro-macro functioning* [59] when the state integration alternates with state disintegration and the system decays with possible transformation of observable virtual process to the evolving information-certain process.

Encoding information units in the IN code-logic, and observer's computation using this code.
*The code, program, and information quality of code-logic*
1. Observer code serves for common external and internal communications, allowing encoding different interactions in universal information measure, and conducts cooperative operations both within and outside which unite the observers.
Each triplet unit generates three symbols from three segments of information dynamics and one impulse-code from the control, composing a minimal *logical code* that encodes this elementary physical information process.
The control joins all three in a single triple unit and transfers this triple code to next triple, forming next level of the IN code. The IN triplet's dynamic space-time connection holds information minimum of three triplet's logic structures.
Each information unit has its unique position in the time-spaced information dynamics, which defines the exact location of each triple code in the IN. Even though the code impulses are similar for each triplet, their time-space locations allows the *discrimination* of each code and the formed logics, distinction both codes and its units.
The IN code includes digital time intervals, encloses code's geometry, which depends on the digits' collected information. The shortened process intervals in the IN condense the observing information in rotating space-time triplets' knots (Figs.4,7), whose nodes cooperate in the IN information structure.

The observing process, chosen by the observer's (0-1) probes, following the information logic integration, determines the IN information code units that encode the IN code-logic.

The information path functional collects the information units, while the IN performs logical computing operations using the doublet-triplet code of the observer's created program, which is specific for each observer, and therefore is self-encrypting. Such operations, performed with the entangled information units memorized in its curvature, model a quantum computation [52, 53,60].

The operations with classical information units, which observer cooperates from quantum information units and runs the units in the IN, model a classical computation.

An observer that unites logic of quantum micro- and macro- information processes enables composing quantum and/or classical computation on different IN levels.

The program holds a distributed space-time processing generated by the observer information dynamics.

Information emanated from different IN nodes encloses distinctive quality measure and logic, which encodes the observer IN *genetic code*. Triplet is elementary *logical* unit holds the IN genetic code which encloses helix geometrical structure (Figs.5, 6) analogous to DNA [54,55, 56, 33,47].

The genetic code can reproduce the encoded system by decoding the IN final node and specific position of each node within IN structure. *This naturally encodes the impulse interaction with environment.*

2. The observed information *specific* quality for each observer IN depends on information density $N_b^{sc}$, defined by the number of information units (bits) that each of this information unit encodes (compresses) from any other source-code. Since each triplet's bit encodes 3 bits, it information density is $N_b^1 = 3$. A following triplet also encodes 3 bits, but each of its bit encodes 3 bits of the previous triplet's bits. Thus, the information density of such two triplets is equal to $N_b^2 = 9$, and so on. Hence, for the $m$-th triplet we have $N_b^m = 3^m$ bits of this $m$-th triplet which encodes $3^m$ bits from all previous triplet's codes. The IN's final node with $m = n/2$ has $N_b^m = 3^{n/2}$, determined by the process' dimension $n$.

The information density, related to the IN's level of its hierarchy, measures also the *value* of information obtained from this level. For such a code, its information density also measures its valueability.

*For example*, an extensive architecture of an ARM chip provides the enhanced code density: it stores a subset of 32-bit instructions as compressed 16-bit instructions and decompresses them back to 32 bits upon execution.

In each particular IN, the triplet elementary logical unit-cells self-organize and compose the observer Logical Structure, satisfying the limitations.

The Emerging the observer's intelligence

The observer intelligence emerges on the path from staring observation, virtual observer, creation microprocess, bits, information macroprocesses, and nested networks (IN) with growing quality of information and the nested logic.

These self-generate the observer selective actions, ability of their prediction, the IN concurrent renovation, and extension to complex self-built IN domains enclosing maximal quality of condense information and its logic.

Emerging the observer's intelligence includes the following information components.

1. The observer's selective actions which evaluate the current information cooperative force initiated by the free information attracting new high-quality information. Such quality delivers a high density-frequency of the related observing information through the selective mechanism. These actions engage acceleration of the observer's information processing, coordinated with the new selection, quick memorizing and encoding each node information with its logic and space-time structure, which minimizes the spending information and complexity. The observer's optimal multiple choices, implement the minimax self-directed strategy through the cooperative force emanated from the IN integrated node.

2. Predicting selection information mechanism carries selective action through the time interval of the impulse control, which requestes the needed information density. Both logical and physical causality minimizes action of the free information and observer "free will" in the optimal prediction. That includes a simple "morality' questions and answers: What is good (Yes) and bad (No) which is the same as information bit. Each of such optimal predicting Bits encloses more information density in IN' higher domain at the same invariant information.

3. The observer's information process, carrying energy, memory and logic of the collected hidden information, conveys the intentional cooperative actions, modeling the selective dynamic efforts that build and organize the observer IN's information space-time dynamic structure. The process information dynamics concurrently renovate the IN by exchanging the requested information with environment, rebuilding the INs, re-encoding, and re-memorizing total recent information.

4. This self-built structure, created under self-synchronized feedback, drives self-organization of the IN and evolving macrodynamics with ability of its self-creation. The free information, arising in each evolving IN level, builds the Observer specific time–space information logical structure.

5. The observer *cognition* emerges from the evolving probabilistic and information logics, intentional ability of requesting, integrating and predicting the observer needed information. The evolving free information builds the Observer specific time–space information logical structure that assembles the growing IN which conserves its cognition.

Such structure integrates the explicit information at each observing level up to the IN highest level, enclosing the observer multilevel cognition.

6. The intelligence is an ability of the observer to build the informational networks and domains, which includes the cognitions. The coordinated selection, involving verification, synchronization, and concentration of the observed information, necessary to build its logical structure of growing maximum of accumulated information, unites the observer's organized intelligence action. The quality of information integrated in the IN node evaluates the information spent on this action. The functional organization integrates the interacting observers' IN levels and domains, which evaluates the amount of quality information memorized in observer IN highest hierarchical level. The cognitive process at each triplet level preempts the memorizing. The quality of information memorized in an ended triplet of the observer hierarchical informational networks and domains measures level of the observer intelligence. Maximal level of emerging intelligence measures maximal cooperative complexity, which enfolds maximal number of the nested INs structures, memorized in the ending node of the highest IN.

All information observers have different levels of intelligence which classify the observer by these levels.

The self-directed strategy develops multiple logical operations of a self-programming and computation which enhance collective logic, knowledge, and organization of diverse intelligent observers.

The intelligent actions and the intelligence of different observers connect their level of knowledge, build and organizes the observers IN's information space-time logical structure. Increasing the INs enfolds growing information density that expands the intelligence, which concurrently memorizes and transmits itself over the time course in an observing time scale. The intelligence, growing with its time-space region, increases the observer life span, which limits a memory of the multiple final IN ending node in the extended region.

Since whole multiple IN information is *limited*, as well as a total time of the IN existence, the IN self-replication arises, which enhances the collective's intelligence, extends and develops them, expanding the intellect's growth.

The self-organized, evolving IN's time-space distributed information structure models *artificial intellect.*

The invariance of information minimax law for any *information observer* preserves their common regularities of accepting, proceeding information and building its information structure.

That guarantees objectivity (identity) of basic observer's individual actions with *common information mechanisms.*

The common mechanism enables creation of *specific* information structures for each particular observed information, with individual goal, preferences, energy, material carriers, and various implementations.

Multiple communications of numerous observers (by sending a message-demand, as quality messenger (qmess) [43], enfolding the sender IN's cooperative force, which requires access to other IN observers allowing the observer to increase the IN personal intelligence level and generate a collective IN's logic of the multiple observers.

This not only enhances the collective's intelligence but also extends and develops them, expanding the intellect's growth. Which attributes define an intelligent observer?
We believe two attributes: levels of cognition and intelligence, each of which requires definition and information measure.

Specific of the Information Cognition.

The Observer logical structure possesses both virtual probabilistic and real information causality and complexity [51,58].

A virtual observer, forming the rotational space-time displacement of the impulse' opposite actions during virtual observation, starts accumulating virtual information through its temporal memorizing and the probabilistic logic, which initiates cognitive movement. After emerging the memorized bit during the observation, the rotation develops information form of double helix movement (Figs.4,5).

The rotating cognitive movement connects the impulse microprocess with the bits in macroprocess, composing triple macrounits through the created free information, which arranges each evolving IN. Then, the growing IN's levels quality information in an IN ending triplet integrates multiple nested IN's information logic in information domains.

The cognitive movement, at forming each nodes and level, processes a temporary loop (Fig.6) which might disappear after the new formed IN triplet is memorized. The rotating process in the *coherent* loop *harmonizes speeds-information frequencies* at different levels analogously to Efimoff's scenario, which can be temporal after new formed IN triplet is memorized. The loop includes Borromini knot and ring. The observer's cognition assembles common units through the multiple resonances at forming the IN-triplet hierarchy, which accept only units that each IN node concentrates and recognizes. The loop rotates the thermodynamic process (cognitive thermodynamics) with minimal Landauer energy, which performs natural memorizing of each bit on all evolution levels.

The cognitive process encodes the merged rotating double spirals, whose sequential knots, memorized on the process' ending reversible sections, compose the evolving information logic.

The cognitive actions model the correlated inter-actions and feed-backs between the IN levels, which controls the highest domain level. Both cognitive process and cognitive actions emerge from the evolving observations, which maintain the cognitive functions' emerging properties and encodes the cognitive logic information language.

Thus, the cognition emerges in two forms: a virtual rotating movement processing temporal memory, and following real information process' mechanisms rotating the double helix geometrical structure, which concurrently places and organizes the observing information bits in the IN nodes, whose sequential knots memorize information causality and logic.

These processes start with the elementary virtual observer and emerging bit at microlevel which memorizes the prehistory and participates in evolving information observer.

Results [57] confirm that cognition arises at quantum level as "a kind of entanglement in time"…"in process of measurement", where…"cognitive variables are represented in such a way that they don't really have values (only potentialities) until you measure them and memorize", even "without the need to invoke neurophysiologic variables", while "perfect knowledge of a cognitive variable at one point in time requires there to be some uncertainty about it at other times". Moreover, this analysis shows the both cognition and intelligence have information nature.

Specific of Information Intelligence and estimation its information values

The causal probabilities, following from Kolmogorov-Bayes probabilities' link, start the Markovian correlation connection with minimum of tree probabilistic events. An observer integrates the observing events in the information networks, which accumulate the nested triple connections, depending on the IN information invariant properties.

Each IN has invariant information geometrical structure and maximal number of nodes-triple bits, whose ability of cooperating more triplet nodes limits a possibility of the IN self-destruction by arising a chaotic movement.

The intelligence measures the *memorized ending node of the observer IN highest levels*, while cognitive process at each triplet level preempts its memorizing. This means each memorizing involves the cognition. The information measure of intelligence is *objective for each particular observer* while the IQ is an *empiric subjective* measure.

The theory shows that an observer, during current observation, can build each IN with maximum 24-26 nodes with average $3^{26}$ bits and enfold maximum of 26 such IN's.

Since each IN following level integrates information from all the IN previous levels, it measures the relative information quality of this level, which exposes information relationships between the levels in the triplet forms.

Because the subsequent relationships have been enclosed by the cognitive rotating mechanism, they formalize a causal comparative meaning getting for the observation.

The *Observer Intelligence* has ability to uncover causal relationships enclosed in the evaluated observer $N_{o1} = 3^{26} \times 26 bits$ networks bits. That requires not only build each of $N_{1I} = 26$ INs but also sequentially enfold them in a final node whose single bit accumulates $N_{oI}$ bits, which evaluate

$$N_{oI} = (3^{26}) \times 26 = 2,541.865.828329 \times 26 = 66,088.511.536.554 \cong 66.1 \times 10^8 = 6.61 \times 10^9 \text{ bits.} \quad (31)$$

However, since each IN node holds single triplet's information, the final IN node' bit keeps the triple causal information relationship with density $D_{oI} = N_{oI} / bit$ per bit.

To support the IN node impulse feedback communication with the requested attracting information, this node requires information density:

$$i_{md} \cong 1.8 \times 10^{14} Nat/\sec = 1.44 \times 1.8 \times 10^{14} bit/\sec, \quad (32)$$

where each such bit accumulates $N_{oI}$. Thus total information density of the observer final IN bit:

$$i_{do} \cong 1.44 \times 1.8 \times 10^{14} \times (3^{26}) \times 26 bit/\sec \quad (33)$$

evaluates the intelligent observer's information density.

With this density, the intelligent observer can obtain maximal information from the EF through the impulse interaction with entropy random process during time observation $T$.

Let us evaluate the EF according to (Sec.3.7,3.8)[33a]:

$$I_e = 1/8\ln[r(T)/r(t_s)] \approx 1/8\ln(T/t_s), T = m_N t_s.$$

Here $m_N$ is a total number of the IN nodes needed to build intelligent observer $t_s$ is time interval of invariant impulse which is also invariant. At $m_N = 26 \times 26$ it allows estimate $I_e = 1/8\ln 26^2 = 11.729 Nat$.

Therefore, the intelligent observer needs $N_i \cong 12$ invariant impulses to build its total IN during time interval of observation $T$.

Comments. The human brain consists of about 86 billion neurons [62], which approximately in 14 times exceeds $N_{oI}$ (8.1), if each single bit of the cognition commands each neuron?

Nonetheless it agree with this estimation, if each neuron builds own IN with about five-six triplets (with levels $3 + 2^4 = 11, or 3 + 2^5 = 13$), while ending triplet bit condenses this $N_{oI}$.

Ability of a neuron building a net concurs with [62] and [61]. If its thrue, then $N_{oI}$ measures information memory of human being. •

According to estimation [63, others] maximal information in Universe approximates

$$I_U \cong 3 \times 10^{29} Nat = 4.328 \times 10^{29} bits, \quad (34)$$

from which each invariant intelligent observer can get $I_{ob} \cong 6.61 \times 10^9 bits$.

To obtain all $I_U$ information, number $M_{ob} \cong 1,527 \times 10^{16}$ of such intelligent observers it is needed.

Each IN triplet node may request $I_m \cong (3.45 - 2.45) bits$ which measures this IN level of quality information that memorizes the node bit.

Depending on each IN's levels $N_{1I} = 26$, such node's level accumulates average information between $I_m bits$ and $N_{om} = 3^{26} bits$.

Quantity $N_{oI}$ (31) measures invariant transformation to build the extreme IN nodes' structure during the observation, which transforms a probable observing process to information process in emerging information observer with intelligence. The initial probability field of random processes, evaluated by entropy functional, contains potential information which an intelligent observer can obtain through the invariant transformation. Information threshold $N_{oI}$ limits level of intelligence the intelligent observer satisfying the minimax variation principle. The intelligent (human) observer can overcome this threshold requiring highest information up to $I_U$. Such an observer that conquers the threshold possess a supper intellect, which can control not only own intellect, but control other observers.

Multiple joint supper intellectual observers can form a super-intellectual system (with $I_U$) controlling Universe, or would destroy themselves and others.

In an intelligent machine, collecting the observing information, the emerging invariant regularities of the mimax law limits the AI observer actions.

The interacting intelligent observers through communication

Since any information intelligent observer emerges during the evolving interactive observation, important issue is interaction of such observers in a common observation, which preserves the invariant information properties.

Suppose an intelligent observer sends a message, containing its information encoding a meaning, which emanates from this intelligent observer's IN node. Another intelligent observer, receiving this information, should be able to read the message, recognize its meaning, select and accept it if this information satisfies the observer needed information quality being memorized using its the DSS code. Fulfillment of these five issues is the subsequent.

Since the DSS code is invariant for all information observers, the sending observer can encode its message in this code, and the observer-receiver can decode and read the message information. The DSS unifies the information language of communicating observers. The code logic and length depend on the sending information, which possibly is collected from the observer–sender's different INs nodes. Recognition of the needed information initiates the observer request.

The observer request for growing quality of needed information measures the specific qualities of free information emanating from the IN distinctive node that need the information compensation.

The recognition involves copying and cooperation of the comparative qualities enclosed in the observer –receiver distinctive INs nodes. (The copies can provide the temporary integral mirrors of the microprocess transitive impulses.)

The message recognition includes cognitive coherence with the reading information, which allows it selection and acceptance. With the considered selective requirements and limitations .

The message acceptance includes cooperation of the message quality with the quality of an IN node enclosed in the observer –receiver IN structure. If the cooperative information coheres in the cognitive loop, the message can be accepted and memorized in the receiver's IN. The cognitive mechanism, capturing the coherent information, rotates it to a succeeding IN level that this meaning accumulates by memorizing it. That allows the intelligent observer to uncover a meaning of observing process using the common message information language and the cognitive acceptance, which are based on the qualities of observing information memorized in the IN hierarchy.

The intelligent observer recognizes and encodes digital images in message transmission, being self-reflective enables understanding the message meaning.

Understanding of receiving message. How does a biological observer having a brain' neurons accept a message?

Understanding the message describes the information formalism, which includes copying copying the accepted message on the cognitive moving helix which temporary memorizes it as triplets' entropies in a virtual IN structure.

This converting mechanism includes a compression of observing image in virtual impulse ending the virtual IN.

The virtual impulse, holding the entropy equivalent of the image information, moves the cognition scanning helix along the observer's INs until its negative curved step-up action, carrying the entropy equivalent of energy, will attract a positive curvature of the IN node bit's step-down action. The forming Bit encloses the equivalent energy's quality measured by its entropy value. When the IN bit's step-down action interacts with the moving image's step-up action, it injects energy capturing the entropy of impulse' ending step-up action. This inter-action models 0-1 bit (Fig.2A, B).

The opposite curved interaction provides a time–space difference (an asymmetrical barrier) between 0 and 1 actions, necessary for creating the Bit. The interactive impulse' step-down ending state memorizes the Bit when the observer interactive process provides Landauer's energy with maximal probability (up to a certainty). Such energy drives the cognitive helix movement having maximal energy quality (minimal entropy production), which can be called "cognitive thermodynamic process". It allows spending minimal cognitive quantity equal to triplet structure Landauer's energy ln 2 for erasure the observing bits and memorizes each bit by the equal neuron information bits. (This means, at forming a triplet, this energy can be spent on memorizing a third bit before it gets asymmetrical structure needed for memorizing.)

If a cooperating information coheres with the cognitive, the message can be accepted and memorized in the receiver's IN. The forming IN bit encloses the equivalent energy's quality measured by its entropy value.

Therefore, the cognitive thermodynamic process practically has not thermodynamic cost, which models of a cognitive software with the minimal algorithmic complexity. The important coordination of an observer external time-space scale with its internal time-space scale happens when an external step-down jump action interacts with observer inner cognitive thermodynamics' time-space interval, which, in the curved interaction measures the difference of these intervals. Understanding the receiving information includes classifying and selecting such information that concurs with this observer's memorized meaning of other comparative images. Thus, cognitive movement, beginning in virtual observation, holds its imaginary form, composing an entropy microprocess, until the memorized IN bit transfers it to an information macro movement. That brings two forms for the cognitive helix process: imaginary reversible with a temporal memory, and real-information moving by the irreversible cognitive thermodynamics, which is memorizing incoming information.

Explaining the mechanism modeling the message acceptance and understanding requires admitting, first, that the developed math-information formalism is considering as software controlling a brain structures–a hardware.

Connecting them requires a converting mechanism, which copies an observation and starts action on hardware.

Those perform different sensors bending neurons, which make a mirror virtual copy of the observing image-message (analogous to transition impulse (Sec.3.6) *on* the cognitive moving helix.

For example, eyes scanning a TV screen, integrate the screen picks in a reflected image, accompany with the ears accumulating sound of the seeing image.

The observer may not need to memorize each currently observed virtual image, which is reflected temporally in some sequence. Accordingly, such multiple virtual copies are formed by temporary triplets units composing a temporal collective IN whose ending node encloses a virtual impulse entropy' bit .

Such virtual IN with temporal memory is forming in a reversible logical process without permanent memory, which comprises a part of observation process (when Kolmogorov-Bayes probabilities link the triple events).

This converting mechanism includes a virtual compression of observing image in virtual impulse ending a virtual IN.

Specifically, the time–space difference between 0 and 1 actions determines formula (Sec.3.6.5.4):

$$\Delta t_{10} = |t_{o1}| - t_o \cong 0.0250 - 0.01847 = 0.00653 Nat.$$

This formula with the formula, connecting time and space measures for interacting impulse $[\tau]/[l] = \pi/2$, allows finding the equivalent difference of the space intervals in Nat: $\Delta l_{10} = 2\Delta t_{10}/\pi, \Delta l_{10} \to 0.00415 Nat$.

When the image bits memorize the observer IN's specific node, this node information quality and its precise position allow the observer to recognize this image information among other distinctive information qualities. The nodes positions already contain the observer INs' memorized information qualities.

Recognizing a collective information image associates with understanding it by that observer enclosed information. Understanding implies that the observer can classify and select such information according to this observer's memorized *meaning* among other comparative images.

The information model of understanding a message includes:

1. Sensor conversion of the observing message-image with building the virtual IN of the message, as a virtual mirror copy of the image collective information, which the IN compresses it in the virtual impulse.

2. Copying on the moving cognitive helix, which scans the observer' INs information enclosed from all IN domain levels.

3. Interaction of a sensor' neuron impulse, initiating yes-no actions, with the virtual impulse of the image through its yes-action, which injects an energy capturing the virtual entropy of the impulse' ending step-up action when the scanning helix cognitive movement contacts the observer IN node that provides this energy.

4. Memorizing the yes-no interactive Bit by the neuron interactive impulse' step-down no-action through the cognitive dynamic interactive process which provides Landauer's energy for erasure the observing image. That builds a mirror's bits memory, which decodes the message-image.

In this neuron-message communication, the neuron yes-action, capturing the virtual impulse's ending step-up action, connects it with this neuron' no-action, which provides step-down action memorizing the message through the cognitive dynamic energy. Thus, the neuron curving interaction connects virtual and real actions, which actually binds the cognitive software with the brain hardware structure.

5. The memorized information bit stops the scanning cognitive mechanism on such IN level, where this information is understood through its observer' IN recognition. That ends the process of understanding of a current message.

Scanning the observer understood meanings allows recovering the message semantic, and then encoding the required one in a sending message. The virtual impulse of cognitive interaction provides logical Maxwell demon, while it transformation to the memorized IN information runs physical IMD.

It presumes that neuron's yes-action starts its impulse entropy microprocess until the neuron' no-action, interacting with the observer macroprocess via the IN node bit by the jumping no-action, memorizes the incoming image in the observer IN structure. Thus, cognitive movement, beginning during virtual observation, holds its imaginary form, composing entropy microprocess, until the memorized IN bit transfers it to an information macro movement.

That brings *two forms* for the cognitive helix process: *imaginary reversible without memory, and real-information moving by the irreversible cognitive thermodynamics memorizing incoming information.*

The imaginary starts with the neuron yes-action and ends with the neuron no-action at the ending state of the neuron impulse, while the real starts with the IN bit yes-action memorizing the accepted bit, which processes the cognitive thermodynamics continuing moving the cognitive helix irreversibly.

The threshold between the imaginary and real cognition holds memory and energy of the cognitive thermodynamics.

This is how the observing quantity and quality of interacting information emerge in observer as the memorized quality encoding the observer cognition.

The observer self-controlling evolution

1. Formal analysis the stages and levels of the evolution, starting from multiple interacting impulses of observable random field holding an energy, shows that each following level enables self –generate next level and self-form a nested time-space pyramidal evolution' stages-an hierarchical network structure of decreasing uncertainty-entropy.

2. The inter-action of the impulses with the field' energy, cutting the impulse entropy, develops its conversion to the emerging bit of information-as and elementary information observer, which self-participates in evolution interaction.
The evolution levels self-build a progressive rotating movement, which sequentially involves the levels and stages in collective rotation that successively adjoins the nearest levels enclosing each previous in the following next.
3. The evolution of reducing uncertainty self-grows a quality of cooperation, which determines the level's time-space location ending each level's network that memorizes it in a quality of the stage information.
The continued interaction delivers new level's information through each level's feedback with other levels along the hierarchy of levels and stages down to the field.
4. Transfer to following stage requires overcoming a threshold between stages that memorizes the stage ending level of quality. Since acquision information through its interactive binding increase the IN parameter $\gamma_{m+1}^{\alpha}$, it decreases the frequency with tendency of growing information quality in evolving IN. That delivers information forces, which enable automatically overcome each following threshold and transfer to next stage of the quality.
5. Each triple dynamically memorized stages, involved in collective rotation, cooperates in a domain with growing cooperative quality of information. The highest domain's stage quality of information measures the observer intelligence, which can progressively grow through delivering increasing interactive cooperative qualities.
6. The rotating mechanism, self-cooperating sequentially the levels and domains information, emerging in the evolution, models the observer cognition which self–forms the observer chain of nested information structure up to developing the growing intelligence. The entire rotation controls the vertex angle of the cone (Fig.4) at a highest level of the domain intelligence with maximal cooperative quality. Cognition at lower levels and stages controls the related angle of rotation in these locations, while each local feedback can change it and renovate all hierarchy.
Since the rotation starts in the observer microprocess, cognition theoretically emerges from this microlevel up to the macrolevels. The cognition measures the cooperative attracting qualities binding each networks level, stage, and the domain. The cognitive mechanism controls observer's all requested information, its acquisition, and structuration along existing and currently building observer hierarchy.
According to Webster dictionary: "cognition is the activities of thinking, understanding, learning, and remembering". In other dictionary: "conscious is intellectual act 'conflict between *cognitions*'.
We define it as an activity moving the observer integration of information up to growing intelligence.
The emerging observer time-space starts the activity, which curves primary interaction, initiating rotation.
Cognitive *mechanism* assembles the observer hierarchy of the current information levels, stages, and domains though the available attracting information (emanated from previous IN or/and supplied by the requested information via the feedback).
Assembling is a cooperative action intents to overcoming the threshold, which the attracting information measures. Current integral information of IPF measures amount of each cognitive action's attracting information.
The observer's cognition models the observer hierarchical rotation mechanism, which enables transferring the observer through the stages to overcome the stage threshold.
The mechanism rotating movement charaterizes its intensity potential $P_{in}(i)$ measured by multiplication the current $(i)$ rotating moment $M(i)$ on angular speed $\omega(i): P_{in}(i) = M(i) \times \omega(i)$.
7. The specific constrains imposed on each level, stage, and domain limit each of its IN.
When the observer attempts to increase information quality by overcoming-destroying its specific constrain, the accidentally arising singularities enable renovating the observer constrain location, bringing new original (personal) level, or stage, and the domain quality distinctive from the evolution of information dynamics within the constrains. Other non-cooperating singularities contribute the random field, which self-closes a current chain of the observer *personal* evolution.

8. Such evolution develops without any preexisting laws following each observer trajectory, which includes all its levels, stages and domains, and potential thresholds between them.

The observer regularity rises in impulse observation from the self-created virtual up to real observers, where each impulse is max-min action transferred to the following through mini-max action. This variation principle is mathematical information form of the law, which encloses the regularities following from it. The extreme trajectory, implementing that law's mathematical form, releases these reqularities in most general form, letting the observer specific regularities self-develop in its self-evolution which self-creates its laws with extending regularities.

These abilities follow from the chain of virtual, logical, and information causalities, which extreme trajectory includes.

Obtaining the extreme trajectory depends on knowing a mathematical master formula (I.I) integrating the observations, whose extreme brings this trajectory. This formula bases on the main common phenomena in natural world – interactions building our Universe, which characterize Second Law of entropy evolution.

The formula integrates entropies of interacting impulses through entropy functional EF on multiple observations, whose interactions in form of information impulses integrate information path functional IPF.

The formula integrates main forms of observable process through Markov diffusion process (MDF), which models a wide variety of physical, chemical, biological, and even economical interactive processes.

Connection EF and IPF with the MDF parameters (in I.I) makes possible the solution of variation problem describing entropy-information form of observer extreme trajectory, which portray the evolution development for each observer.

9. Each threshold of the evolution level, or stage limits, separates them, and filters the random *varieties*.

The evolution *fitness* distincts the stages which connects the thresholds.

10. The observers, reaching potential threshold on the time-space locality along the process trajectory, but unable to overcome it, will settle between the thresholds and eventually disintegrate.

Evolution automatically selects the observer remaining on the trajectory and eliminates others.

The observer, overcoming such threshold, fits another observer located on this stage, and *adapts* the threshold quality information allowing the evolution development on that stage through the observer *adaptive feedback* above.

11. Interactive acquisition, bringing increasing quality information, allows automatically self–overcoming thresholds decreasing the observer *diversity*.

12. The evolution *hierarchy* defines the evolving nested hierarchical structure of the IN levels, stages and domains.

Evolution *stability* depends on each memorized information of the stage. Observer which unable to cross the stage' threshold stays stable within its stage. That keeps *diversity of the selective and stable observers.*

13. Information attraction, measured quantity of information requested by the IN stage, determines the evolution *potential* of this evolutionary stage. Information complexity of evolution dynamics [58, 59] measures density of collective information enfolded in the IN stage, which defines the information value of the stage' potential of cooperation.

14. Self-encoding information units in the IN code-logic and observer's computation, using this code, serves for common external and internal communications, allowing encoding different interactions in universal information language and conduct cooperative operations both within and outside the domains and observer. That unites the observers.

15. The emergence of time, space, and information follows the emerging evolution information dynamics creating multiple evolving observers with intellect and information mechanism of cognition.

*These results formalize Observer's regularities in a comprehensive information-physical theory, connecting the virtual quantum world with the physical classic and relativistic world.*

***Summary of the nested evolutionary levels in impulse observation self-creating law of evolution:***
-Reduction the process entropy under probing impulse, observing by Kolmogorov-Bayesian probabilities link, that increases each posterior correlation;

-Impulse maxmin-minimax principle allowing primary to separate off the probes not decreasing the process entropy, which supports the decrease of process uncertainty;

-Start the observer's time intervals, temporal memory of connected correlations, the space intervals at the entaglement, and self-forming rotating movement. Starting the virtual self-observation limits a threshold of the impulse's connection (Sec.II.2) which decreases number of potentially multiple virtual observers not overcoming the threshold;

-The impulse cutoff correlation sequentially converts the cutting entropy to information that memorizes the probes logic, participating in next probe-conversions;

-The states of cutting correlations hold hidden process' inner connections, which initiate growing correlations up to running the superposition, starting microprocess with entanglement and conversion of the cutting entropy in information. Both entangled anti-symmetric fractions appear simultaneously with starting space interval. The correlations binding this couple or triple with maximal probability are extremely strong. The correlated conjugated entropies of the entangled rotating virtual impulses are no separable and no real action between them is possible;

-The interaction curves the inter-active impulse along which runs the cutting entropy.
That creates the inner impulse rotating movement. Cutting the impulse time-correlation reveals the connected imaginary time correlations and related conjugated entropies moving in the rotating microprocess;

-The imaginary microprocess ends with entangling entropy volume, information microrocess emerges with providing an energy, killing that entropy and memorizing the classical Bit by the end of external impulse. Or information Bit, as the memorized two qubits, can be produced through interaction, which generates the qubits contained by a material -device (a conductor-transmitter) that preserves curvature of the transitional impulse inside a closed device;

-The microprocess within a Bit's formation, connecting both imaginary entropy and information parts in rotating movement, also binds multiple units of Bits in collective movement of information macroprocess, which integrates information path functional (IPF, while the cutting correlations connect the EF-IPF);

Evaluation of the frequency creation of both the opposite curvature of interacting impulses and total path to emerging Bit.

-The information marodynamics are reversible within each EF-IPF extremal segment, imposing the dynamic constraint on IMD Hamiltonian; irreversibility rises at each constrain termination between the segments.
The IMD Lagrangian integrates both the impulses' and constrains' information on time space-intervals;

-A flow of the moving cutoff Bits forms a unit of the information macroprocess (UP), whose size limits the unit's starting maximal and ending minimal information speeds, attracting new UP by its free information.
Each UP, selected automatically during the minimax attracting macro-movement, joins two cutoff Bits with third Bit, which delivers information for next cutting Bit;

-Minimum three self-connected Bits assemble optimal UP-basic triplet whose free information requests and binds new UP triplet, which joins three basics in a knot that accumulates and memorizes triplets' information;

-During macro-movement, multiple UP triples adjoin the time-space hierarchical network (IN) whose free information's request produces new UP at higher level' knot-node and encodes it in triple code logic. Each UP has unique position in the IN hierarchy, which defines exact location of each code logical structures. The IN node hierarchical level classifies quality of assembled information, while the currently ending IN node integrates information enfolding all IN's levels;

-New information for the IN delivers the requested node information's interactive impulse impact on the needed external information, which cutoff memorized entropy of Data. The appearing new quality of information concurrently builds the IN temporary hierarchy, whose high level enfolds information logic that requests new information for the running observer's IN, extending the logic up to the IN code;

-The emergence of current IN level indicates observer's information surprise thru the IN feedback's interaction with both external observations and internal IN's information, which self-renovate its information;

-The time-space information geometry shapes the Observer asymmetrical structure enclosing multiple INs;

-The IPF maximum, integrating unlimited number of Bits' units with finite distances, limits the total information carrying by the process' Bits and increases the Bit information density in a rising the process dimensions. While each Bit preserves its information, the growing information, condensed in the integrated Bit with a finite impulse geometrical size, strengthens the Bit information density, running up to finite IPF maximal information at infinite process dimension;

-The macroprocess integrates both imaginary entropy between impulses of the imaginary microprocesses and the cutoff information of real impulses, which sequentially convert the collected entropy in information physical process during the macro-movement;

-The observation processes, they entropy-information and micro-macroprocesses are Observer-dependent, information of each particular Observers is distinct;

-The invariant information minimax law leads to equivalent information regularities for different Observers. By observing even the same process, each Observer gets information that needs its current IN during its optimal time-space information dynamics, creating specific (personal) information process;

-The equations of information macrodynamics, as the EF extremal trajectories, describe the IPF information macroprocess, which averages all the microprocesses and holds regularity of observations under the maxmin-minimax impulses. At infinitive dimension of process $n \to \infty$, the information macroprocess is extremals of both EF and IPF, while EF theoretically limits IPF. The limited number of the process macrounits, which free information assembles, leads to limited free information connecting the impulse's microprocesses and macroprocesses;

-When the information Observers unify-integrate their multiple information reality facts in EF-IPF variation problem, the VP solution determines regularities of the multiple observation, which become independent of each particular Observer. Such an information-physical law is established when the EF Hamiltonian changes maximum to minimum [Sec.I.3.8]. It minimizes the EF, maximizes probability on the EF trajectories, and reaches causal deterministic reality. The multiple randomly applied deterministic (real) impulses, cutting all process correlations, transforms the initial random process to a limited sequence of independent states;

-The macro-movement in rotating time–space coordinate system forms Observer's information structure confining its multiple INs that determine the Observer time of inner communication with time scale of accumulation information;

-Each Observer owns the time of inner communication, depending on the requested information, and time scale, depending on density of the accumulated information;

-The current information cooperative force, initiated by free information, evaluates the observer's *selective* actions attracting new high-quality information. Such quality delivers high density-frequency of related observing information through the selective mechanism. These actions engage acceleration of the observer's information processing, coordinated with the new selection, quick memorizing and encoding each node information with its logic and space-time structure, which minimizes the spending information and IN cooperative complexity;

-The observer optimal multiple choices, needed to implement the minimax self-directed strategy, evaluates the cooperative force emanated from the IN integrated node;

-The self-built information structure, under the self-synchronized feedback, drives self-organization of the IN and evolution macrodynamics with ability of its self-creation;

-The free information, arising in each evolving IN, self-builds the Observer specific time–space information logical structure that conserves its "conscience" as intentional ability to request and integrate the explicit information in the observer IN highest level, which measures the Observer Information Intelligence;

-The coordinated selection, involving verification, synchronization, and concentration of the observed information, necessary to build its logical structure of growing maximum of accumulated information, unites the observer's organized intelligence action. The IN hierarchical level's amount quality of information evaluates functional organization of the intelligent actions spent on this action. Intelligence of different observer integrates information of their IN's ended nodes;

-The IN node's hierarchical level cooperates the communicating observers' level of integrated knowledge;

- Increasing the IN enfolded information density accelerates grow the intelligence, which concurrently memorizes and transmits itself over the time course in an observing time scale;
- The intelligence, growing with its time interval, increases the observer life span;
- The self-organized, evolving IN's time-space distributed structure models artificial intellect.